%% file: lawson-review.tex
\documentclass[%
 aip,pop,groupedaddress,
 amsmath,amssymb,
 reprint,%
]{revtex4-1}

\usepackage{graphicx}
\usepackage{dcolumn}
\usepackage{bm}

\usepackage[utf8]{inputenc}
\usepackage[T1]{fontenc}
\usepackage{mathptmx}
\usepackage{etoolbox}

\usepackage{longtable}
\usepackage{rotating}
\usepackage{tabularx}
\usepackage{siunitx}
\usepackage{amssymb}
\usepackage{amsmath}
\usepackage{physics}
\usepackage{verbatim}
\usepackage{makecell}
\usepackage{bm}
\usepackage{float}

\makeatletter
\def\@email#1#2{%
 \endgroup
 \patchcmd{\titleblock@produce}
  {\frontmatter@RRAPformat}
  {\frontmatter@RRAPformat{\produce@RRAP{*#1\href{mailto:#2}{#2}}}\frontmatter@RRAPformat}
  {}{}
}%
\makeatother
\begin{document}


\title{Progress toward fusion energy breakeven and gain as measured against the Lawson criterion}
\author{Samuel E. Wurzel}
\email[]{sam.wurzel@hq.doe.gov}
\author{Scott C. Hsu}%
\affiliation{Advanced Research Projects Agency-Energy, U.S. Department of Energy, \mbox{Washington, DC 20585}}


\begin{abstract}
The Lawson criterion is a key concept in the pursuit of fusion energy, relating the fuel density $n$,
pulse duration $\tau$ or energy confinement time $\tau_E$, and fuel temperature $T$ to the
energy gain $Q$ of a fusion plasma.  
The purpose of this paper is to explain and review the Lawson criterion and to provide a compilation of achieved parameters for a broad 
range of historical and contemporary
fusion experiments. Although this paper focuses on the Lawson criterion, it is only one of many equally important factors
in assessing the progress and ultimate likelihood
of any fusion concept becoming a commercially viable
fusion-energy system.  Only experimentally measured or inferred values of
$n$, $\tau$ or $\tau_E$, and $T$ that have been published in the 
peer-reviewed literature are included in this paper, unless noted otherwise. For
extracting these parameters, we discuss methodologies that are necessarily 
specific to different fusion approaches (including magnetic, inertial, and 
magneto-inertial fusion).  This paper is intended to serve as a reference for fusion researchers and a tutorial for all others interested in fusion energy.
\keywords{fusion energy, nuclear fusion, Lawson criterion, triple product}
\end{abstract}

\maketitle

%

\section{\label{sec:intro}Introduction}

In 1955, J.~D. Lawson identified a set of necessary physical conditions for a ``useful'' fusion system.\cite{1955Lawson} 
By evaluating the energy gain $Q$, the ratio of energy released by fusion reactions to the delivered energy for heating and sustaining the fusion fuel, Lawson concluded that for a pulsed system, energy gain is a function of temperature $T$ and the product of fuel density $n$ and pulse duration $\tau$ (Lawson used $t$).
When thermal-conduction losses are included in a steady-state system (extending Lawson’s analysis), the power gain is a function of $T$ and the product of $n$ and energy confinement time $\tau_E$.
We call both these products, $n\tau$ and $n\tau_E$, the {\em Lawson parameter}.
The required temperature and Lawson parameter for self heating from charged fusion products to exceed all losses is known as the {\em Lawson criterion}.
A fusion plasma that has reached these conditions is said to have achieved
{\em ignition}. Although ignition is not required for a 
commercial fusion-energy system, higher values of energy gain will generally yield more attractive economics, all other things being equal. If the energy applied to heat and sustain the plasma can be recovered in a useful form, the requirements on energy gain for a useful system are relaxed.

Lawson's analysis was declassified and published in 1957\cite{Lawson_1957} and has formed the scientific
basis for evaluating the {\em physics progress} of fusion research toward the key milestones of plasma energy breakeven and gain. 
Over time, the Lawson criterion has been cast into other formulations,
e.g., the {\em fusion triple product}\cite{mcnally73,mcnally77}
($nT\tau_E$) and ``p-tau''  (pressure $p$ times $\tau_E$), which have the same dimensions (with units of m$^{-3}$\,keV\,s or atm\,s) and combine all the relevant parameters 
conveniently into a single value. However,
these single-value parameters do not map to a unique
value of $Q$, whereas unique combinations of
$T$ and $n\tau$ (or $n\tau_E$) do.
Various plots of the Lawson parameter, triple product, and ``p-tau'' versus year achieved or versus $T$ have been published for subsets of experimental
results,\cite{Braams_Stott_2002,Wesson_2011,Parisi_Ball_2019,FESAC_2008} but to our knowledge there did not exist a comprehensive compilation of 
such data in the peer-reviewed literature that spans the major
thermonuclear-fusion approaches of magnetic confinement fusion (MCF),
inertial confinement fusion (ICF), and magneto-inertial fusion (MIF)\@.  This paper fills that gap.

The motivation to catalog, define our methodologies for inferring, and establish credibility for a compilation of these parameters stems from the prior development of the Fusion Energy Base (FEB) website (\url{http://www.fusionenergybase.com}) by the first author. FEB is a free resource with a primary mission of providing objective information to those,
especially private investors, interested in fusion energy. This paper provides access to the many included plots, tables, and codes, while also providing context for understanding the history of fusion research\cite{Bromberg_1982,Clery_2013,Dean_2013} and the tremendous scientific progress that has been made in the 65$+$ years since Lawson's report.

\begin{figure}[b!]
\includegraphics[width=8cm]{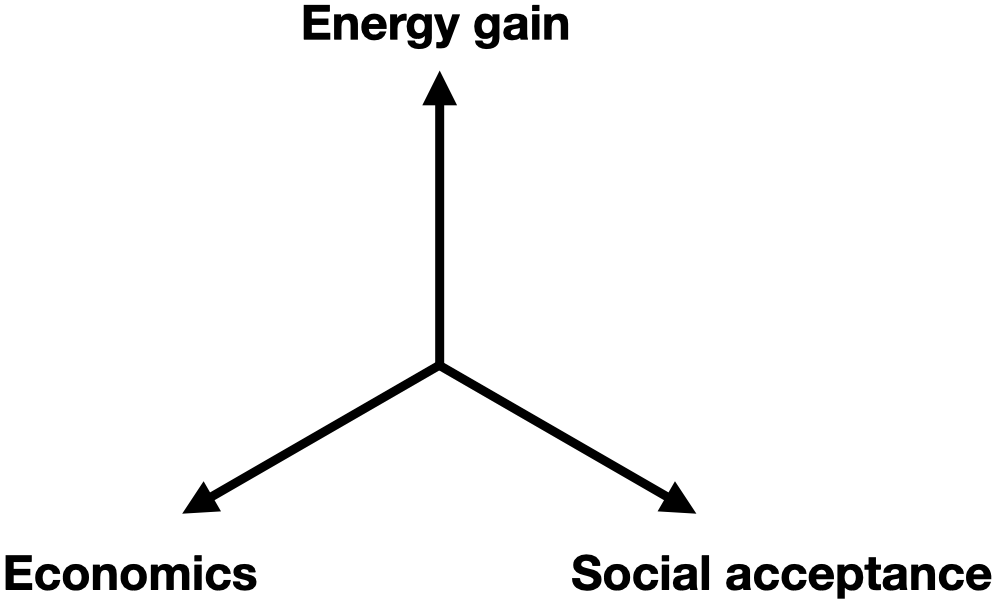}
\caption{Progress towards commercially viable fusion energy requires progress along three equally important axes. This paper focuses only on the axis of energy gain.}
\label{fig:fusion_energy_trinity}
\end{figure}

The combination of $T$ and $n\tau$ (or $n\tau_E$) is a scientific
indicator of how far or near a fusion experiment is from energy breakeven and gain. Achieving high values of these
parameters is tied predominantly to plasma physics and related
engineering challenges
of producing stable plasmas, heating them to fusion temperatures,
and exerting sufficient control. Since the 1950s, these challenges
have driven the development of the entire scientific discipline of
plasma physics, which has dominated fusion-energy research
to this day.
{\em However, we emphasize that there are many additional 
considerations, entirely independent of but equally
important as the Lawson criterion, in evaluating the
remaining technical and socio-economic risks of any fusion approach and the likelihood of any approach ultimately becoming a commercially viable fusion-energy system}.  These include the feasibility, safety, and complexity of the engineering and materials subsystems and fuel cycle 
that impact a commercial fusion system's economics\cite{Handley2021} and
social acceptance,\cite{hoedl19}
as illustrated conceptually in Fig.~\ref{fig:fusion_energy_trinity}. The issues of
RAMI (reliability, accessibility, maintainability, and inspectability)\cite{Maisonnier18} and government regulation\cite{NRC2021,UK2021} impact both the economics and social acceptance.
This paper discusses only the progress of fusion energy along the axis of energy gain,
and we caution the reader not to over-emphasize nor under-emphasize any one axis.

Although we do not further emphasize it 
in this paper, a different scientific metric called the {\em Sheffield
parameter}\cite{sheffield85,FESAC_2008} aims to embody
both the required physics performance (like
the Lawson parameter) and the ``efficiency'' of achieving that performance for MCF concepts. The Sheffield parameter can be thought of as a
normalized triple product by explicitly including
the parameter $\beta$, which is a measure
of how much plasma pressure (related to the triple product) can
be confined for a given magnetic field (which affects cost
and engineering difficulty).

Because of these additional considerations, fusion approaches that have achieved the highest values of $T$ and $n\tau$ (or $n\tau_E$),
i.e., tokamak-based MCF\cite{Wesson_2011} and laser-driven ICF,\cite{Nuckolls_1972,Atzeni04} may not necessarily become the first widely deployed commercial fusion-energy systems.  In fact, most private fusion companies focusing on developing
commercial fusion systems have opted for fusion 
approaches with lower demonstrated values to-date of temperature and Lawson parameter because of the expectation
that the required economics and social acceptance may be more readily
achievable.  Further discussion of these other considerations are beyond the scope of this paper but are
discussed elsewhere in the fusion literature.\cite{Kaslow94,Woodruff12,Maisonnier18,FESAC_2008}

This paper is organized as follows. Section~\ref{sec:datatables} defines the key variables used in the paper and provides plots of the compiled parameters.  
Section~\ref{sec:lawson_criterion} provides a review and mathematical
derivations of the Lawson criterion and the multiple
definitions of fusion energy gain used by fusion researchers. Section~\ref{sec:methodology} provides a physics-based justification for the approximations required to compare fusion energy gain across a wide range of fusion experiments and approaches. 
Readers primarily interested in seeing and
using the data without getting entangled in the details can largely ignore
Secs.~\ref{sec:lawson_criterion} and \ref{sec:methodology}\@.
Section~\ref{sec:conclusion} provides a summary and conclusions.
The appendices provide supporting information, including
data tables of the compiled parameters, additional plots,
and consideration of advanced fusion fuels (D-D, D-$^3$He, p-$^{11}$B).

\section{Variable definitions and plots}
\label{sec:datatables}
This section provides variable definitions (Table~\ref{tab:glossary}), and
plots of compiled Lawson parameters, fuel temperatures, and triple products. In many places (especially Secs.~\ref{sec:intro},
\ref{sec:lawson_criterion}, and \ref{sec:conclusion}), we use the generic variables
$n$, $T$, $\tau$, $Q$ for economy. However, 
in most of the paper and as indicated
in Table~\ref{tab:glossary}, all these variables have
more precise and differentiated versions with various subscripts.
The energy unit keV is used for temperature
variables throughout this paper,
and therefore the Boltzmann constant $k$ is not explicitly shown.
\input{table_1}

\begin{figure*}[p]
\centerline{
\includegraphics[width=17.5cm]{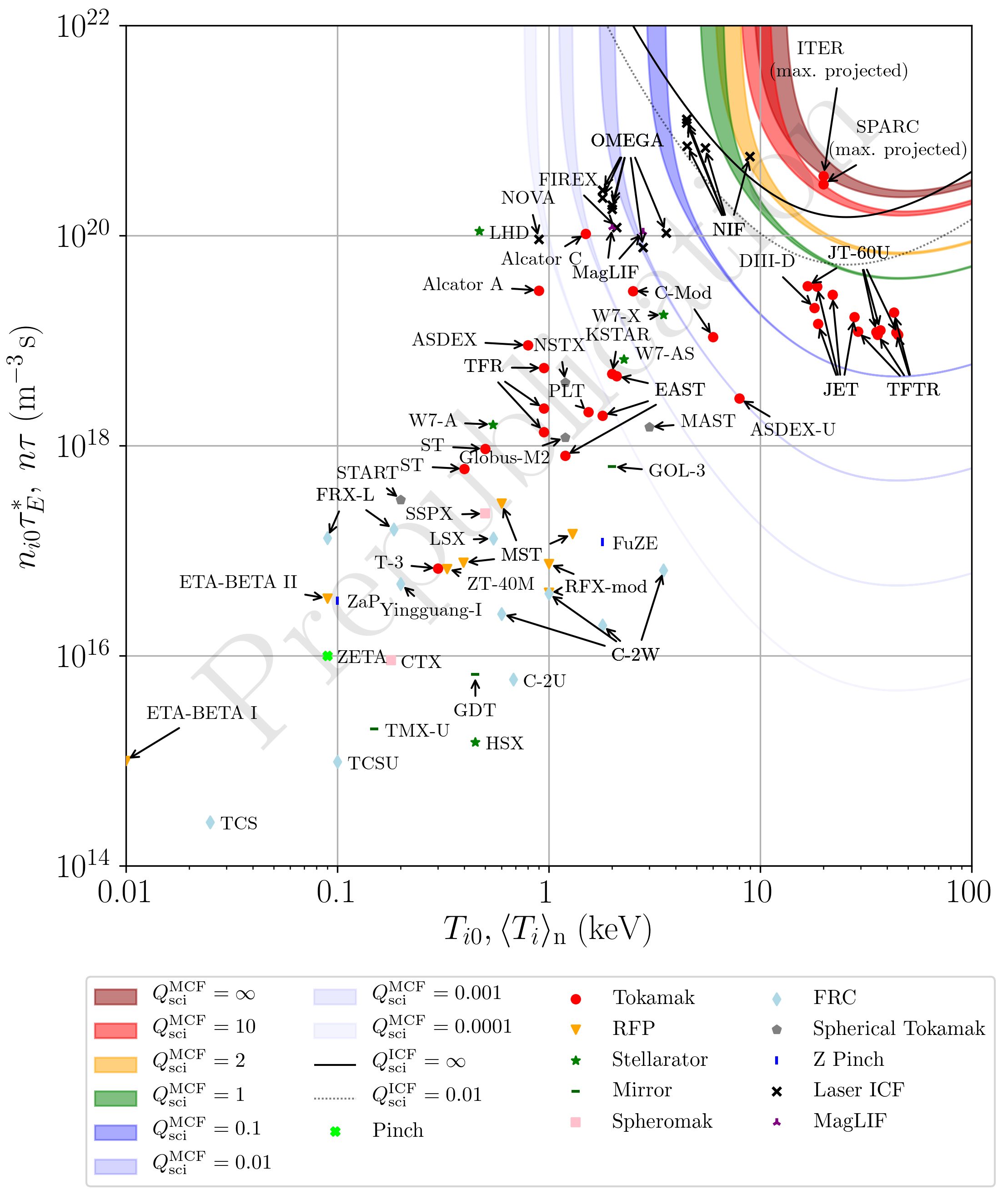}}
\caption{\label{fig:scatterplot_ntauE_vs_T} Experimentally inferred Lawson parameters ($n_{i0}\tau^*_E$ for MCF and $n\tau$ for ICF) of fusion experiments vs.\ $T_{i0}$ for MCF and $\langle T_i \rangle_{\rm n}$ for ICF (see Sec.~\ref{sec:lawson_criterion} for definitions of these quantities), extracted from the published literature (see Tables \ref{tab:mainstream_mcf_data_table}, \ref{tab:alternates_mcf_data_table}, and \ref{tab:icf_mif_data_table})\@. The various contours in the upper right correspond to the required Lawson parameters and ion temperatures required to achieve the indicated values of scientific gain $Q^{\rm MCF}_{\rm sci}$ for MCF (colored contours) and $Q^{\rm ICF}_{\rm sci}$ for ICF (solid and dotted black contours), assuming representative density and temperature profiles, external-heating absorption efficiencies, and D-T fuel. For experiments that do not use D-T, the contours represent a D-T-equivalent value of $Q_{\rm sci}$. The finite widths of the $Q_{\rm sci}^{\rm MCF}$ contours represent a range of assumed impurity levels. See the rest of the paper for details on how individual data points are extracted and how the $Q_{\rm sci}^{\rm MCF}$ and $Q_{\rm sci}^{\rm ICF}$ contours are calculated. Note that $Q_{\rm sci}^{\rm MCF} \gtrsim 20$ and $Q_{\rm sci}^{\rm ICF} \gtrsim 100$ are likely needed for practical fusion energy; see Sec.~\ref{sec:Qeng} and Eq.~(\ref{eq:engineering_gain_Q_scientific}) for discussion and definition, respectively, of $Q_{\rm eng}$.}
\end{figure*}

\begin{figure*}[p]
\centerline{
\includegraphics[width=17.5cm]{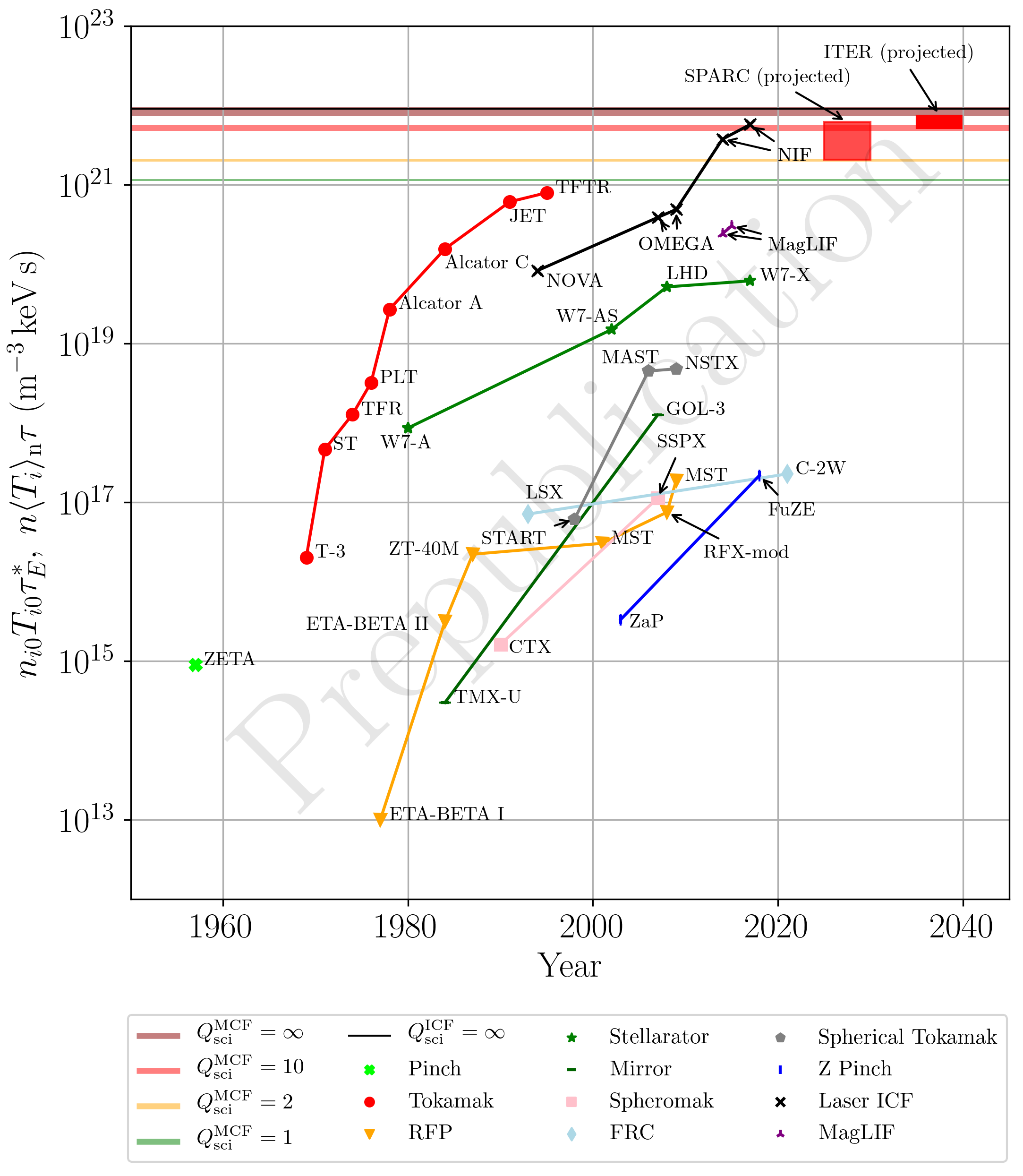}}
\caption{\label{fig:scatterplot_nTtauE_vs_year}Triple products ($n_{i0}T_{i0}\tau^*_E$ for MCF and $n \langle T_i \rangle_{\rm n} \tau$ for ICF; see Sec.~\ref{sec:lawson_criterion} for definitions of these quantities) that set a record for a given concept vs.\ year achieved. Record values for different concepts are shown to illustrate the progress towards energy gain of different concepts over time.
The horizontal lines labeled $Q_{\rm sci}^{\rm MCF}$ represent the minimum required triple product to achieve the indicated values of $Q_{\rm sci}^{\rm MCF}$, assuming $\eta_{\rm abs}=0.9$.
The horizontal line labeled $Q_{\rm sci}^{\rm ICF}=\infty$ represents the required triple product to achieve ignition and propagating burn for ICF, assuming $T_i=4$~keV and $\eta_{\rm abs}=0.006$. The projected triple-product ranges for SPARC and ITER are bounded above by their projected peak triple products
and below by the stated mission of each experiment (i.e.,
$Q_{\rm fuel}^{\rm MCF}=2$ for SPARC and $Q_{\rm fuel}^{\rm MCF}=10$ for ITER). Note that 
$Q_{\rm sci}^{\rm MCF} \gtrsim 20$ and $Q_{\rm sci}^{\rm ICF} \gtrsim 100$ are likely needed for practical fusion energy; see 
Sec.~\ref{sec:Qeng} and Eq.~(\ref{eq:engineering_gain_Q_scientific}) for
discussion and definition, respectively, of $Q_{\rm eng}$.
The NIF shot from August 8, 2021 does not appear in this plot because it did not achieve a record triple product, despite achieving a record $Q_{\rm sci}^{\rm ICF}$ for ICF\@. This highlights the main limitation of the triple product, i.e., it does not map to a unique value of gain (see Secs.~\ref{sec:fusion_triple_product_and_p_tau} and \ref{sec:ICF_methodology} for further explanation).}
\end{figure*}

Figure~\ref{fig:scatterplot_ntauE_vs_T} plots
achieved Lawson parameters versus $T_i$ for MCF, MIF, and ICF experiments, overlaid with contours of {\em scientific energy gain} $Q_{\rm sci}$, which is the fusion energy released divided by the energy
delivered to the plasma fuel (in the case of MCF) or the target (in the case of ICF)\@.  See the remainder of the paper for details on how the relevant data are
extracted from the primary literature, the 
mathematical definition of $Q_{\rm sci}$, and 
how the effects of non-uniform spatial profiles, impurities, heating efficiency, and other experimental details are treated.
Figure~\ref{fig:scatterplot_nTtauE_vs_year} shows record triple products achieved by different fusion concepts versus year
achieved (or anticipated to be achieved) relative to horizontal lines representing various values of $Q_{\rm sci}$\@.

Typically, MCF uses $\tau_E$ and ICF uses $\tau$ in their respective Lawson-parameter and triple-product definitions. Although $\tau_E$ and $\tau$ have
different physical meanings (see Secs.~\ref{sec:MCF_extending_lawson} and \ref{sec:ICF_extending_lawson}, respectively), they lead to analogous measures of energy breakeven and gain, allowing for
MCF and ICF to be plotted together in Figs.~\ref{fig:scatterplot_ntauE_vs_T}, \ref{fig:scatterplot_nTtauE_vs_year}, and \ref{fig:scatterplot_nTtauE_vs_T}.
We caution the reader that sometimes Lawson parameters and triple products
may be overestimated by concept advocates, especially
in unpublished materials, because $\tau$ is used incorrectly in place of $\tau_E$.

\section{Lawson Criterion, Lawson Parameter, Triple Product, and Energy Gain}
\label{sec:lawson_criterion}

In this section, we provide a detailed review of the derivation of the Lawson criterion, following Lawson's original papers.\cite{1955Lawson,Lawson_1957} We then introduce the mathematical 
definitions of the Lawson parameter in the context of idealized MCF and ICF scenarios, derive the fusion triple product, and define three forms of fusion energy gain used by fusion researchers.

\begin{figure}[htbp!]
\includegraphics[width=8cm]{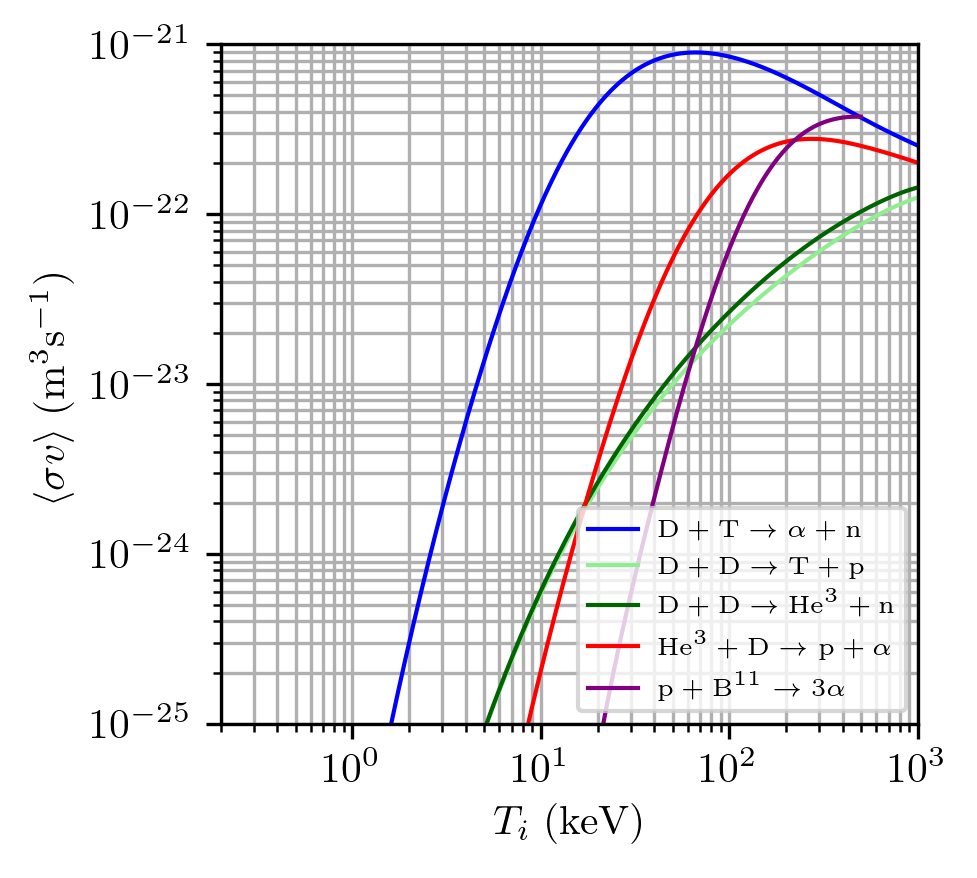}
\caption{\label{fig:reactivities}Thermal fusion reactivities $\langle \sigma v \rangle$ 
vs.\ $T_i$ for fusion reactions shown in the legend. All reactivities are calculated by numerical integration of velocity-averaged cross sections from Ref.~\onlinecite{Bosch_Hale_1992} with the exception of p-$^{11}$B, which is calculated from the parametrization of Ref.~\onlinecite{Nevins_2000}. Note that the two D-D branches are nearly on top of each other.}
\end{figure}

Lawson considered the deuterium-tritium (D-T) and deuterium-deuterium (D-D) fusion reactions:
\begin{align}
    \mathrm{D} + \mathrm{T} & \rightarrow \alpha~(\mathrm{3.5~MeV}) + \mathrm{n}~(\mathrm{14.1~MeV})\\
    \mathrm{D} + \mathrm{D} &
    \underset{50\%}{\rightarrow}
    \mathrm{T}~(\mathrm{1.01~MeV}) + 
    \mathrm{p}~(\mathrm{3.02~MeV}) \\
    \mathrm{D}+\mathrm{D} &
    \underset{50\%}{\rightarrow} 
    ^3\!\!\mathrm{He}~(0.82~\mathrm{MeV}) + 
    \mathrm{n}~(2.45~\mathrm{MeV}),
\end{align}
where $\alpha$ denotes a charged helium ion ($^4$He$^{2+}$), p denotes a proton,
n denotes
a neutron, and 1~MeV $=1.6\times 10^{-13}$~J\@. The fusion reactivities 
$\langle \sigma v\rangle$ for thermal ion distributions for these reactions, as well as the additional reactions,
\begin{align}
   \mathrm{D} + ^3\!\!\mathrm{He} & \rightarrow \mathrm{\alpha~(\mathrm{3.6~MeV})+ p}~(\mathrm{14.7~MeV}) \\
  \mathrm{p} + ^{11}\!\!\mathrm{B} &\rightarrow \mathrm{3 \alpha}~(\mathrm{8.7~MeV}),
\end{align}
are shown in Fig.~\ref{fig:reactivities}.

As did Lawson, this paper assumes {\em thermal} populations of ions and electrons,
i.e., Maxwellian velocity distributions characterized by a temperatures $T_i$ and $T_e$, respectively. Throughout this paper, we assume that ions and electrons are in thermal equilibrium with each other such that $T=T_i=T_e$. Non-equilibrium fusion approaches, where $T_i > T_e$, must account for the energy loss channel and timescale of energy transfer from ions to electrons.\cite{Rider_1995} Analysis of such systems is not included in this paper. Furthermore, this paper does not consider non-thermal ion or electron populations
such as those with beam-like distributions.  The latter typically
must contend with reactant slowing at a much faster rate than the fusion rate.
The inherent difficulty (though not necessarily impossibility) for non-thermal fusion approaches to achieve $Q_{\rm sci}>1$ is discussed in Ref.~\onlinecite{Rider_1997}.

Lawson's original papers considered two distinct fusion operating conditions. The first is a steady-state scenario in which the charged fusion products are confined and contribute to self heating. The second is a pulsed scenario in which the charged fusion products escape and energy is supplied over the duration of the pulse.
Lawson's analysis did not address {\em how} the fusion plasma is confined and assumed an ideal scenario without thermal-conduction losses in both cases.

\subsection{Lawson's first insight: ideal ignition temperature}
\label{sec:lawson_first}

\begin{figure}[tb!]
\includegraphics[width=8cm]{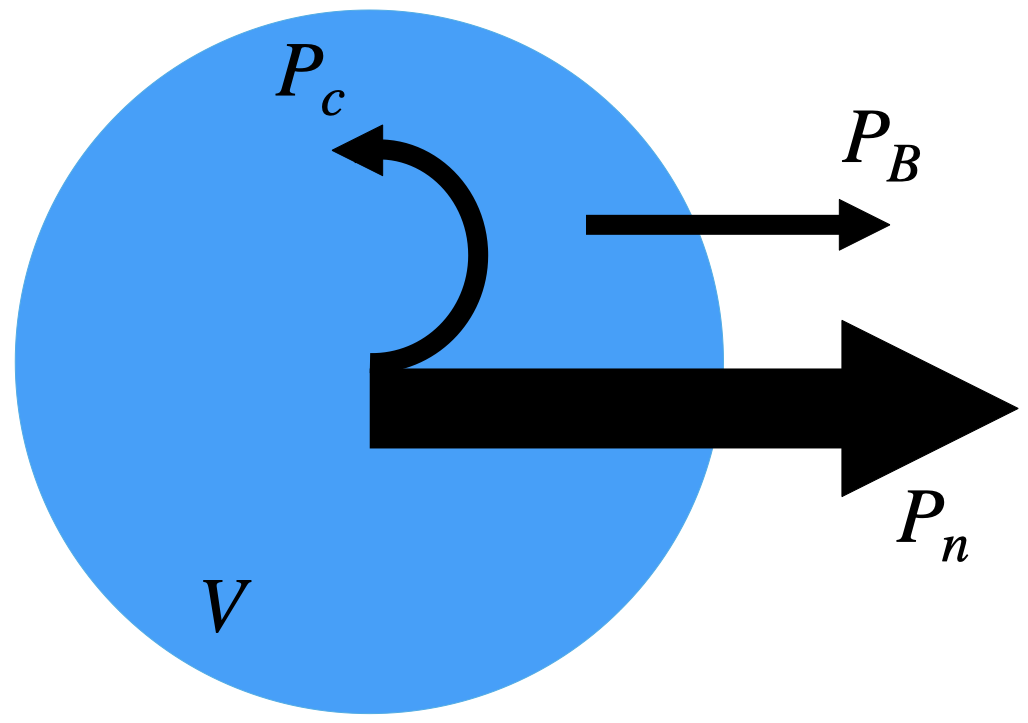}
\caption{The steady-state scenario corresponding to Lawson's first insight. Self heating from charged fusion products $P_c$ appears as bremsstrahlung power $P_B$ in a steady state plasma of volume $V$. Fusion power emitted as neutrons $P_n$ escapes the plasma and does not contribute to self-heating. An unspecified, idealized confinement mechanism is assumed, and thermal-conduction is ignored.}
\label{fig:lawsons_1st}
\end{figure}

Lawson's first insight was that a self-sustaining, steady-state fusion system without external heating must, at a minimum, balance radiative power losses with self heating from the charged fusion products, as illustrated conceptually in Fig.~\ref{fig:lawsons_1st}.
The power released by charged fusion products in a plasma of volume $V$ is
\begin{equation}
\label{eq:fusion_power_density}
    P_c = f_c P_F = f_c \frac{n_{1}n_{2}}{1+\delta_{1,2}} \langle \sigma v \rangle_{1,2} \epsilon_{F} V,
\end{equation}
where $n_1$ and $n_2$ are the number densities of the reactants,
$\delta_{1,2}=1$ in the case of identical reactants (e.g., D-D), and $\delta_{1,2}=0$ otherwise (e.g., D-T)\@.


The power emitted by bremsstrahlung radiation is
\begin{equation}
\label{eq:bremsstrahlung_power_density}
    P_B = C_B n_{i} n_{e} Z^2 T_e^{1/2} V,
\end{equation}
where $C_B$ is a constant and $Z=1$ in a hydrogenic plasma.
Entering values of density in m$^{-3}$, temperature in keV, volume in m$^3$, and setting $C_B=5.34\times 10^{-37}$~W\,m$^3$\,keV$^{-1/2}$ gives $P_B$ in watts. 

\begin{figure}[tb!]
\includegraphics[width=8cm]{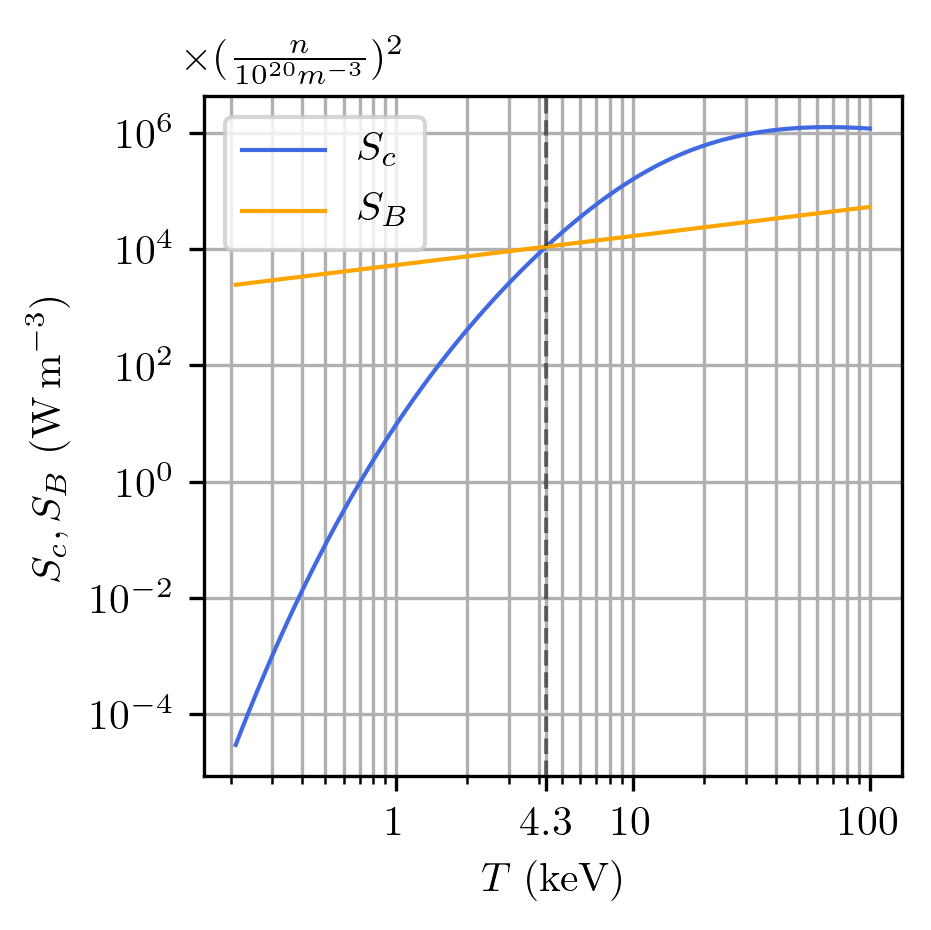}
\caption{Power produced per unit volume $S_c$ in charged D-T fusion products ($\alpha$ particles) and power lost to bremsstrahlung per unit volume $S_B$
vs.\ $T$ in a D-T plasma. 
When $T<4.3$~keV, $S_B>S_c$ and ignition is not possible (assuming $T=T_{e}=T_{i}$).}
\label{fig:ideal_ignition}
\end{figure}

If the fusion plasma is to be completely self heated by charged fusion products (i.e., $\alpha$, T, p, or He$^3$ in the above reactions), then $P_c \ge P_B$ is
required in order for the plasma to reach ignition (ignoring conduction losses for
the moment).
In the case of an equimolar D-T fusion plasma, i.e., $n/2 = n_{1} = n_{2}$,
where $n$ is the total ion number density and $Z=1$, and given the assumption 
$T = T_i = T_e$, the condition $P_c \ge P_B$
becomes
\begin{equation}
\label{eq:ideal_ignition}
\frac{1}{4} f_c n^{2} \langle \sigma v \rangle_{DT} \epsilon_{F} V \ge C_B n^{2} T^{1/2} V.
\end{equation}
Dividing both sides by $V$ and plotting the resulting fusion power density $S_c = P_c/V$ (left-hand side) and bremsstrahlung power density $S_B = P_B/V$ (right-hand side) versus $T$ in Fig.~\ref{fig:ideal_ignition} shows that
$T\ge 4.3$~keV is required for $S_c \ge S_B$. This temperature is known as the ideal ignition temperature because, under the idealized scenario of perfect confinement, ignition occurs at this temperature. Note that because $n^2$ cancels on both sides of Eq.~(\ref{eq:ideal_ignition}), the ideal ignition temperature is independent of density. In Appendix~\ref{sec:mitigating_brems}, we discuss and show how the ignition temperature
could be modified if bremsstrahlung radiation losses are mitigated.

\subsection{Lawson's second insight: dependence of fuel energy gain on $T$ and $n\tau$}
\label{sec:lawson_second}

\begin{figure}[tb!]
\includegraphics[width=8cm]{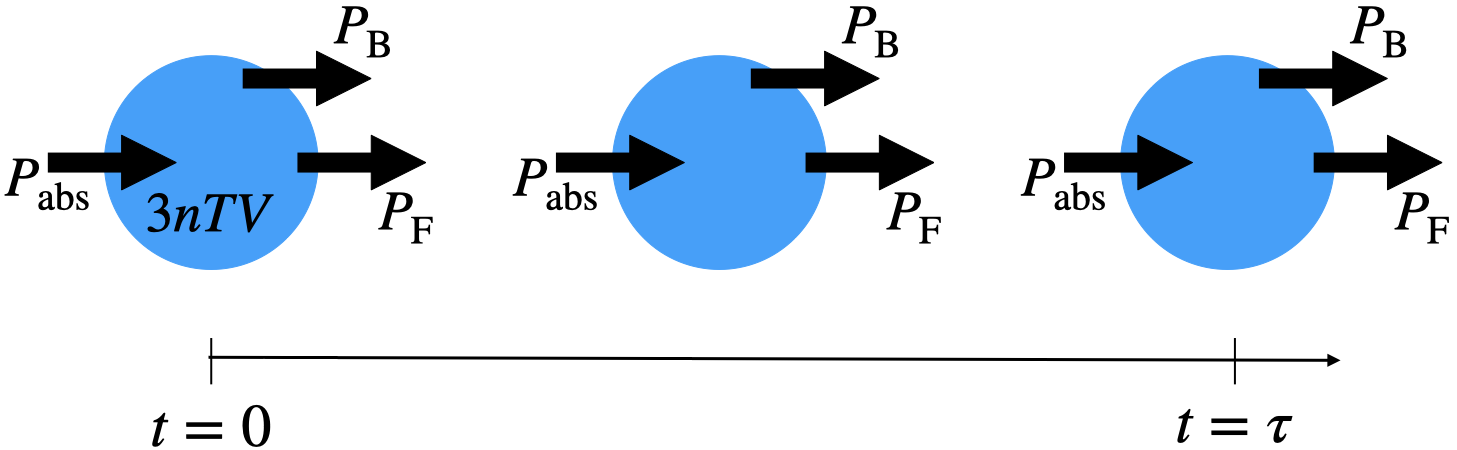}
\caption{Pulsed scenario corresponding to Lawson's second insight. At time $t=0$, the plasma temperature is instantaneously raised to $T$ and maintained for a duration $\tau$ by externally applied and absorbed power $P_{\rm abs}$. All fusion products escape (no self-heating), and thermal conduction is neglected (ideal confinement).
Absorbed power $P_{\rm abs}$ appears as bremsstrahlung power $P_B$ during the pulse duration.}
\label{fig:lawsons_2nd}
\end{figure}

Lawson's second insight involves a pulsed scenario where
a plasma is heated instantaneously to a temperature $T$ and maintained at that temperature
for time $\tau$, as illustrated conceptually in Fig.~\ref{fig:lawsons_2nd}.
In this scenario, bremsstrahlung radiation and all fusion reaction products escape, and heating must come from an external source during duration $\tau$. Idealized confinement is assumed, and thermal-conduction losses are ignored.

\begin{figure}[tb!]
\includegraphics[width=8cm]{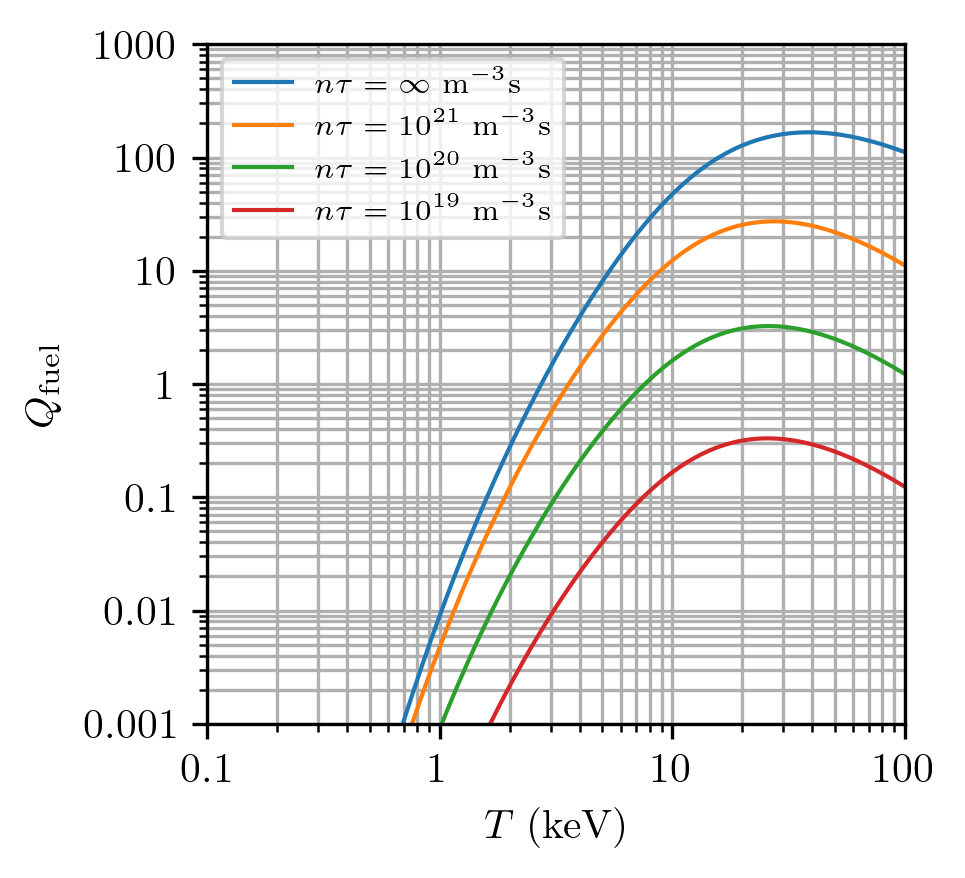}
\caption{Plot of $Q_{\rm fuel}$ vs. $T$ for indicated values of $n\tau$,
assuming no self heating and no thermal-conduction losses.}
\label{fig:Q_vs_T}
\end{figure}

We define the fuel gain
$Q_{\rm fuel}$ (Lawson used $R$) as the ratio of energy released in fusion products to the applied external energy that is {\em absorbed} by the entire fuel over the duration $\tau$ of the pulse.
This absorbed energy is the sum of the instantaneously deposited energy $\frac{3}{2}(n_e+n_i)TV = 3nTV$ (assuming $T=T_i=T_e$ and $n=n_i=n_e$) and the energy applied and absorbed over the pulse duration, $\tau P_{\rm abs}$. To maintain constant $T$ over duration $\tau$, $P_{\rm abs} = P_B$ is required,
and the fuel gain is therefore,
\begin{equation}
\label{eq:Lawson_Q_fuel}
Q_{\rm fuel} = \frac{\tau P_F}{\tau P_B+3nTV} = \frac{P_F/(3n^{2}TV)}{P_B/(3n^{2}TV)+1/(n\tau)}.
\end{equation}
Because both $P_F$ and $P_B$ are proportional to $n^{2}V$ and functions
of $T$ [see Eqs.~(\ref{eq:fusion_power_density}) and (\ref{eq:bremsstrahlung_power_density})], the $n^2V$ dependence
cancels out, and $Q_{\rm fuel}$ is solely a function of $T$ and $n\tau$,
\begin{equation}
\label{eq:Lawson_Q_fuel_explicit}
Q_{\rm fuel}= \frac{\langle \sigma v \rangle \epsilon_F/12T}{C_B/3T^{1/2} + 1/(n\tau)}. 
\end{equation}
Figure~\ref{fig:Q_vs_T} plots $Q_{\rm fuel}$ as a function of $T$ for the indicated values of $n\tau$, illustrating that even without self-heating, $Q_{\rm fuel}\gg 1$ is theoretically possible. Lawson noted that a ``useful'' system would require $Q_{\rm fuel} > 2$, assuming that fusion energy and bremsstrahlung could be converted to useful energy with an efficiency of 1/3, and remarked on the severity of the required $T$ and $n\tau$.

In this section, we have assumed that at time $t=\tau$ the external heating is turned off and none of the applied energy is recaptured. Lawson noticed, however, that if a fraction $\eta$ (Lawson used $f$) of the thermal energy at the conclusion of the pulse duration is recovered and converted into a useful form of energy (e.g., electrical or mechanical) that could offset the externally applied energy, the quantity $n\tau$ in Eq.~(\ref{eq:Lawson_Q_fuel_explicit}) is replaced by $n\tau/(1-\eta)$. The utilization of energy recovery to relax the requirements on $n\tau$ for achievement of energy gain is discussed further in Sec.~\ref{sec:Qeng}.

\subsection{Extending Lawson's second scenario: effect of self heating and relationship between characteristic times $\tau$ and $\tau_E$}
\label{sec:extending_lawsons_second_scenario}
\begin{figure}[b!]
\includegraphics[width=8cm]{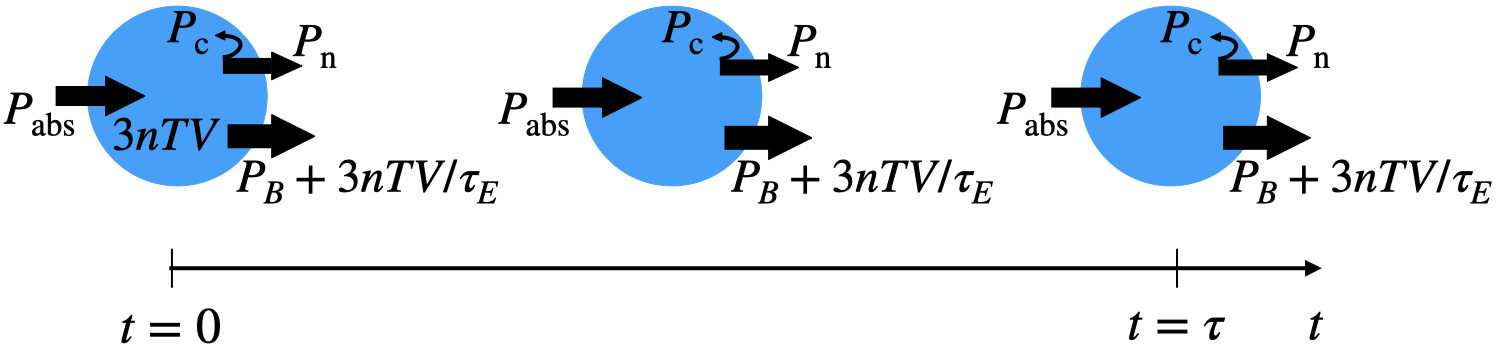}
\caption{Extension of Lawson's second scenario. At time $t=0$, the plasma temperature is instantaneously raised to $T$ and maintained for a duration $\tau$ by absorbed external power $P_{\rm abs}$ and self-heating power $P_c$. The sum of absorbed
external heating and self-heating appear as bremsstrahlung $P_B$ and thermal conduction $3nTV/\tau_E$.}
\label{fig:lawsons_generalized}
\end{figure}

In an effort to capture experimental realities, we extend Lawson's second scenario to include thermal-conduction losses and self heating from charged fusion products, as illustrated in Fig.~\ref{fig:lawsons_generalized}.
The rate of energy leaving the plasma via thermal conduction is characterized by an {\em energy confinement time} $\tau_E$, which is the time for energy equal to the thermal energy $3nTV$ to exit the plasma.
The power balance over the duration of the constant-temperature pulse is
\begin{equation}
P_{\rm abs} + P_{c} = P_{B} + 3nTV/\tau_E.
\label{eq:power_balance_steady_state}
\end{equation}

Applying a similar analysis to that of the previous section, we obtain
\begin{equation}
Q_{\rm fuel} = \frac{\langle \sigma v \rangle \epsilon_F/12T}{C_B/3T^{1/2} -f_c\langle \sigma v \rangle \epsilon_F / 12T + 1/(n\tau_{\rm eff})},
\label{eq:Q_fuel_generalized}
\end{equation}
where 
\begin{equation}
    \label{eq:effective_lawson_parameter}
    n\tau_{\rm eff} = n\frac{\tau \tau_E}{\tau + \tau_E}.
\end{equation}
The relationship between the two characteristic times $\tau$ and $\tau_E$ is like two resistors in parallel, i.e., 
it is the smaller of the two that limits the value of $\tau_{\rm eff}$.
If $\tau \ll \tau_E$, the confinement duration $\tau$ limits $Q_{\rm fuel}$ because there is limited time to overcome the initial energy investment of
raising the plasma temperature.
If $\tau_E \ll \tau$, the energy confinement time $\tau_E$ limits $Q_{\rm fuel}$ because the rate of energy leakage from thermal conduction places higher demand
on external and self heating.
If the two characteristic times are of similar magnitude, then both play a role in limiting $Q_{\rm fuel}$.

\begin{figure}[tb!]
\includegraphics[width=8cm]{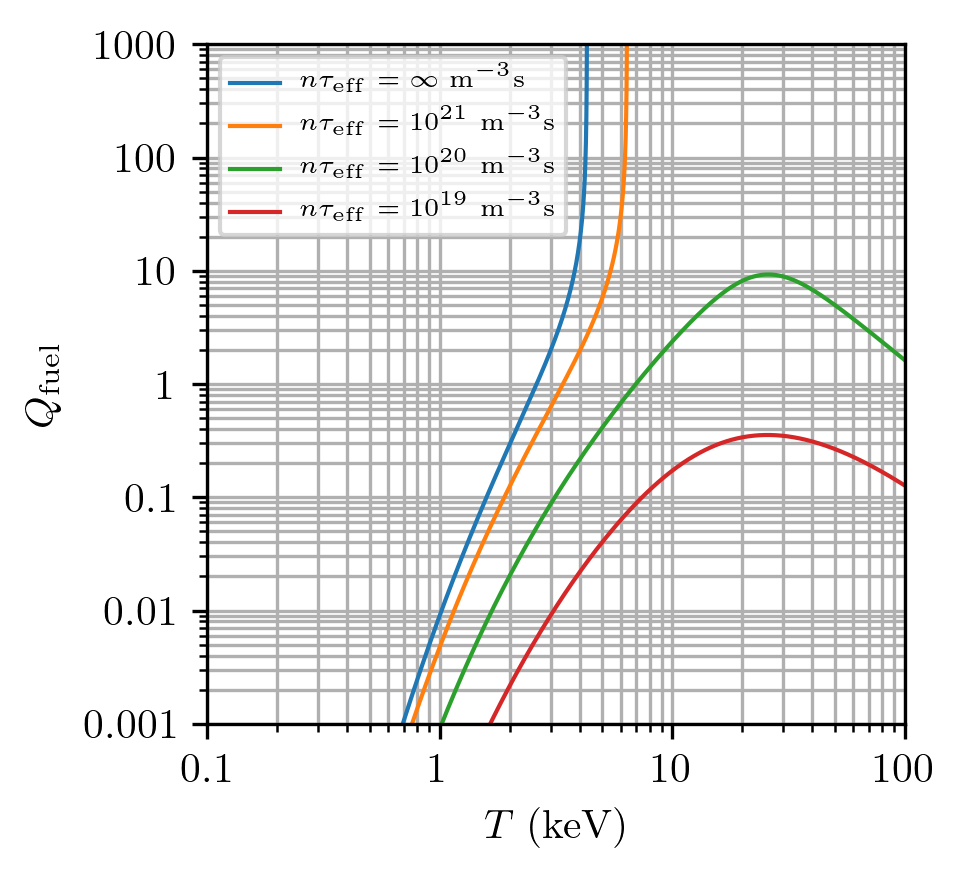}
\caption{Plot of $Q_{\rm fuel}$ vs.\ $T$ for indicated values of effective Lawson parameter $n\tau_{\rm eff}$, for a pulsed scenario that includes self heating from charged fusion products and thermal conduction. Self heating reduces the demands on externally applied and absorbed heating. Above a threshold of $T$ and $n\tau_{\rm eff}$, $Q_{\rm fuel}$ increases without bound, corresponding to ignition.}
\label{fig:Q_vs_T_extended}
\end{figure}

Figure \ref{fig:Q_vs_T_extended} plots $Q_{\rm fuel}$ versus $T$ for the indicated values of $n\tau_{\rm eff}$, illustrating that self heating enables ignition 
($Q_{\rm fuel}\rightarrow \infty$) above a threshold of $T$ and $n\tau_{\rm eff}$, made possible by the reduction of the denominator of Eq.~(\ref{eq:Q_fuel_generalized}) by amount $f_c\langle \sigma v \rangle \epsilon_F / 12T$. We explore these thresholds in subsequent sections.

\subsection{Scientific energy gain and breakeven}
\label{sec:steady_state_model_with_external_heating}

Because external-heating efficiency varies widely across fusion concepts, and because the absorption efficiency is intrinsic to the physics of each concept, we define $P_{\rm ext}$ as the heating power applied {\em at the boundary} of the plasma (in the case of MCF) or the target assembly (in the case of ICF). This definition of
$P_{\rm ext}$ encapsulates all {\em physics} elements of the experiment.
The boundary can typically be regarded as the vacuum vessel for all concepts,
where $P_{\rm ext}$ could be electromagnetic waves for MCF, laser beams for ICF, or electrical current and voltage for MIF\@.
The previously introduced $P_{\rm abs}$ is the fraction $\eta_{\rm abs}$ of $P_{\rm ext}$ that is actually
{\em absorbed by the fuel}, i.e., $P_{\rm abs}=\eta_{\rm abs}P_{\rm ext}$.
The previously defined {\em fuel gain} is
\begin{equation}
\label{eq:fuel_gain}
Q_{\rm fuel} = \frac{P_F}{P_{\rm abs}},
\end{equation}
and the newly defined {\em scientific gain} is
\begin{equation}
\label{eq:scientific_gain}
Q_{\rm sci} = \frac{P_F}{P_{\rm ext}} = \eta_{\rm abs}Q_{\rm fuel} < Q_{\rm fuel}.
\end{equation}
Whereas $Q_{\rm fuel}$ ignores the plasma-physics losses
of the absorption of
heating energy into the fuel (e.g., neutral-beam shine-through in MCF or reflection of laser light via laser-plasma
instabilities in ICF), $Q_{\rm sci}$ accounts
for all plasma-physics-related losses between the vacuum vessel and the fusion fuel. Therefore, $Q_{\rm sci}$ is the better metric for assessing remaining {\em physics risk} of a fusion concept.

{\em Scientific breakeven} is historically defined as $Q_{\rm sci}=1$,
which is an important milestone in the development of fusion energy because it signifies that very significant (but not all) plasma-physics challenges have been retired. Scientific breakeven has not yet been achieved, although D-T tokamak experiments such as TFTR and JET from the 1990s and the NIF experiment of August 8, 2021\cite{LLNL2021} have come close ($Q_{\rm sci} = 0.27$ for TFTR,\cite{McGuire1995} $Q_{\rm sci} = 0.64$ for JET,\cite{Keilhacker_1999} and $Q_{\rm sci} \sim 0.7$ for NIF\cite{ScienceNews2021})\@.
Because $\eta_{\rm abs}$ is much closer to unity in MCF experiments, the MCF community often uses $Q$ to refer to $Q_{\rm fuel}$ or $Q_{\rm sci}$ interchangeably.

\subsection{Idealized, steady-state MCF: $\tau_E \ll \tau$}
\label{sec:MCF_extending_lawson}
MCF relies on strong magnetic fields to confine fusion fuel, minimize thermal-conduction losses, and trap the charged fusion products for self heating. By the time that Lawson's report was declassified in 1957, the UK, US, and USSR were all actively developing MCF experiments that included externally applied heating.

\begin{figure}[tb!]
\includegraphics[width=8cm]{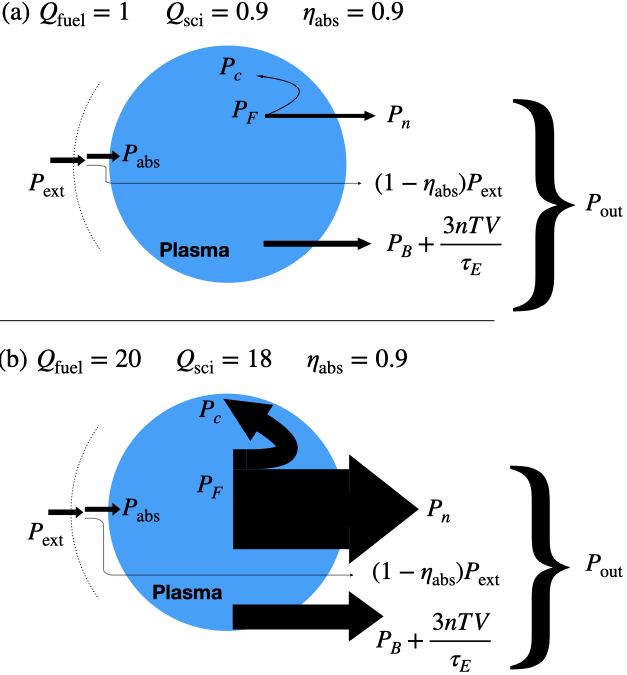}
\caption{Conceptual illustrations of the steady-state power balance for two hypothetical steady-state MCF scenarios.
The dotted line represents the boundary, e.g., vacuum chamber, between the physics and engineering aspects of the experiment.}
\label{fig:power_balances}
\end{figure}

Adapting the extension of Lawson's second insight to this scenario, we consider the power balance of an externally heated and self-heated, steady-state plasma. Figure~\ref{fig:power_balances} illustrates this scenario for two different values of energy gain.
The power balance and fuel gain of the plasma are described by Eqs.~(\ref{eq:power_balance_steady_state}) and (\ref{eq:Q_fuel_generalized}),
respectively, in the limit of steady-state operation, i.e., $\tau \rightarrow \infty$.

\begin{figure}[ht!]
\includegraphics[width=8cm]{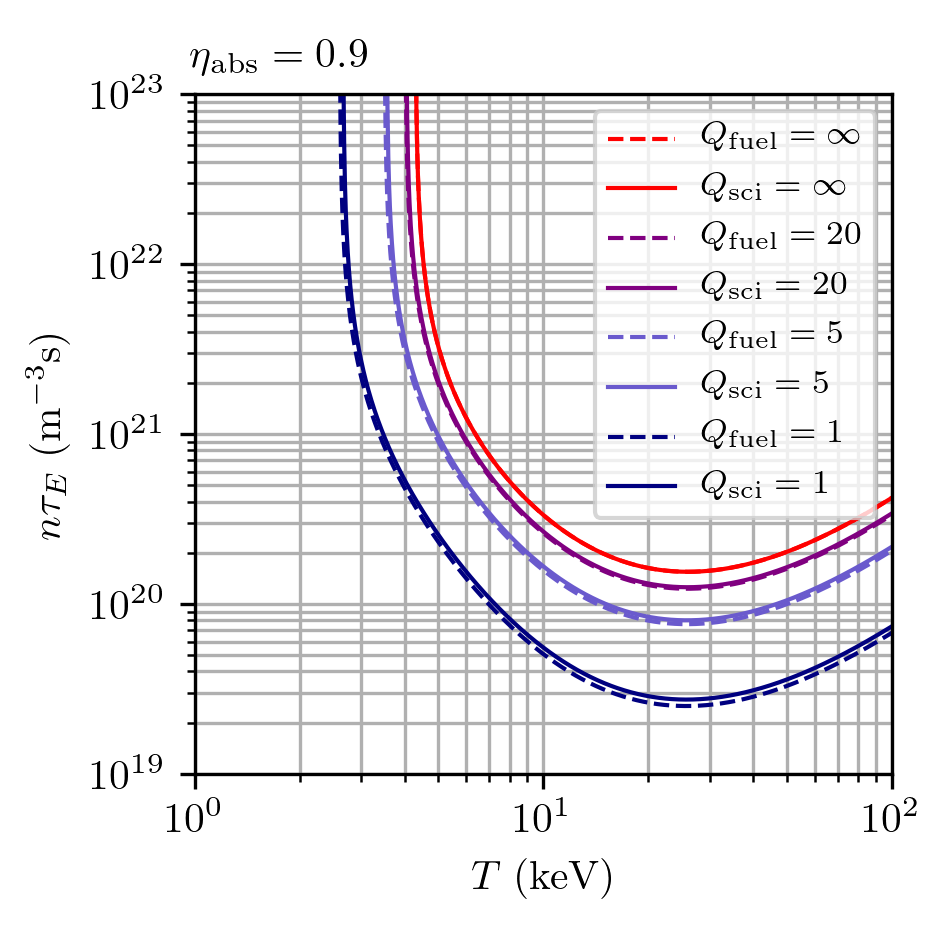}
\caption{Lawson parameter $n\tau_E$ vs.\ $T$ required to achieve indicated values of $Q_{\rm fuel}$ (dashed lines) and $Q_{\rm sci}$ (solid lines), assuming $\eta_{\rm abs}=0.9$ (representative of MCF)\@. Because $\eta_{\rm abs}$ is close to unity for MCF, $Q_{\rm fuel}$ and $Q_{\rm sci}$ are nearly coincident (the ignition contours are exactly coincident) and are often used interchangeably and referred to as $Q$.}
    \label{fig:MCF_ntau_contours_q_fuel_q_sci}
\end{figure}

To more clearly observe the requirements on $n\tau_E$ and $T$ to achieve certain values of $Q_{\rm fuel}$, we solve Eq.~(\ref{eq:Q_fuel_generalized}) for $n \tau_E$ in the steady-state limit ($\tau\gg\tau_E$),
\begin{equation}
\label{eq:MCF_Lawson_parameter_Q_fuel}
n\tau_E = \frac{3T}{(f_c+ Q_{\rm fuel}^{-1})\langle \sigma v  \rangle \epsilon_{F}/4 - C_{B} T^{1/2}}.
\end{equation}
Plotting this expression in Fig.~\ref{fig:MCF_ntau_contours_q_fuel_q_sci} (dashed lines) for D-T fusion shows that a threshold value of $n\tau_E$, which varies with $T$, is required to achieve a given value of $Q_{\rm fuel}$. Table~\ref{tab:minimum_lawson_parameter_table} lists the minimum values of Lawson parameter and corresponding temperature required to achieve $Q_{\rm fuel}=1$ and $Q_{\rm fuel}=\infty$ for the indicated reactions.
Thus far, spatially uniform profiles of all quantities are assumed, and geometrical effects and impurities are ignored.  Later in the paper, we consider the effects of nonuniform
spatial profiles, different geometries (e.g., cylinder, torus, etc.), and impurities. 
\input{table_2}

To more clearly observe the requirements on $n\tau_E$ and $T$ to achieve certain values of $Q_{\rm sci}$, we replace $Q_{\rm fuel}$ with $Q_{\rm sci}/\eta_{\rm abs}$ in Eq.~(\ref{eq:MCF_Lawson_parameter_Q_fuel}), 
\begin{equation}
\label{eq:MCF_Lawson_parameter_Q_sci}
n\tau_E = \frac{3T}{(f_c+ \eta_{\rm abs} Q_{\rm sci}^{-1})\langle \sigma v  \rangle \epsilon_{F}/4 - C_{B} T^{1/2}}.
\end{equation}
The ignition contours are identical for $Q_{\rm fuel}=\infty$ and $Q_{\rm sci}=\infty$.
For MCF experiments, where $\eta_{\rm abs}$ is close to unity ($\eta_{\rm abs} \sim 0.9$), non-ignition
$Q_{\rm sci}<\infty$ contours are shifted relative to their respective $Q_{\rm fuel}$ contours only very slightly toward the ignition contour ($Q_{\rm fuel}, Q_{\rm sci}=\infty$), as seen in Fig.~\ref{fig:MCF_ntau_contours_q_fuel_q_sci} (solid lines).

The {\em Lawson criterion}, where $P_{\rm abs} \rightarrow 
0$ and $Q_{\rm fuel} \rightarrow \infty$ in Eqs.~(\ref{eq:power_balance_steady_state})
and (\ref{eq:fuel_gain}), respectively, is satisfied for values of
$n\tau_E$ and $T$ on or above the $Q_{\rm fuel}, Q_{\rm sci}=\infty$ 
curves
in Fig.~\ref{fig:MCF_ntau_contours_q_fuel_q_sci}. In this ignition regime, the plasma is entirely self heated by charged fusion products, and external heating is zero.
While the minimum Lawson parameter required for ignition occurs at
$T \approx 25$~keV, MCF approaches aim for $T \approx 10$--20~keV because the pressure required to achieve high gain is minimized in this lower-temperature range (as discussed in Sec.~\ref{sec:fusion_triple_product_and_p_tau}).

\subsection{Idealized ICF: $\tau \ll \tau_E$}
\label{sec:ICF_extending_lawson}
\begin{figure}[b!]
\includegraphics[width=8cm]{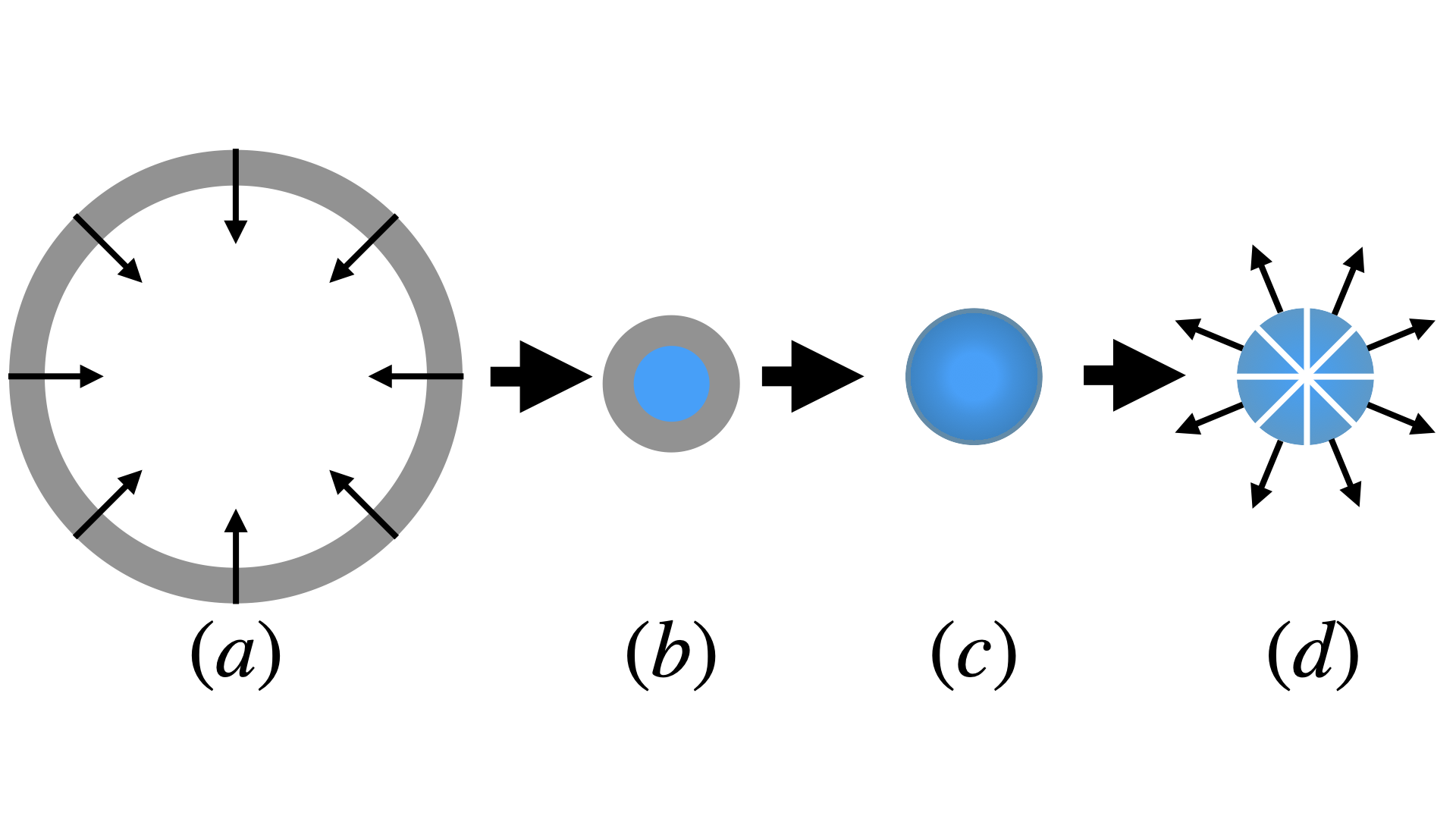}
\caption{Conceptual schematic of idealized ICF (a) compression, (b) hot-spot ignition, (c) propagating burn of the cold, dense shell, and (d) disassembly.}
\label{fig:conceptual_icf_basic}
\end{figure}

ICF relies on the inertia of highly compressed fusion fuel to provide a duration to fuse a sufficient amount of fuel to overcome the energy invested in compressing the fuel assembly. 
In 1971, the concept of using lasers to compress and heat a fuel pellet was declassified, first by the USSR and later that year by the US.\cite{Kidder_1998}
In 1972, \textcite{Nuckolls_1972} described the direct-drive laser ICF concept, where lasers ablate the surface of a hollow fuel pellet outward, driving the inner surface toward the center. In this scenario the kinetic energy of the inward-moving material is converted to thermal energy of a central, lower-density ``hot-spot'' that ignites. The fusion burn propagates outward through the surrounding denser fuel shell, which finally disassembles.  The four-step, ``central hot-spot ignition'' process is illustrated in 
Fig.~\ref{fig:conceptual_icf_basic}.
Laser indirect-drive ICF bathes the fuel pellet in X-rays generated by the interactions between lasers and the inside of a ``hohlraum'' (a metal enclosure surrounding the fuel pellet) to similar effect.

To adapt the extension of Lawson's second insight, we consider the energy balance of the hot spot over duration $\tau$, during which it is inertially confined [Fig.~\ref{fig:conceptual_icf_basic}(b)]. The sequence of events that leads to energy delivered to the hot spot are:
\begin{enumerate}
    \item The laser energy strikes the fuel pellet (or hohlraum);
    \item A fraction $\eta_{\rm abs}$ of the laser energy is absorbed by the fuel in the form of kinetic energy $E_{abs}$ of the imploding fuel shell;
    \item The imploding shell with energy $E_{abs}$ does $p\,\textrm{d}V$ work on the hot spot of volume $V$, resulting in hot-spot thermal energy $E_{hs}=\eta_{\rm hs} E_{abs}$;
    \item If sufficiently high temperature and Lawson parameter are achieved, additional energy $\tau P_c$ is delivered to the hot spot by charged fusion products.
\end{enumerate}

We describe the fuel gain of the hot-spot by applying the following assumptions and modifications to Eq.~(\ref{eq:Q_fuel_generalized}). In this simplified model, we neglect bremsstrahlung and thermal-conduction losses, i.e., $C_B \rightarrow 0$ and $\tau_E \rightarrow \infty$. While both processes are present in the hot spot, the cold, dense shell is largely opaque to bremsstrahlung and partially insulates the hot spot. In practice (which we also ignore here), both loss mechanisms have the effect of ablating material from the inner shell wall into the hot spot, increasing density and decreasing temperature while maintaining a constant pressure.\cite{Betti_2010}
To account for the fraction $\eta_{\rm hs}$ of the shell kinetic energy that is deposited in the hot-spot, the definition of $Q_{\rm fuel}$ becomes, 
\begin{equation}
    Q_{\rm fuel} = \frac{\tau P_F}{E_{\rm abs}} = \frac{\tau P_F}{E_{\rm hs}/\eta_{\rm hs}}.
    \label{eq:Q_fuel_energy_balance}
\end{equation}
We assume that the charged fusion products generated in the hot spot deposit all their energy within the hot spot.

\begin{figure}[b!]
\includegraphics[width=8cm]{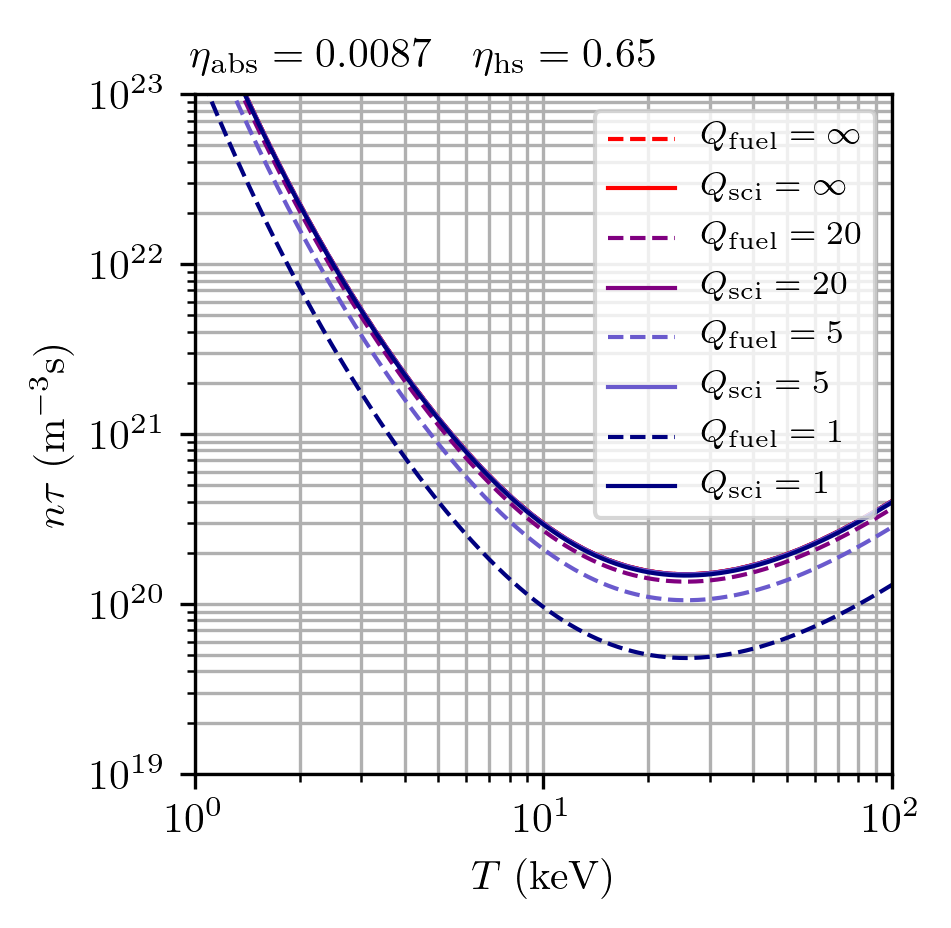}
\caption{Lawson parameter $n \tau$ vs.\ $T$ required to achieve indicated values of $Q_{\rm fuel}$ (dashed lines) and $Q_{\rm sci}$ (solid lines), assuming $\eta_{\rm abs} \eta_{\rm hs}=0.006$ (representative of indirect-drive ICF)\@. Because $\eta_{\rm abs} \eta_{\rm hs} \ll 1$ for laser ICF, contours of $Q_{\rm sci}$ are shifted to be nearly on top of the the ignition contour relative to their $Q_{\rm fuel}$ counterparts, illustrating that ignition is effectively required to achieve scientific breakeven for laser ICF.}
\label{fig:ICF_ntau_contours_q_fuel_q_sci}
\end{figure}

 To more clearly observe the requirements on $n\tau$ and $T$ to achieve certain values of $Q_{\rm fuel}$, we solve Eq.~(\ref{eq:Q_fuel_generalized}) for $n\tau$ with the above limits and modifications,
\begin{equation}
\label{eq:ICF_Lawson_parameter_Q_fuel}
n\tau = \frac{12T}{(f_c+ \eta_{\rm hs} Q_{\rm fuel}^{-1})\langle \sigma v  \rangle \epsilon_{F}},
\end{equation}
Plotting this expression in Fig.~\ref{fig:ICF_ntau_contours_q_fuel_q_sci} (dashed lines) for D-T fusion shows that a threshold value of $n\tau$, which varies with $T$, is required to achieve a given value of $Q_{\rm fuel}$ in an ICF hot spot. We have assumed $\eta_{\rm hs}=0.65$ based on NIF shot N191007.\cite{Zylstra_2021}
Thus far, reductions in $\tau$ due to instabilities, impurities, losses due to bremsstrahlung and thermal conduction, and the requirements to initiate a propagating burn in the cold, dense shell have been ignored. Later in this paper, we consider some of these effects.

Similarly to the MCF example, the required Lawson parameter and temperature required to reach a certain value of $Q_{\rm sci}$ can be evaluated by replacing $\eta_{\rm hs}Q_{\rm fuel}^{-1}$ with $\eta_{\rm abs} \eta_{\rm hs}Q_{\rm sci}^{-1}$ in Eq.~(\ref{eq:ICF_Lawson_parameter_Q_fuel}), 
\begin{equation}
\label{eq:ICF_Lawson_parameter_Q_sci}
n\tau = \frac{12T}{(f_c+ \eta_{\rm abs} \eta_{\rm hs} Q_{\rm sci}^{-1})\langle \sigma v  \rangle \epsilon_{F}}.
\end{equation}
For ICF experiments, where $\eta_{\rm abs} \eta_{\rm hs}$ is very low (e.g., $\eta_{\rm abs} \eta_{\rm hs} \sim 0.006$ for indirect-drive ICF), non-ignition
($Q_{\rm sci}<\infty$) contours are shifted relative to their respective $Q_{\rm fuel}$ contours strongly toward the ignition contour ($Q_{\rm sci}=\infty$), as seen in Fig.~\ref{fig:ICF_ntau_contours_q_fuel_q_sci} (solid lines). For this reason, ignition is effectively required to achieve scientific breakeven in ICF\@.
While the minimum Lawson parameter required for ignition occurs at $T \approx 25$~keV, laser-driven ICF approaches aim for hot-spot $T \approx 4$~keV (prior to the onset of significant
fusion leading to further increases in $T_i$) due to the limits of achievable implosion speed, which sets the maximum achievable temperature due to $p\,\mathrm{d}V$ heating alone.

Note that our definition of $Q_{\rm fuel}$ for ICF differs slightly from the standard definition of ICF fuel gain, $G_{\rm f}$, which is the ratio of fusion energy to total energy content of the fuel immediately before ignition.\cite{Atzeni04} The Lawson parameter of an ICF hot spot is usually framed in terms of the hot-spot $\rho_{\rm hs} R_{\rm hs}$, where $\rho_{\rm hs}$ and $R_{\rm hs}$ are the hot-spot mass density and radius, respectively.\cite{Atzeni04}  For the purposes of having a Lawson parameter and fuel gain that parallel the MCF case, we proceed with our definition of ICF 
$Q_{\rm sci}$, which is the same as the standard definition of ICF target gain $G$.\cite{Atzeni04} 

The condition for hot-spot ignition for a D-T plasma is,
\begin{equation}
\label{eq:ICF_Lawson_parameter_ignition}
(n\tau)_{\rm ig} = \frac{12T}{\langle \sigma v  \rangle \epsilon_{\alpha}},
\end{equation}
where $\epsilon_{\alpha}$ is the energy of the charged alpha-particle fusion product in the D-T fusion reaction.
More generally, ``ignition'' has many different meanings in the ICF context.\cite{Tipton_2015} The 1997 National Academies review of ICF\cite{NAS_1997} addressed the lack of consensus around the definition of ICF ignition by defining ignition as fusion energy produced exceeding the laser energy (i.e., $Q_{\rm sci} > 1$). More recently, the hot-spot conditions needed to initiate propagating burn in the colder, dense fuel shell (another definition of ignition) have been quantified.\cite{Christopherson_2020}
 These details are discussed further in Sec.~\ref{sec:ICF_methodology}.

\subsection{Fusion triple product and ``p-tau''}
\label{sec:fusion_triple_product_and_p_tau}
The triple product ($nT\tau_E$) and p-tau ($p \tau_E$) are commonly used by the MCF community to quantify fusion performance in a single value. While less common in the ICF community, $p \tau$ is sometimes used, and triple product ($nT\tau$) is typically used
only in the context of comparing ICF to MCF\@.\cite{Betti_2010} 
In a uniform plasma with $n=n_i=n_e$ and $T=T_i=T_e$,
the relationship between triple product and p-tau in both embodiments is $n T \tau = \frac{1}{2} p \tau$ and $n T \tau_E = \frac{1}{2} p \tau_E$.

\begin{figure}[tb!]
\includegraphics[width=8cm]{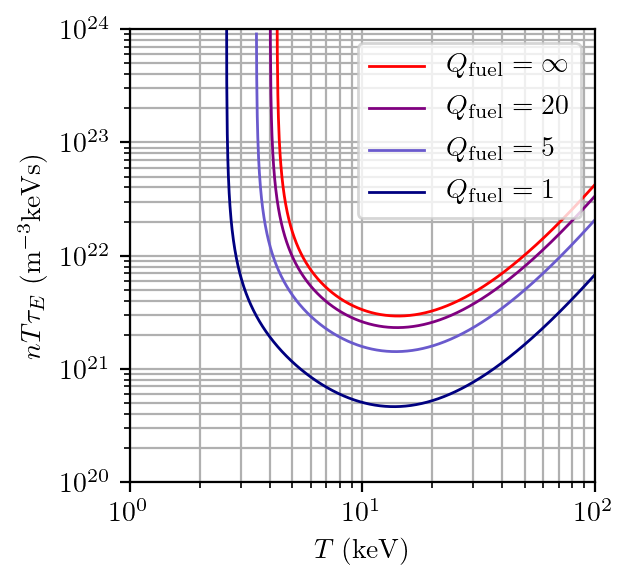}
\caption{Triple product vs.\ $T$ required to achieve indicated values of $Q_{\rm fuel}$ for MCF [Eq.~(\ref{eq:triple_product_steady_state})].}
\label{fig:MCF_nTtau_contours_q_fuel}
\end{figure}

An expression for the MCF triple product is obtained by multiplying both sides of Eq.~(\ref{eq:MCF_Lawson_parameter_Q_fuel}) by $T$,
\begin{equation}
\label{eq:triple_product_steady_state}
    nT\tau_E = \frac{3T^{2}}{(f_c+ Q_{\rm fuel}^{-1})\langle \sigma v  \rangle(T) \epsilon_{F}/4 - C_{B}T^{1/2}}.
\end{equation}
Figure~\ref{fig:MCF_nTtau_contours_q_fuel} shows the
$n T \tau_E$ required to achieve a specified value of $Q_{\rm fuel}$ as a function of $T$
(see also Table~\ref{tab:minimum_triple_product_table}).
Note that the minimum triple product needed to achieve ignition occurs at a lower $T$ than that of the minimum Lawson parameter. This lower $T$ is a better approximation of the intended $T$ of MCF experiments because it corresponds to the minimum pressure required to achieve a certain value of $Q_{\rm fuel}$, and pressure (rather than Lawson parameter) is a
more-direct experimental limitation of MCF\@.

\input{table_3}

We emphasize the limitation of the triple product (or ``p-tau'') as a metric: it
does not correspond to a unique value $Q_{\rm fuel}$ or $Q_{\rm sci}$ unless $T$ is specified. While $n$ and $\tau$ in the Lawson parameter may be traded off in equal proportions, $T$ must
be within a fixed range for an appreciable number of fusion reactions to occur.
Appendix~\ref{sec:other_formulations} provides a plot of achieved triple products and temperatures analogous to Fig.~\ref{fig:scatterplot_ntauE_vs_T}.
Appendix~\ref{sec:advanced_fuels} provides plots of $nT\tau_E$ vs.\ $T$ for D-D, D-$^3$He, and p-$^{11}$B fusion.

\subsection{Engineering gain}
\label{sec:Qeng}
The previously defined $Q_{\rm sci}$ [Eq.~(\ref{eq:scientific_gain})] is the ratio of power released in fusion reactions $P_F$ to applied external heating power $P_{\rm ext}$ (see Fig.~\ref{fig:power_balances}),
encapsulating the physics of plasma heating, thermal and radiative losses, and fusion energy production.
Based on conservation of energy in Fig.~\ref{fig:power_balances}, we can rewrite
\begin{equation}
    \label{eq:Qsci_complete}
    Q_{\rm sci} = \frac{P_{\rm out} - P_{\rm ext}}{P_{\rm ext}} = \frac{P_F}{P_{\rm ext}},
\end{equation}
which is equivalent to Eq.~(\ref{eq:scientific_gain}).


\begin{figure}[!b]
    \includegraphics[width=8cm]{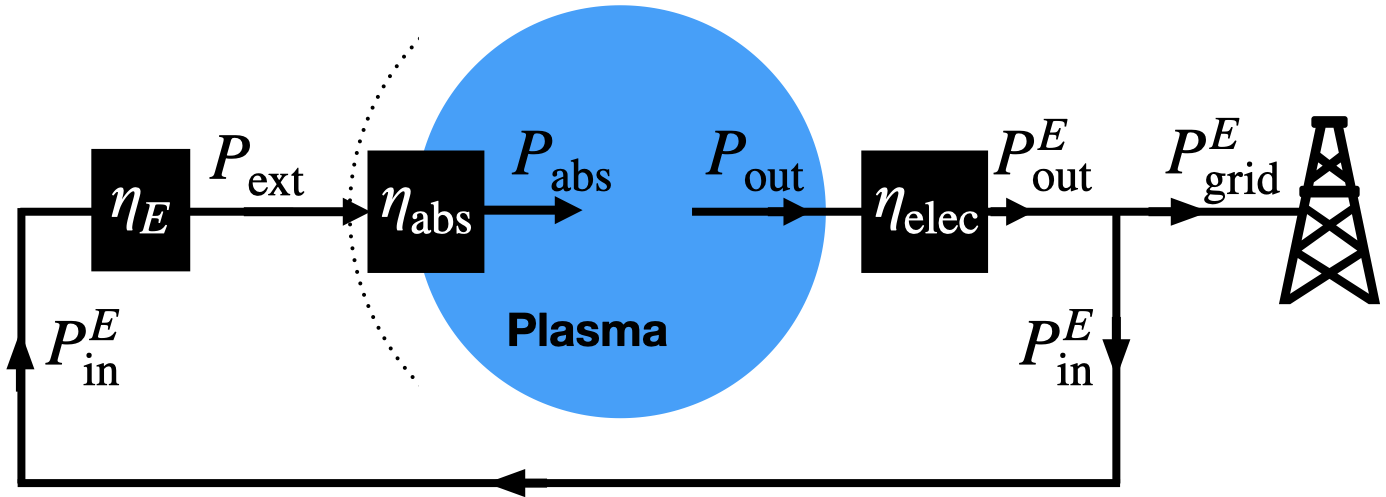}
    \caption{Conceptual schematic of a fusion power plant which recirculates electrical power. In this system $Q_{\rm eng} =P_{\rm grid}^{E}/P_{\rm in}^{E}$.}
    \label{fig:conceptual_plant}
\end{figure}

Similarly, the engineering gain, 
\begin{equation}
    Q_{\rm eng} = \frac{P_{\rm out}^{E} - P_{\rm in}^{E}}{P_{\rm in}^{E}} = \frac{P^{E}_{\rm grid}}{P_{\rm in}^{E}},
\end{equation}
is the ratio of
electrical power $P^{E}_{\rm grid}$ (delivered to the grid)
to the input (recirculating) electrical power $P_{\rm in}^{E}$ used to heat, sustain, control, and/or assemble the fusion plasma\cite{Freidberg_2007} (see Fig.~\ref{fig:conceptual_plant}). Some fusion designs do not recirculate electrical power but rather recirculate mechanical power
(see Appendix~\ref{sec:mechanical_recirculating power}).
For the case of electrical recirculating power it is straightforward to show that
\begin{equation}
    \label{eq:engineering_gain}
    Q_{\rm eng} = \eta_{\rm elec}\eta_E(\eta_{\rm abs} Q_{\rm fuel} +1) - 1,
\end{equation}
where $\eta_E$, $\eta_{\rm abs}$, and $\eta_{\rm elec}$ are the efficiencies of 
going from $P_{\rm in}^{E}\rightarrow P_{\rm ext}$, $P_{\rm ext}\rightarrow P_{\rm abs}$, and $P_{\rm out} \rightarrow P_{\rm out}^{E}$,
respectively. Note that we have included the portion of $P_{\rm ext}$ that is {\em not} absorbed by the plasma, i.e., $(1-\eta_{\rm abs})P_{\rm ext}$, in $P_{\rm out}$;
this is shown in Fig.~\ref{fig:power_balances} but not explicitly shown in Fig.~\ref{fig:conceptual_plant}.

Finally, the ``wall-plug'' gain,
\begin{equation}
    Q_{\rm wp} = \frac{P_F}{P_{\rm in}^E} = \eta_E \frac{P_F}{P_{\rm ext}} = \eta_E \eta_{\rm abs}\frac{P_F}{P_{\rm abs}},
\end{equation}
relates the total fusion power to the power drawn from the grid (i.e., the wall plug) to assemble,
heat, confine, and control the plasma. This is a useful energy gain metric for
all contemporary fusion experiments because they are not yet generating electricity.
We regard the eventual demonstration of $Q_{\rm wp}=1$ (not $Q_{\rm fuel}$ or $Q_{\rm sci}=1$) as the so-called ``Kitty Hawk moment'' for fusion energy.

Direct conversion from charged fusion products to electricity could be realized with advanced fusion fuels (e.g., D-$^3$He and p-$^{11}$B), which produce nearly all of their fusion energy in charged products. This could raise $\eta_{\rm elec}$ from approximately 40\% to $>80$\% and enable significantly higher $Q_{\rm eng}$ for a given $Q_{\rm fuel}$ or $Q_{\rm sci}$.

For D-T fusion with a tritium-breeding blanket, the \mbox{$^6$Li(n,$\alpha$)T} reaction to
breed tritium is exothermic (releasing 4.8~MeV per reaction),
thus amplifying $P_{\rm out}$ by a factor
of approximately 1.15 depending on the blanket design.
For the purposes of this paper, this factor
can be considered to be absorbed into $\eta_{\rm elec}$.

\begin{figure}[!tb]
    \includegraphics[width=8cm]{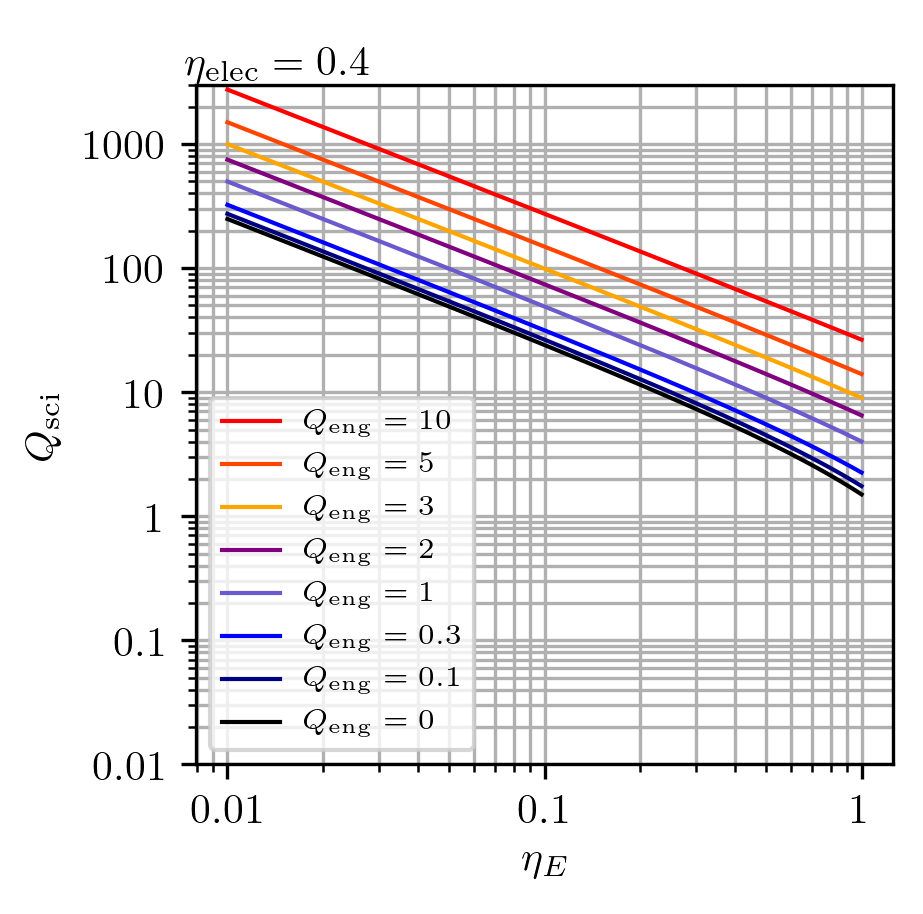}
    \caption{Required combinations of $Q_{\rm sci}$ and $\eta_E$ in the system shown in Fig.~\ref{fig:conceptual_plant} to permit values of $Q_{\rm eng}$ ranging from zero (i.e., $P_{\rm grid}^{E}=0$) to ten (i.e., $P_{\rm grid}^{E}=10P_{\rm in}^{E}$), where $\eta_{\rm elec}=0.4$ is assumed.}
    \label{fig:Qeng}
\end{figure}

Using $Q_{\rm sci} = \eta_{\rm abs} Q_{\rm fuel}$, we can rewrite Eq.~(\ref{eq:engineering_gain}) as
\begin{equation}
    \label{eq:engineering_gain_Q_scientific}
    Q_{\rm eng} = \eta_{\rm elec}\eta_E(Q_{\rm sci} +1) - 1.
\end{equation}
Because $Q_{\rm sci}$ encapsulates all the {\em plasma-physics aspects}
of both the absorption efficiency $\eta_{\rm abs}$ and fuel gain $Q_{\rm fuel}$, it is instructive to plot the required combinations of $Q_{\rm sci}$ and $\eta_E$, assuming $\eta_{\rm elec}=0.4$
(representative of a standard steam cycle and blanket gain), to
achieve certain values of $Q_{\rm eng}$ (see
Fig.~\ref{fig:Qeng}). A convenient rule-of-thumb is that the gain-efficiency product must 
exceed 10 for practical fusion energy, i.e., $Q_{\rm sci} \eta_E \ge 10$ (corresponding
to $Q_{\rm eng}\approx 3$ in Fig.~\ref{fig:Qeng}), but of course the
actual requirement depends on the required economics of the fusion-energy system.

While the value of $\eta_{\rm elec}$ would be around 0.4 for a standard steam cycle for D-T fusion (and higher if an advanced
power cycle is used), the values of $\eta_E$ and $\eta_{\rm abs}$ vary considerably depending on the class of fusion concept (see Table \ref{tab:efficiency_table}).
\input{table_4}
For MCF/MIF, $\eta_E > 0.5$ is expected (conservatively), meaning that $Q_{\rm sci} \gtrsim 20$ is required.  For laser-driven ICF, $\eta_E\sim 0.1$ is expected, meaning that $Q_{\rm sci} \gtrsim 100$ is required.  For an eventual fusion power plant, the required
$Q_{\rm sci}$ and $Q_{\rm eng}$ will depend on a number of factors including but not limited to
market constraints (e.g., levelized cost of
electricity and desired value of $P^{E}_{\rm grid}$) and the maximum achievable values of $\eta_E$, $\eta_{\rm elec}$.

\begin{figure}[!tb]
    \includegraphics[width=8cm]{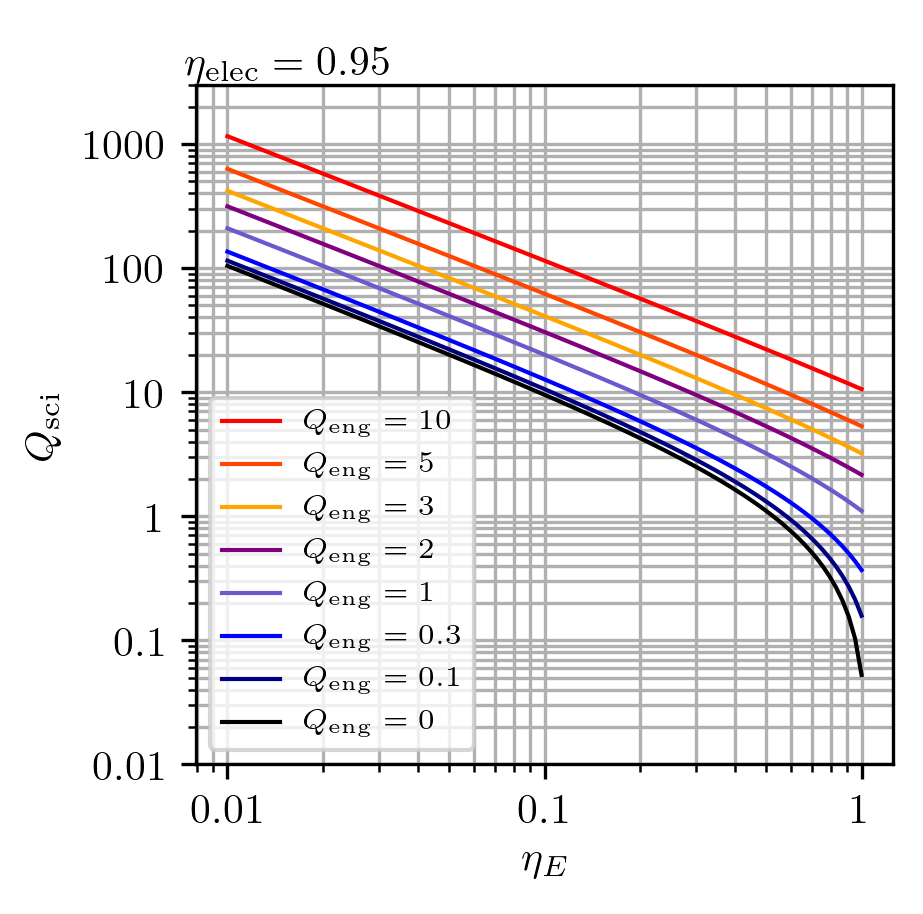}
    \caption{Required combinations of $Q_{\rm sci}$ and $\eta_E$ in the system shown in Fig.~\ref{fig:conceptual_plant} to permit values of $Q_{\rm eng}$ ranging from zero (i.e., $P_{\rm grid}^{E}=0$) to ten (i.e., $P_{\rm grid}^{E}=10P_{\rm in}^{E}$), where $\eta_{\rm elec}=0.95$ is assumed. Note that at high $\eta_E$ and $\eta_{\rm elec}$, net electricity generation ($Q_{\rm eng} > 0$) is possible with $Q_{\rm sci} < 1$.}
    \label{fig:Qeng_high_efficiency}
\end{figure}

In Sec.~\ref{sec:lawson_second}, we noted Lawson's observation (in the context of his second scenario) that if a fraction $\eta$ of the plasma energy at the conclusion of the pulse is recovered as electrical or mechanical energy, the requirement on $n\tau$ to achieve a given value of $Q_{\rm fuel}$ is reduced by a factor $1/(1-\eta)$. In principle, this can be extended to recover $P_{\rm out}$ with an efficiency $\eta_{\rm elec}$ and reinject the recirculating fraction with efficiency  $\eta_{E}$, thus relaxing the requirements on $Q_{\rm sci}$ to achieve a given $Q_{\rm eng}$. This is shown in Fig.~\ref{fig:Qeng_high_efficiency}, which assumes a high recovery fraction $\eta_{\rm elec}=0.95$. If we also assume a high electricity to heating efficiency $\eta_E=0.9$, $Q_{\rm eng}=0.3$ (corresponding to net electricity) can be achieved with $Q_{\rm sci}=0.5$. While it may appear counter-intuitive that net electricity can be generated in a system 
with $Q_{\rm sci}<1$, a high $\eta_{\rm elec}$ and $\eta_E$ mean that most of the recovered heating energy recirculates while most of the fusion energy is used for electricity generation.
The lower-right quadrant Fig.~\ref{fig:Qeng_high_efficiency} (corresponding to high re-injection efficiency) illustrates that that net electricity generation (i.e., $Q_{\rm eng} > 0$) is possible at values of scientific gain below break-even (i.e., $Q_{\rm sci} < 1$).

\section{Methodologies for Inferring Lawson Parameter and Temperature}
\label{sec:methodology}
It is not trivial to infer the component
values of the Lawson parameter and temperature achieved in real  experiments.
Simplifying approximations must be made with certain caveats, both
across (e.g., MCF vs.\ ICF) and within classes (e.g., tokamaks vs.\ mirrors
within MCF) of fusion experiments. In this section, we describe the
methodologies that we use to infer the component
values of achieved Lawson parameters and temperatures
for different fusion classes and concepts, and how the values can be meaningfully compared against each other.
For all values reported here, we require that experimentally inferred values occur within a single shot or across multiple well-reproduced shots. An example that we would disqualify would be to combine the highest $T_i$ achieved in one shot with the highest $n_i$ and $\tau_E$ from a qualitatively different shot.

\subsection{MCF methodology}
\label{sec:mcf_methodology}

The analysis presented in Sec.~\ref{sec:MCF_extending_lawson} assumes that $T_i=T_e$ and $n_i=n_e$, and that they are spatially uniform and time independent. In real experiments, these assumptions are generally
not valid.
Because diagnostic capabilities are finite, only a subset of the complete data (i.e., spatial profiles and time evolutions) are ever measured and published.
Although many experiments were not aiming to maximize $n_i$, $T_i$, and $\tau_E$
as the goal,
we include these experiments because they provide historical context. Furthermore, the data reported from one experiment may not be easily compared to data reported from another due to differences in definitions. In the remainder of this section, these issues are discussed, and uniform definitions are developed.

\subsubsection{Effect of temporal profiles}
Within a particular experiment, the maximum values of $n_i$, $T_i$, and $\tau_E$ may occur at different times. Where possible we choose the values of these quantities at a single point in time during a ``flat-top'' time period, the duration of which must exceed $\tau_E$. Even though the total pulse duration of some MCF experiments may be of similar magnitude to $\tau_E$, we only consider $\tau_E$ in the Lawson parameter for MCF experiments (as opposed to the expression for $\tau_{\rm eff}$ in Eq.~\ref{eq:effective_lawson_parameter}) because we consider the progress towards energy gain in MCF to be limited by thermal-conduction losses and not pulse duration.

In the literature, tables of parameters are commonly published that report the values of many parameters during such a flat-top time period. Following this convention, Tables~\ref{tab:mainstream_mcf_data_table} and \ref{tab:alternates_mcf_data_table} list parameters relevant to our analysis. The reported parameters are $T_{i0}$, $T_{e0}$, $n_{i0}$, $n_{e0}$, and $\tau_{E}^*$. Not all experiments have published temporal evolution of these quantities. 
In the absence of such data, we use the values reported with the understanding that it is unknown if they occurred simultaneously during the shot (although, as discussed in the previous paragraph, they must occur in the same shot or in shots intended to be the same). This deficiency primarily occurs in experiments prior to 1970 or in small experiments with limited diagnostic capabilities and $n_i T_i \tau_E < 10^{16}$~m$^{-3}$\,keV\,s.

\subsubsection{Effect of spatial profiles}

To quantify the effect of nonuniform temperature and density spatial profiles on the requirements to achieve a certain value of $Q_{\rm fuel}$, which we denote as $\langle Q_{\rm fuel} \rangle$ (brackets refer to volume-averaging over nonuniform profiles), the power balance of Eq.~(\ref{eq:power_balance_steady_state}) becomes
\begin{equation}
\label{eq:power_balance_steady_state_volume_average}
f_c\langle S_F \rangle V + P_{\rm abs} = \langle S_B \rangle V + \frac{3\langle n T \rangle}{\tau_E} V, 
\end{equation}
where power {\em densities} are denoted with variables $S$, and we assume $n=n_e=n_i$ (i.e., hydrogenic plasma without impurities) and $T=T_e=T_i$ everywhere. Reported/inferred values of $P_{\rm abs}$ and $\tau_E$ are already global, volume-averaged quantities.

To quantify the profile effect on $S_F$, we introduce
\begin{equation}
\lambda_{F}=\langle S_F \rangle/S_{F0},
\label{eq:lambda_F}
\end{equation}
where $S_{F0}$ is the fusion power density with spatially uniform $T_{i0}$ and $n_{i0}$, and $\langle S_F \rangle$ is the volume-averaged fusion power density of the 
nonuniform-profile case with peak values $T_{i0}$ and $n_{i0}$.
Similarly,
\begin{equation}
 \label{eq:lambda_B}
\lambda_{B}=\langle S_B \rangle/S_{B0} 
\end{equation}
and
\begin{equation}
 \label{eq:lambda_kappa}
\lambda_{\kappa}=\frac{\langle n T \rangle}{n_0 T_0}, 
\end{equation}
which quantify the nonuniform-profile modifications to the bremsstrahlung power density and thermal energy density, respectively.

The result is a modified version of Eq.~(\ref{eq:power_balance_steady_state}), where profile effects are captured in the terms $\lambda_{F}$, $\lambda_{B}$, and $\lambda_{\kappa}$,
\begin{equation}
\label{eq:power_balance_steady_state_volume_average_lambdas}
\lambda_{F} f_c S_{F0} V + P_{\rm abs} = \lambda_{B} S_{B0} V + \frac{3 \lambda_{\kappa} n_0 T_0} {\tau_E} V. 
\end{equation}
From this power balance of the nonuniform-profile case, the peak value of the Lawson parameter $n_0 \tau_E$ required to achieve a particular value of $\langle Q_{\rm fuel} \rangle$ as a function of $T_0$ is
\begin{equation}
\begin{split}
\label{eq:n_tau_E_vs_T_profile}
    & n_0 \tau_E = \\
    & \frac{3 \lambda_{\kappa} T_0}{\lambda_{F}(f_c+\langle Q_{\rm fuel}\rangle^{-1})\langle \sigma v  \rangle \epsilon_{F} / 4 - \lambda_{B} C_B T_0^{1/2}},
\end{split}
\end{equation}
where 
\begin{equation}
    \langle Q_{\rm fuel} \rangle = \frac{\lambda_{F} S_{F0}}{P_{\rm abs}/V} \equiv \lambda_{F} Q_{\rm fuel}.
    \label{eq:volume_averaged_Q}
\end{equation}

We adopt the approach of using the same peak (rather than average)
values of density
and temperature when evaluating $Q_{\rm fuel}$ (uniform
spatial profiles) versus $\langle Q_{\rm fuel} \rangle$ (nonuniform spatial profiles), for
the practical reasons that peak values are more commonly reported in the literature and that profiles are often not reported. When using
the same peak rather than profile-averaged values, spatially nonuniform profiles increase rather than decrease the requirements on peak density and temperature for achieving a given $Q_{\rm fuel}$.


Next we consider representative profiles in order to quantify the differences
between $Q_{\rm fuel}$ and $\langle Q_{\rm fuel}\rangle$ for cylindrical and toroidal geometries.
A wide variety of temperature and density profiles have been observed in fusion experiments. These profiles are typically modeled as functions of normalized radius $x=r/a$, where $a$ is the device radius for cylindrical systems and the minor radius for toroidal systems
with circular cross section.  
Commonly used and flexible models of density and temperature profiles are
\begin{equation}
    n(x) = n_{0}(1-x^{2})^{\nu_{n}}~\mathrm{and} ~T(x) = T_{0}(1-x^{2})^{\nu_{T}},
    \label{eq:parabolic_profiles}
\end{equation}
where $n_0$ and $T_0$ are the central/peak ion or electron densities and temperatures, respectively. The values of $\nu_{n}$ and $\nu_{T}$ adjust the sharpness of the peaks of the profiles. In the limit $\nu_{n} \rightarrow 0$ and $\nu_{T} \rightarrow 0$, the peak is infinitely broad and we recover the uniform-profile case. This approach accommodates a wide range of profiles.\cite{Kesner_1976,Khosrowpour_2016}

From Eqs.~(\ref{eq:fusion_power_density}) and (\ref{eq:lambda_F}),
\begin{equation}
	\lambda_{F} =  \frac{\langle S_F \rangle}{S_{F0}} = 
    \frac{\langle n_i^2 \langle \sigma v \rangle (T_i(x)) \rangle}{n_{i0}^2 \langle \sigma v \rangle (T_{i0})},
\end{equation}
where the $T_i$ dependence of $\langle \sigma v\rangle$ is shown explicitly, resulting in $\lambda_{F}$ being a function of the $T_i$ profile\@. 
From Eqs.~(\ref{eq:bremsstrahlung_power_density}) and (\ref{eq:lambda_B}),
\begin{equation}
    \lambda_{B} = \frac{\langle S_B\rangle}{S_{B0}}= 
    \frac{\langle n_e^2 T_e^{1/2} \rangle}{n_{e0}^2 T_{e0}^{1/2}}.
\end{equation}

For a cylinder or large-aspect-ratio torus (i.e.,
$R/a \gg 1$, where $R$ and $a$ are the major and minor radii, respectively) with circular cross section and the profiles of Eq.~(\ref{eq:parabolic_profiles}), we use the expressions in Appendix~\ref{sec:volume_averaging} to obtain
\begin{equation}
\label{eq:lambda_F_eval}
	\lambda_{F} = \frac{\int_0^1 (1-x^2)^{2 \nu_{n}} \langle \sigma v\rangle (T_i(x))2x\mathrm{d}x}{\langle \sigma v\rangle(T_{i0})},
\end{equation}
which may be evaluated numerically for any tabulated or parameterized values of $\langle \sigma v \rangle (T_i)$,
\begin{equation}
\lambda_{B} =  \int_0^1 (1-x^2)^{2\nu_{n} + \nu_{T} / 2} 2x\mathrm{d}x  =  \frac{2}{4 \nu_{n} + \nu_{T} + 2},
\end{equation}
and
\begin{equation}
\lambda_{\kappa} = \int_0^1 (1-x^2)^{(\nu_{n} + \nu_{T})} 2x \mathrm{d}x = \frac{1}{1+\nu_{n}+\nu_{T}}.
\end{equation}

For a torus with circular cross section and arbitrary values of $R/a$, $\lambda_{F}$, $\lambda_{B}$, and $\lambda_{\kappa}$ must be evaluated numerically (see Appendix~\ref{sec:volume_averaging})\@.
For profiles with large Shafranov shift, i.e., magnetic axis shifted toward larger
$R$, the reduction of fusion power due to profile effects (and hence $\lambda_{F}$) is mitigated because the high-temperature region occupies a larger fraction of the total volume. Therefore the profiles considered here represent a likely worst-case scenario and provide a lower bound on $\lambda_{F}$.

\begin{figure}[htbp]
\begin{flushleft}(a)\end{flushleft}
   \includegraphics[width=8cm]{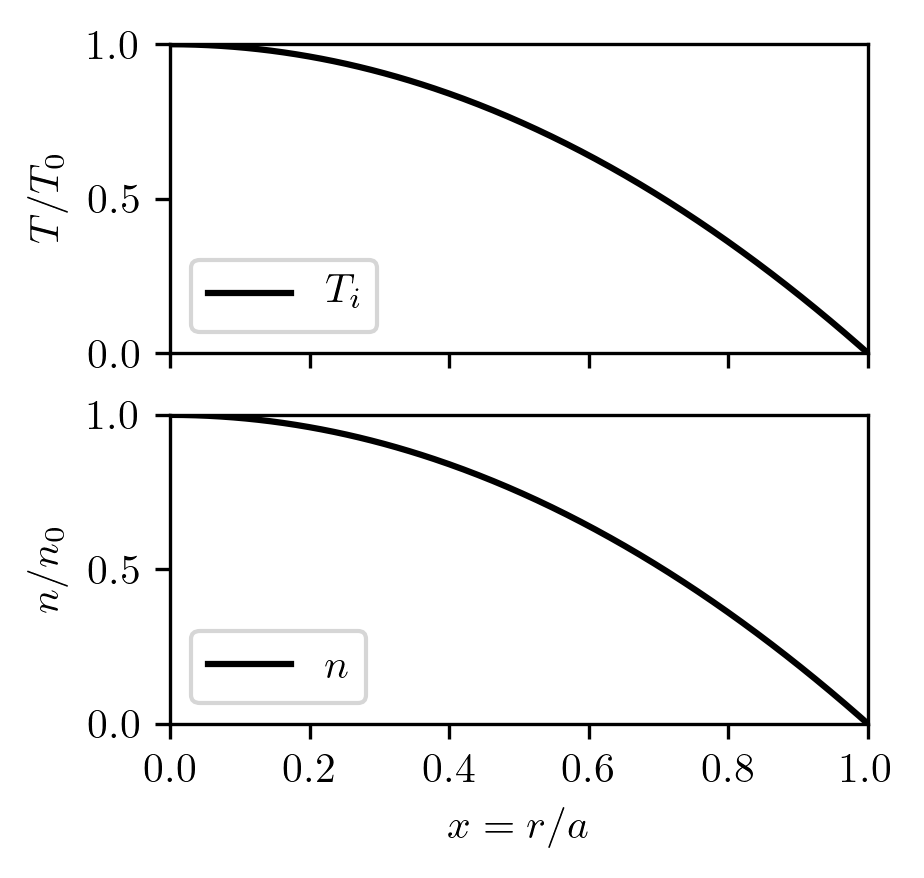}
\begin{flushleft}(b)\end{flushleft}
    \includegraphics[width=8cm]{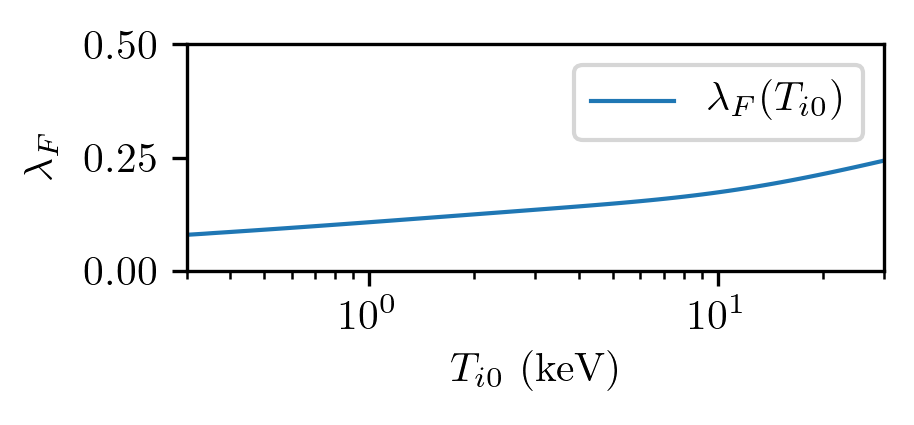}
\begin{flushleft}(c)\end{flushleft}
    \includegraphics[width=8cm]{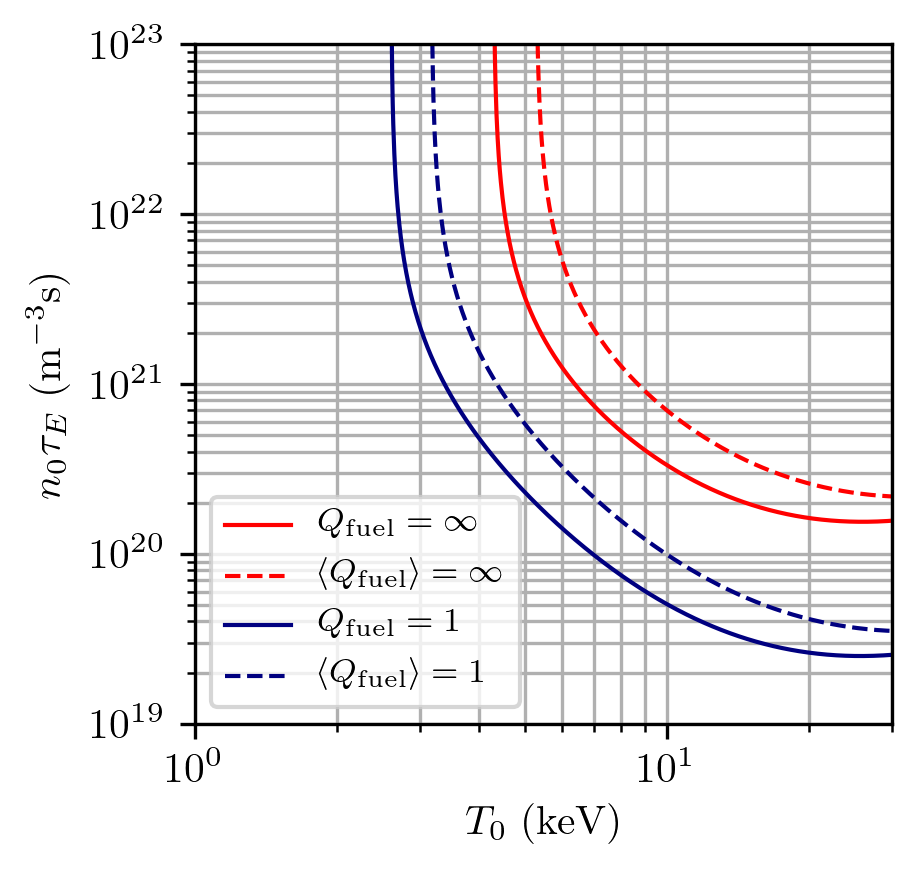}
    \caption{(a)~Normalized parabolic profiles (with $\nu_{T}=1$ and $\nu_{n}=1$) of $T=T_i=T_e$ and $n = n_i = n_e$. (b)~Parameter $\lambda_{F}$ vs.\ $T_{i0}$ ($\lambda_{B}=0.286$ and $\lambda_{\kappa}=0.333$ for these profiles). (c)~Peak Lawson parameter vs.\ $T_0$ for the parabolic profiles (dashed lines) shown in (a) and uniform plasma (solid lines), for $Q_{\rm fuel}=1$ (blue) and $Q_{\rm fuel}=\infty$ (red).}
\label{fig:parabolic_profiles}
\end{figure}

\begin{figure}[htbp]
\begin{flushleft}(a)\end{flushleft}
    \includegraphics[width=8cm]{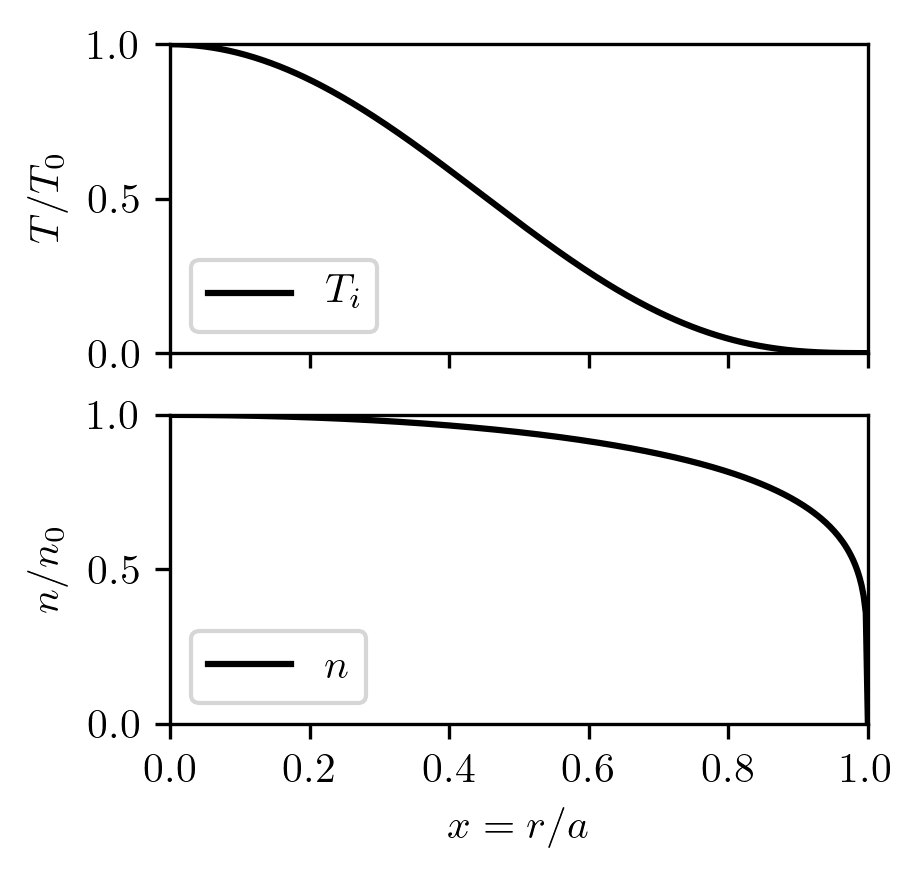}
\begin{flushleft}(b)\end{flushleft}
    \includegraphics[width=8cm]{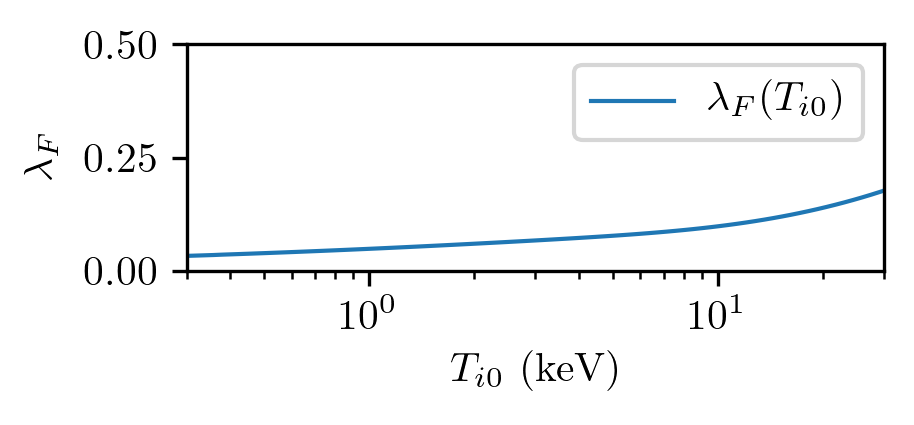}
\begin{flushleft}(c)\end{flushleft}
    \includegraphics[width=8cm]{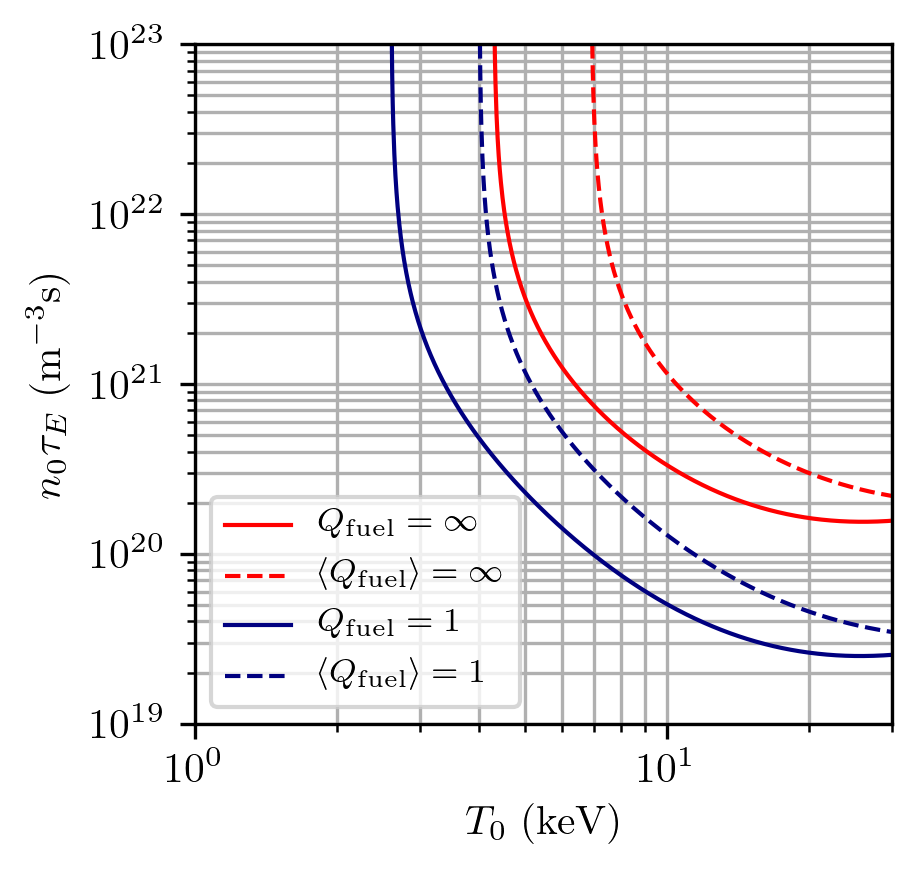}
    \caption{(a)~Normalized peaked and broad profiles (with $\nu_{T}=3$ and $\nu_{n}=0.2$) of $T=T_i=T_e$ and $n = n_i = n_e$. (b)~Parameter $\lambda_{F}$ vs.\ $T_{i0}$ ($\lambda_{B}=0.345$ and $\lambda_{\kappa}=0.238$ for these profiles). (c)~Lawson parameter vs.\ $T_{i0}$ for the profiles (dashed lines) shown in (a) and uniform plasma (solid lines), for $Q_{\rm fuel}=1$ (blue) and $Q_{\rm fuel}=\infty$ (red).}
\label{fig:peaked_broad_profiles}
\end{figure}

To demonstrate the effect of nonuniform 
profiles on the contours of
$\langle Q_{\rm fuel}\rangle$ compared
to $Q_{\rm fuel}$, we consider two sets of profiles. The first is a parabolic profile with $\nu_{T}=1$ and $\nu_{n}=1$, which is a simple approximation of the profiles in tokamaks.\cite{Wesson_2011}
The second is a more strongly peaked temperature profile with $\nu_{T}=3$ and a broader density profile with $\nu_{n}=0.2$, which are representative of profiles in the advanced tokamak or reversed-field pinch.\cite{Chapman_2002} For both sets of profiles, we assume $T=T_i=T_e$ and $n=n_i=n_e$ (impurity-free hydrogenic plasma).
Figures~\ref{fig:parabolic_profiles} and \ref{fig:peaked_broad_profiles} show these two
sets of profiles, respectively, along with their corresponding values of $\lambda_{F}$ vs.\ $T_{i0}$ and resulting
adjustments to the $Q_{\rm fuel}$ contours.
For both sets of profiles (Figs.~\ref{fig:parabolic_profiles} and \ref{fig:peaked_broad_profiles}), nonuniform profiles [dashed lines in panel (c)] increase the peak Lawson parameter needed to achieve a particular value of $\langle Q_{\rm fuel} \rangle$ for temperatures below approximately 50~keV\@. Additionally, the ideal ignition temperature, defined by Eq.~(\ref{eq:ideal_ignition}), is increased.  At high temperatures approaching 100~keV, where fusion power exceeds bremsstrahlung by a large factor (see Fig.~\ref{fig:ideal_ignition}), the adjustment is equal to the ratio $\lambda_{\kappa} / \lambda_{F}$, which is close to unity in the case of the parabolic profiles, and drops below unity in the case of the peaked and broad profiles.
At intermediate temperatures, $\lambda_{F}$, $\lambda_{B}$, and $\lambda_{\kappa}$ all contribute to the modification of
$\langle Q_{\rm fuel}\rangle$ compared to $Q_{\rm fuel}$.

\subsubsection{Effect of impurities (and non-hydrogenic plasmas)}

Real fusion experiments must contend with the effect of ions with charge state
$Z>1$. These may be from helium ash, impurities from the first wall, or advanced fuels.
These impurities increase the bremsstrahlung radiation by a factor
\begin{equation}
\label{eq:Z_eff}
Z_{\rm eff} = \frac{\Sigma_i n_i Z_i^2}{n_e},
\end{equation}
where $i$ is summed over all ion species in the plasma. Additionally, impurities increase the electron density relative to the ion density by a factor
of the mean charge state of the entire plasma,
\begin{equation}
    \bar{Z} = n_e / n_i,
\end{equation}
which reduces $n_i$ and therefore $P_F$ at fixed pressure.

Using these definitions along with the generalized expression
for bremsstrahlung,
\begin{equation}
\label{eq:generalized_bremsstrahlung_power_density}
P_B = C_B n_e T_e^{1/2} \Sigma_i (Z_i^2 n_i),
\end{equation}
Eq.~(\ref{eq:n_tau_E_vs_T_profile}) becomes
\begin{equation}
\begin{split}
\label{eq:n_tau_E_vs_T_profile_impurities}
    & n_{i0} \tau_E = \\
    & \frac{(3/2)\lambda_{\kappa} (1 + \bar{Z}) T_0}{\lambda_{F}(f_c + \eta_{\rm abs} \langle Q_{\rm sci}\rangle^{-1})\langle \sigma v  \rangle \epsilon_{F} / 4 - \lambda_{B} C_B Z_{\rm eff} \bar{Z}^2 T_0^{1/2}},
\end{split}
\end{equation}
and 
\begin{equation}
\begin{split}
\label{eq:n_T_tau_E_vs_T_profile_impurities}
    & n_{i0} T_0 \tau_E = \\
    & \frac{(3/2)\lambda_{\kappa} (1 + \bar{Z}) T_0^2}{\lambda_{F}(f_c + \eta_{\rm abs} \langle Q_{\rm sci}\rangle^{-1})\langle \sigma v  \rangle \epsilon_{F} / 4 - \lambda_{B} C_B Z_{\rm eff} \bar{Z}^2 T_0^{1/2}},
\end{split}
\end{equation}
where $\lambda_{F}$, $\lambda_{B}$, and $\lambda_{\kappa}$ are unchanged because $Z_{\rm eff}$ and $\bar{Z}$ are treated as volume-averaged quantities. We have also replaced the $\langle Q_{\rm fuel} \rangle ^{-1}$ term with $\eta_{\rm abs} \langle Q_{\rm sci} \rangle ^{-1}$, which allows us to include the effect of absorption efficiency. 

\begin{figure}[htb!]
    \includegraphics[width=8cm]{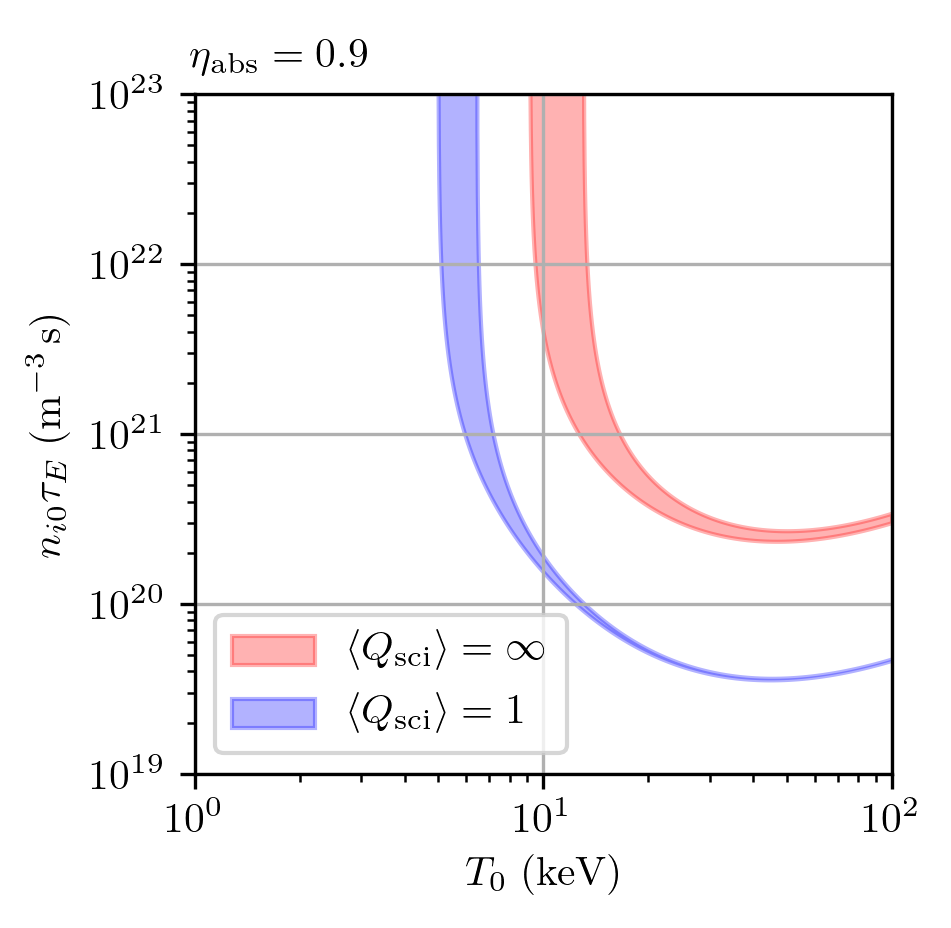}
    \caption{Finite-width $\langle Q_{\rm sci} \rangle$ contours 
    vs.\ peak Lawson parameter and $T_0$ bounded by low-impurity ($Z_{\rm eff}=1.5$, $\bar{Z}=1.2$) and high-impurity ($Z_{\rm eff}=3.4$, $\bar{Z}=1.2$) cases,
    for peaked and broad spatial profiles ($\nu_{T}=3$, $\nu_{n}=0.2$). These assumptions are made in plotting the $Q_{\rm sci}$ contours of Figs.~\ref{fig:scatterplot_ntauE_vs_T}, \ref{fig:scatterplot_nTtauE_vs_year}, and \ref{fig:scatterplot_nTtauE_vs_T}.}
    \label{fig:nTtauE_vs_T_peaked_and_broad_bands}
\end{figure}

Each experiment has different values of $\lambda_{F}$, $\lambda_{B}$, $\lambda_{\kappa}$, $\bar{Z}$, $Z_{\rm eff}$, and $\eta_{\rm abs}$, and therefore
each experiment has different $\langle Q_{\rm sci} \rangle$ contours.  It is not feasible to show unique $\langle Q_{\rm sci} \rangle$ contours for each experiment in Figs.~\ref{fig:scatterplot_ntauE_vs_T}, \ref{fig:scatterplot_nTtauE_vs_year}, and \ref{fig:scatterplot_nTtauE_vs_T}.
Figure~\ref{fig:nTtauE_vs_T_peaked_and_broad_bands} shows finite-width $\langle Q_{\rm sci} \rangle$ contours of the peaked and broad profiles whose lower and upper limits correspond to low-impurity ($Z_{\rm eff}=1.5$, $\bar{Z}=1.2$) and high-impurity ($Z_{\rm eff}=3.4$, $\bar{Z}=1.2$) models, respectively. These impurity levels correspond to the range of impurity levels considered for SPARC\cite{Rodriguez-Fernandez_2020} and ITER.\cite{Mukhovatov_2003} For both the high and low-impurity models, we assume $T=T_i=T_e$ and $\eta_{\rm abs}=0.9$.
The finite ranges of $\langle Q_{\rm sci} \rangle$ aim to account for the main features and uncertainties of a future experimental device that will achieve $\langle Q_{\rm sci}\rangle>1$, and therefore we
show finite-width $\langle Q_{\rm sci}\rangle$ contours in Fig.~\ref{fig:scatterplot_ntauE_vs_T} (despite the
$Q_{\rm sci}$ labels in the legend). 
We emphasize that the finite width of the $\langle Q_{\rm sci}\rangle$ contours
are merely illustrative of the effects of profiles and impurities and of the approximate values of $\langle Q_{\rm sci}\rangle$ that might
be achieved by SPARC or ITER\@. To predict $\langle Q_{\rm sci}\rangle$ with higher precision would require detailed analysis and simulations.

\subsubsection{Inferring peak from volume-averaged values}
\label{sec:inferring_peak_from_average}

When only volume-averaged values of density and temperature
are reported, we infer the peak values from an estimated value of the peaking, $T_0 / \langle T \rangle$ and $n_0 / \langle n \rangle$, respectively. Detailed empirical models of peaking exist for predicting the profiles of future experiments.\cite{Angioni_2007,Greenwald_2007,Takenaga_2008,Angioni_2009} However, for the purposes of this paper, we have chosen peaking values on a per-concept basis, the values of which are indicated in Table~\ref{tab:mcf_peaking_values_table}. Only concepts for which peak values must be inferred from reported volume-averaged values, along with citations for those values, are listed in Table~\ref{tab:mcf_peaking_values_table}.
In Tables~\ref{tab:mainstream_mcf_data_table} and~\ref{tab:alternates_mcf_data_table}, we append a superscript asterisk ($^*$) to peak values inferred from reported volume-averaged quantities.
\input{table_5}

\subsubsection{Inferring ion quantities from electron quantities}
\label{sec:inferring_ion_quantities_from_electron_quantities}

When only $T_e$ and not $T_i$ is reported, we cannot assume $T_i=T_e$ in calculating the triple product without further consideration. If the thermal-equilibration time is much shorter than the plasma duration, and assuming there are
no other effects that would give rise to $T_i \ne T_e$, then
we can assume $T_i = T_e$. In these cases we append a superscript dagger ($^\dagger$) to the inferred value of $T_i$ in Tables~\ref{tab:mainstream_mcf_data_table} and~\ref{tab:alternates_mcf_data_table}.
In cases where both $T_i$ and $T_e$ are reported in MCF experiments, we use the reported $T_i$.

When only $n_e$ but not $n_i$ is reported, we assume $n_i=n_e$ for D-T and D-D plasmas. In such cases we append a superscript double dagger ($^\ddagger$) to the inferred value of $n_i$ in Tables~\ref{tab:mainstream_mcf_data_table} and~\ref{tab:alternates_mcf_data_table}.

\subsubsection{Accounting for transient heating}
\label{sec:accounting_for_transient_effects}

All experiments experience a transient start-up phase during which a portion of the heating power goes into raising the plasma thermal energy $W_{p}=3nTV$ (assuming
$T=T_i=T_e$ and $n=n_i=n_e$).
There are two self-consistent approaches for deriving an expression for $Q_{\rm fuel}$ that accounts for the effect of transient heating $\mathrm{d}W_p/\mathrm{d}t$. In the remainder of this subsection, we closely follow Ref.~\onlinecite{Meade1997}.

The first approach is to group the transient term with $P_{\rm abs}$ in the instantaneous power balance which effectively treats the transient term as a reduction in the externally applied and absorbed heating power,
\begin{equation}
\label{eq:power_balance_transient_JET_JT-60}
    \left(P_{\rm abs} - \dv{W_{p}}{t}\right) + P_{c} = P_B + \frac{3nTV}{\tau_{E}}.
\end{equation}
In this approach, the definition of $Q_{\rm fuel}$ is modified, i.e., 
\begin{equation}
    \label{eq:Qsci_JET_JT-60}
    Q^*_{\rm fuel} = \frac{P_F}{P_{\rm abs}-\mathrm{d}W_p/\mathrm{d}t}.
\end{equation}
From here, we derive an expression for the Lawson parameter
following the same steps as Sec.~\ref{sec:MCF_extending_lawson},
which results in an analogous expression to Eq.~(\ref{eq:MCF_Lawson_parameter_Q_fuel})
but with $Q_{\rm fuel}$ replaced by $Q_{\rm fuel}^*$,
\begin{equation}
\label{eq:triple_product_transient_JET_JT-60}
n \tau_E = \frac{3T}{(f_c+ Q_{\rm fuel}^{*-1})\langle \sigma v  \rangle \epsilon_{F}/4 - C_{B} T^{1/2}}.
\end{equation}
From Eq.~(\ref{eq:power_balance_transient_JET_JT-60}), 
\begin{equation}
\label{eq:tau_E_transient_JET_JT-60}
    \tau_E = \frac{W_p}{P_{\rm heat} - \mathrm{d}W_p/\mathrm{d}t},
\end{equation}
where
\begin{equation}
\label{eq:P_heat}
P_{\rm heat} = P_{\rm abs} + P_c - P_B.
\end{equation}
This approach, defined by  Eqs.~(\ref{eq:Qsci_JET_JT-60})--(\ref{eq:P_heat}), is the one used by JET and JT-60.

The second approach is to treat the transient heating term as a ``loss'' term alongside thermal conduction, i.e., 
\begin{equation}
\label{eq:power_balance_transient_TFTR}
    P_{\rm abs} + P_{c} = P_B + \left(\frac{3nTV}{\tau_{E}} + \dv{W_{p}}{t}\right).
\end{equation}
We then define a modified energy confinement time $\tau_E^*$ which characterizes thermal conduction and transient heating power.
\begin{equation}
    \frac{3nTV}{\tau_E^*} = \frac{3nTV}{\tau_E} + \dv{W_{p}}{t}.
\end{equation}
Combining the latter with Eqs.~(\ref{eq:P_heat}) and (\ref{eq:power_balance_transient_TFTR}),
\begin{equation}
\label{eq:tau_E_transient_lawson_tftr}
   \tau_E^* = \frac{W_p}{P_{\rm heat}}.
\end{equation}
From this point, we derive an expression for the Lawson parameter following the same steps as Sec.~\ref{sec:MCF_extending_lawson}, which results in 
an analogous expression to Eq.~(\ref{eq:MCF_Lawson_parameter_Q_fuel}) but with $\tau_E$ replaced by $\tau_E^*$,
\begin{equation}
\label{eq:triple_product_transient_lawson_tftr}
n \tau_E^* = \frac{3T}{(f_c+ Q_{\rm fuel}^{-1})\langle \sigma v  \rangle \epsilon_{F}/4 - C_{B} T^{1/2}}.
\end{equation}
In this formulation, the definition of instantaneous $Q_{\rm fuel}$ is unchanged from the steady-state value of Eq.~(\ref{eq:fuel_gain}),
and fuel breakeven occurs at $Q_{\rm fuel}=1$, regardless of the value of $\mathrm{d}W_p/\mathrm{d}t$.
This approach, defined by Eqs.~(\ref{eq:tau_E_transient_lawson_tftr}), 
(\ref{eq:triple_product_transient_lawson_tftr}), and
(\ref{eq:scientific_gain}), is the one used by TFTR and consistent with Lawson's original formulation.

For the JET/JT-60 approach, fuel breakeven does not necessarily occur at $Q_{\rm fuel}^*=1$ but rather occurs at a value of $Q_{\rm fuel}^*$ that depends on the value of $\mathrm{d}W_p/\mathrm{d}t$. The TFTR/Lawson approach keeps the definition of instantaneous $Q_{\rm fuel}$ the same as the steady-state $Q_{\rm fuel}$, and fuel breakeven always occurs at $Q_{\rm fuel}=1$ regardless of the transient-heating value.
Because a key objective of this paper is to chart the progress
of many different experiments toward and beyond $Q_{\rm fuel}=1$,
we use the TFTR/Lawson definition for which $Q_{\rm fuel}=1$ means the same thing across different MCF experiments.
In practice, this means we use $\tau_E^*$
and Eq.~(\ref{eq:triple_product_transient_lawson_tftr}) for all MCF
experiments. When
$\tau_E$ and $\mathrm{d}W_p/\mathrm{d}t$ are reported and $\mathrm{d}W_p/\mathrm{d}t$ is nonzero (e.g., JET and JT-60), we calculate and use $\tau_E^*$, indicating such cases with a superscript hash ($^{\#}$) in Tables~\ref{tab:mainstream_mcf_data_table} and~\ref{tab:alternates_mcf_data_table}\@. Some TFTR publications report $\tau_E$, requiring the conversion step, and thus we append a superscript hash for those cases as well.

\subsection{ICF methodology}
\label{sec:ICF_methodology}
Direct measurements of plasma parameters are more challenging for ICF\@. Commonly measured parameters in ICF are fuel areal density $\rho R$ (via neutron downscattering), $T_i$ and ``burn duration'' (via neutron time-of-flight), and neutron yield (via various types of neutron detectors). Some experiments report an inferred stagnation pressure $p_{\rm stag}$ based on statistical analysis of other measured quantities and simulation databases.

\begin{figure}[tb!]
\centerline{\includegraphics[width=8cm]{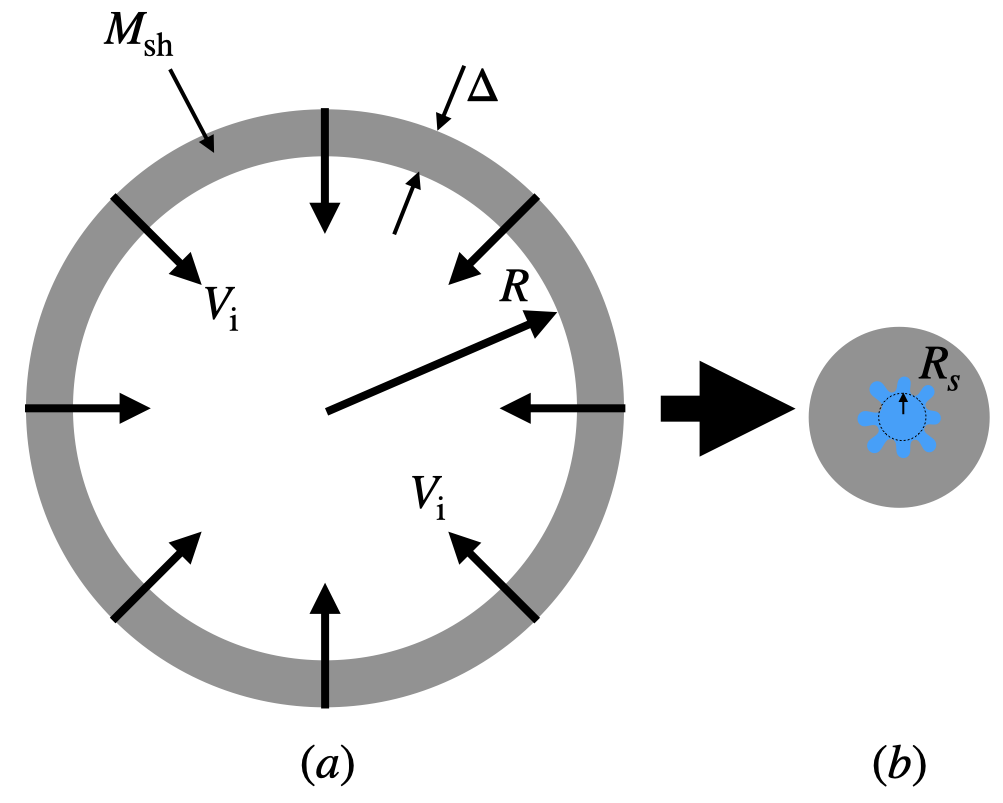}}
\caption{Representation of an ICF capsule implosion and hot-spot creation with instability growth: (a)~dense fuel shell, 
with radius $R$ and thickness $\Delta$, at maximum shell velocity $V_i$ during implosion, (b) fuel assembly at stagnation with the ``hot spot'' (blue) with effective radius $R_s$, surrounded by the cold, dense fuel (grey). Rayleigh-Taylor instabilities are shown. If the hot spot reaches high-enough $n_i \tau$ and $T_i$,
then it can potentially generate enough fusion energy to initiate a
propagating burn into the surrounding cold shell.}
\label{fig:conceptual_icf_detailed}
\end{figure}

Identifying the requirements for ignition of an ICF capsule is difficult. The analysis presented in Sec.~\ref{sec:ICF_extending_lawson} assumes an idealized ICF scenario. Real ICF experiments must contend with instabilities, impurities, non-zero bremsstrahlung and thermal-conduction losses, and other factors that make it more difficult to achieve ignition. For the highest-performing ICF experiments considered here (NIF, OMEGA), a two-stage approach to ignition is pursued, i.e., ignition of a central lower-density ``hot spot'' followed by propagating burn into the surrounding colder, denser fuel, as depicted in Fig.~\ref{fig:conceptual_icf_detailed}. Because of the low value of $n_{abs}$ inherent in these experiments, this two-stage process is required to achieve $Q_{\rm sci}>1$. Therefore, we consider both ignition of the hot spot and a propagating burn in the dense fuel when we refer to ``ignition'' in this section.

Below we describe two methodologies used in this paper for 
inferring the Lawson parameter $n\tau$ and triple product $nT\tau$
for cases in which pressure is or is not experimentally inferred, respectively.

\subsubsection{Inferring Lawson parameter and triple product without reported inferred pressure}

For ICF experiments that do not report experimentally inferred values of
fuel pressure (i.e., rows with ``--'' in the $p_{\rm stag}$ column of Table~\ref{tab:icf_mif_data_table}), we employ the methodology of \textcite{Betti_2010} to infer $n_i\tau$ from other measured ICF experimental quantities. Here, we state the key logic and equation of this methodology for
the convenience of the reader, but we refer the reader to Ref.~\onlinecite{Betti_2010} for further details, equation derivations, and justifications. It is important to
note that Ref.~\onlinecite{Betti_2010} makes a simplifying assumption that
thermal-conduction and radiation losses are negligible (on the
timescale of the fusion burn) because of the insulating effects of
the dense shell of an ICF target capsule, meaning that Lawson parameters and
triple products inferred
via this method should be considered as upper bounds.

The ICF-capsule shell 
is modeled as a thin shell with thickness $\Delta \ll R$, where
$R$ is the shell radius, as illustrated in Fig.~\ref{fig:conceptual_icf_detailed}.
A fraction of the peak kinetic energy of the shell is assumed to be converted to
thermal pressure in the hot spot at stagnation. An {\em upper bound} on $\tau$
is obtained based on the time it takes for the stagnated shell (at peak compression) to expand a distance of order its inner radius $R_s$\@. Significant
3D effects arising from Rayleigh-Taylor-instability spikes and bubbles
at the interface of the shell and hot spot reduce the effective
hot-spot volume by a ``yield-over-clean'' factor $YOC^\mu$, where
$\mu\sim 0.4$--0.5 is inferred from two 
simulation databases.\cite{Chang2010} With these
and other simplifying assumptions, \textcite{Betti_2010} obtain
\begin{equation}
n T \tau~{\rm (3D)} \approx 4[(\rho R)_{\rm tot(n)}^{\rm no \; \alpha} T_n^{\rm no \; \alpha}]^{0.8} YOC^{\mu}~[\mathrm{atm\,s}],
\label{eq:ICF_triple_product_3D}
\end{equation}
with measured total areal density $(\rho R)_{\rm tot(n)}^{\rm no\; \alpha}$ in g\,cm$^{-2}$, and measured ``burn-averaged'' ion 
temperature $T_n^{\rm no\; \alpha}$ in keV\@.
The superscript ``$\rm no \; \alpha$'' refers to experimental measurements made when $\alpha$ heating is not an appreciable effect (and $\alpha$ heating is turned off in simulations). 
For ICF experiments without reported values of hot-spot pressure,
Eq.~(\ref{eq:ICF_triple_product_3D}) is used to plot
achieved ICF values of Lawson parameters and triple products, where the unit [atm\,s] is multiplied by 
$6.333\times 10^{20}$~keV\,m$^{-3}$\,atm$^{-1}$ to convert to [m$^{-3}$\,keV\,s]. Dividing the triple product by $T$ gives the Lawson parameter $n\tau$.

\subsubsection{Inferring Lawson parameter from inferred pressure and confinement dynamics}

When the inferred stagnation pressure $p_{\rm stag}$ and the duration of fuel stagnation $\tau_{\rm stag}$ are reported, the pressure times the confinement time $\tau$ can be calculated directly. However, following Christopherson,\cite{Christopherson_2019}
three adjustments are made to $\tau_{\rm stag}$, which is defined as the full-width half-maximum (FWHM) of the neutron-emission history
(i.e., ``burn duration''),
to obtain an approximation for $\tau$. The first adjustment is that,
for marginal ICF ignition, only alphas produced before bang time (time of maximum neutron production) are useful to ignite the hot spot because, afterward, the shell is expanding and the hot spot is cooling, reducing the reaction rate; this introduces a factor of 1/2. The second adjustment is that only a fraction of fusion alphas are absorbed by the hot spot; this factor is estimated to be 0.93. The third adjustment is that, to initiate a propagating burn of the surrounding fuel, an additional factor of 0.71 is applied to account for the dynamics of alpha heating of the cold shell. Applying these three corrections results in $\tau \approx \tau_{\rm stag}/3$ and
\begin{equation}
n T \tau \approx \frac{1}{2}p_{\rm stag} \tau \approx \frac{1}{6}p_{\rm stag} \tau_{\rm stag}.
\label{eq:ICF_triple_product}
\end{equation}
The only exception to this approach is the FIREX experiment, for which we estimate the value of $p_{\rm stag} \tau$ directly from the reported values.

\subsubsection{Adjustments to the required values of Lawson parameter and temperature required for ignition}
The ignition requirement derived in Sec.~\ref{sec:ICF_extending_lawson} ignores a number of factors that increase the requirements for ignition of an ICF capsule. We consider these effects to be incorporated in reductions to $\tau$ in the previous subsection. Thus, no further adjustments are made to the contours of constant $Q_{\rm sci}$ defined by Eq.~(\ref{eq:ICF_Lawson_parameter_Q_sci}).

\subsubsection{Differences between ICF and MCF}
It is not straightforward to compare the achieved Lawson parameters and triple-product values between ICF and MCF\@.
While a quantitative approach can be taken via the ignition parameter $\chi$ described in Ref.~\onlinecite{Betti_2010}, the approach taken here is qualitative and is reflected in the different 
$Q_{\rm sci}$ contours for ICF and MCF in Figs.~\ref{fig:scatterplot_ntauE_vs_T}, \ref{fig:scatterplot_nTtauE_vs_T}, and \ref{fig:scatterplot_nTtauE_vs_year}. 

Firstly, the achieved triple product for ICF is higher than for MCF in part because
of two assumptions made in their inference.
Following Ref.~\onlinecite{Betti_2010}, we assume in ICF that there are no bremsstrahlung radiation losses due
to trapping by the pusher (with a high-enough areal density to be opaque to x-rays)
and that the fuel hot-spot pressure is spatially uniform. These assumptions lead to 
higher values for the inferred Lawson parameter and triple product.

Secondly, whereas $P_{\rm ext}$ and $P_{\rm abs}$ differ
by only a factor of order unity in MCF,\cite{Freidberg_2007}
they differ by a factor of $\gtrsim 50$ in ICF (see Table \ref{tab:efficiency_table}).
This is due to the low conversion efficiency from applied laser energy to absorbed fuel energy.
Thus, while both MCF\cite{Keilhacker_1999} and ICF\cite{ScienceNews2021} have achieved $Q_{\rm sci} \sim 0.7$, ICF has necessarily achieved a higher value of $Q_{\rm fuel}$ compared to MCF\@.

Note further that the horizontal line representing $Q_{\rm sci}^{ICF}=\infty$ in Fig.~\ref{fig:scatterplot_nTtauE_vs_year}
(corresponding
to the $nT\tau$ value of the contour at 4~keV) is
at a higher value than the minimum $nT\tau$ value of the corresponding contour in Fig.~\ref{fig:scatterplot_nTtauE_vs_T}. This is because $T_i$ in laser ICF experiments (prior to onset of significant fusion) is limited by the maximum implosion velocity at which the shell becomes unstable, corresponding to a maximum $T_i$ of about 4~keV\@. Thus, marginal onset
of ignition corresponds to the required $nT\tau$ value at approximately 4~keV\@. In the case of NIF N210808, which exceeded the threshold
for onset of ignition\cite{Christopherson_2021}, $T_i$ increased due to self heating and $\tau$ decreased because of the increased pressure. These effects resulted in a slightly {\it lower} triple product compared with previous non-ignition results, which is visible in Fig.~\ref{fig:scatterplot_nTtauE_vs_T}.

\subsection{MIF/Z-pinch methodology}

\subsubsection{MagLIF}
The Magnetized Liner Inertial Fusion (MagLIF) experiment\cite{slutz10pop} compresses a cylindrical liner surrounding a pre-heated and axially pre-magnetized plasma. The Z-machine at Sandia National Laboratory supplies a large current pulse to the liner along its long axis, compressing it in the radial direction.

While the solid liner makes diagnosing MagLIF plasmas more difficult, it is still possible to extract the parameters needed to estimate the
Lawson parameter and triple product. The burn-averaged $T_i$ at stagnation is measured by neutron time-of-flight diagnostics. The spatial configuration of the plasma column at stagnation is imaged from emitted x-rays. From this spatial configuration and a model of x-ray emission, the effective fuel radius is inferred.
The stagnation pressure is inferred from a combination of diagnostic signatures. Given the plasma volume, burn duration, and temperature, the pressure was inferred by setting the pressure and mix levels to simultaneously match the x-ray yield and neutron yield.
In the emission model used to determine the spatial extent of the stagnated plasma, the pressure in the stagnated fuel is assumed to be spatially constant and the temperature and density profiles are assumed to be inverse to each other.\cite{McBride_2015} For our purposes, we infer an average $n_i$ from the stagnation pressure and the measured burn-averaged $T_i$.

Finally, the burn time, the duration during which the fuel assembly is inertially confined and hard x-rays (surrogates for fusion neutrons) are emitted, is measured. This duration is an upper bound on $\tau$, and in practice $\tau$ is estimated to be equal to it.
Data for MagLIF are shown in Table~\ref{tab:icf_mif_data_table} and plotted in Figs.~\ref{fig:scatterplot_ntauE_vs_T}, \ref{fig:scatterplot_nTtauE_vs_year}, and \ref{fig:scatterplot_nTtauE_vs_T}.

\subsubsection{Z pinch}
\begin{figure}[htbp!]
\includegraphics[width=8cm]{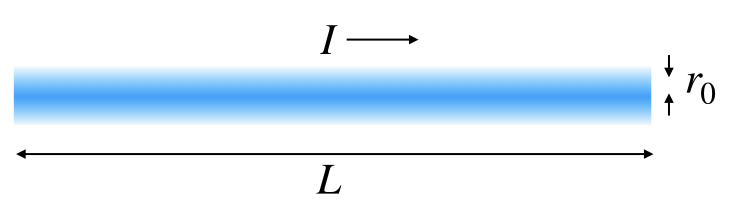}
\caption{A representation of a Z-pinch plasma of length $L$, effective radius $r_0$, and electrical current $I$. $V_p$ is the voltage difference between the left and right side of the plasma.}
\label{fig:z_pinch}
\end{figure}

Z-pinch experiments were one of the earliest approaches to fusion because no external magnetic field is required for confinement. This simplifies the experimental setup and reduces costs. Figure~\ref{fig:z_pinch} shows a representative diagram of a Z-pinch plasma. While fusion neutrons were detected in some of the earliest Z-pinch experiments, those fusion reactions were found to be the result of plasma instabilities generating non-thermal beam-target fusion events (see pp.~91--93 of Ref.~\onlinecite{Bishop_1958}), which would not scale up to energy breakeven.
More recently, however, stabilized Z-pinch experiments have provided evidence
of sustained thermonuclear neutron production.\cite{Zhang_2019,shumlak20}

\begin{figure}[tb!]
\includegraphics[width=8cm]{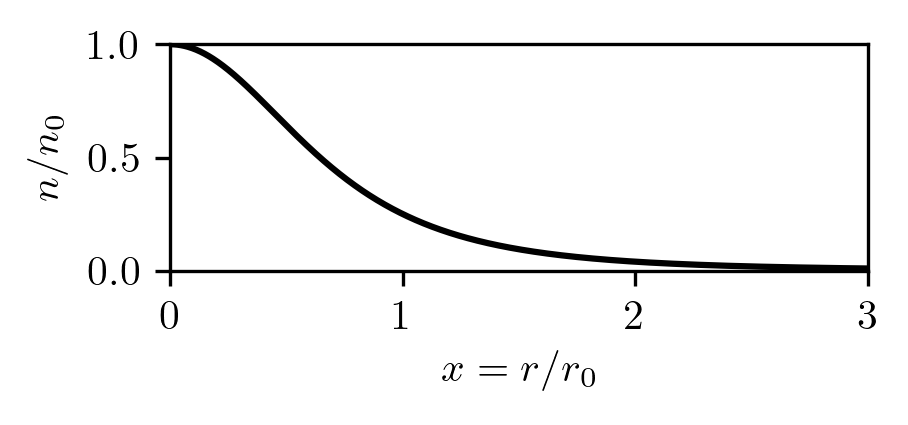}
\caption{Bennett-type density profile. In contrast to the parabolic profiles, the plasma extends beyond the effective radius $r_0$.}
\label{fig:bennett_profiles}
\end{figure}

Z-pinch plasmas exhibit profile effects perpendicular to the direction of current flow so the profile considerations discussed in Section \ref{sec:mcf_methodology} apply to Z~pinches as well. The radial density profile of Z~pinches is typically described by a Bennett-type profile\cite{Bennett_1934} of the form $n(r)=n_0/[1+(r/r_0)^2]^2$ and illustrated in Fig.~\ref{fig:bennett_profiles}.

Assuming $T = T_i = T_e$,  $n = n_i = n_e$, and a uniform profile for the plasma temperature, the thermal energy of a Z-pinch plasma can be estimated as
\begin{equation}
\begin{split}
    W_p = & 3T \int_V n(\bm{r}) \mathrm{d}^3 \bm{r} \\
        = & 3LT \int_0^\infty n(r) 2 \pi r \mathrm{d}r \\ 
        = & 6L  \pi n_0 T \int_0^\infty \frac{r}{(1+(r/r_0)^2)^2} \mathrm{d}r \\
        = & 3 \pi r_0^2 L n_0 T.
\end{split}
\end{equation}
The power applied is
\begin{equation}
    P_{\rm abs}=IV_p,
\end{equation}
where $I$ is the Z-pinch current and $V_p$ is the voltage across the plasma driving the current along the long axis. Assuming no self heating and that thermal conduction is the primary source of energy loss, the $\tau_E^*$ for the stabilized Z-pinch is
\begin{equation}
    \tau_E^*=\frac{3 \pi r_0^2 L n_0 T_0}{I V_p},
\end{equation}
and the Lawson parameter for a stabilized Z-pinch is 
\begin{equation}
    n_0 \tau_E^*=\frac{3 \pi r_0^2 L n_0^2 T_0}{I V_p}.
\end{equation}
However, in practice $V_p$ may not be measured directly, and the voltage across the power supply driving the Z-pinch may overestimate $V_p$. Therefore, evaluations of $\tau_E^*$ that substitute the power supply voltage for $V_p$ (as done for FuZE\cite{Zhang_2019,shumlak20}) provide only a lower bound on $\tau_E^*$. An upper bound on $\tau_E^*$ is the flow-through time of the Z-pinch. Our reported value is the lower of the two.

In other Z-pinch approaches like the dense plasma focus (DPF), fusion yields occur from a combination of non-Maxwellian ion
energy distributions and thermal ion populations.\cite{Krishnan_2012} Because thermal temperatures and $\tau_E^*$ are typically not well characterized in such approaches, it is not
feasible to report a reliable, achieved Lawson parameter or triple product.  Furthermore, fusion concepts with strong beam-target components 
may not be scalable to $Q_{\rm fuel}>1$.\cite{Rider_1997}

\subsubsection{Other MIF approaches}
\label{sec:mif_methodology}

For other MIF approaches,\cite{Wurden16} e.g., liner or flux compression of FRCs or spheromaks, it is difficult to rigorously measure $\tau_E^*$ due to limited access. A few attempts to quantify $\tau_E^*$ based on measurable or calculable parameters, such as particle confinement time
$\tau_N$, have been proposed.\cite{Steinhauer_2018} In particular, we estimate $\tau_E^*$ of FRCs to be $\tau_N/3$ (for both
MIF and MCF)\@.

\section{Summary and Conclusions}
\label{sec:conclusion}

The combination of achieved Lawson parameter $n\tau$ or $n\tau_E$ and fuel temperature
$T$ of a thermonuclear-fusion concept are a rigorous scientific indicator
of how close it is to energy breakeven and gain.
In this work, we have compiled the achieved
Lawson parameters and $T$ of a large number of fusion
experiments (past, present, and projected) from around the world.
The data are provided in multiple tables and figures.  Following Lawson's
original work, we provided a detailed review, re-derivation, and extension
of the mathematical expressions underlying the Lawson parameter
(and the related triple product) and
four ways of measuring energy gain ($Q_{\rm fuel}$, $Q_{\rm sci}$, $Q_{\rm wp}$, and $Q_{\rm eng}$), and explained
the physical principles upon which these quantities are based.  Because
different fusion experiments report different observables, we explained
precisely how we infer both electron and ion densities and temperatures and
the various definitions of confinement time that are used in the Lawson-parameter and triple-product values that
we report, including accounting for the effects of spatial profile shapes
(through a peaking factor) and a range in the level of impurities in the
plasma fuel.  All data reported in this paper are based on the published literature or are expected to be published shortly.

The key results of this paper are encapsulated in
Figs.~\ref{fig:scatterplot_ntauE_vs_T}, \ref{fig:scatterplot_nTtauE_vs_year}, and \ref{fig:scatterplot_nTtauE_vs_T}, which show that
(1)~tokamaks 
and laser-driven ICF have achieved the highest Lawson
parameters, triple products, and $Q_{\rm sci}\sim 0.7$; 
(2)~fusion concepts have demonstrated rapid advances in Lawson
parameters and triple products early in
their development
but slow down as values approach what is needed for $Q_{\rm sci}=1$;
(3)~private fusion companies
pursuing alternate concepts are now exceeding the
breakout performance of early tokamaks; and
(4)~at least three experiments may achieve $Q_{\rm sci}>1$
within the foreseeable future, i.e., NIF and SPARC in the 2020s and ITER by 2040.

The reason
for item (2) in the preceding paragraph is commonly attributed to the fact that
experimental facilities became extremely expensive (e.g., \$3.5B for NIF according
to the U.S. Government Accountability Office, and
exceeding US\$25B for ITER) for making continued and required advances toward
energy gain.  However, 
there are two reasons
that other approaches or experiments might potentially achieve {\em commercially relevant} energy breakeven and gain on a faster timescale.  Firstly, most of the other
paths being pursued (i.e., privately funded development paths
for tokamaks, stellarators, alternate concepts, and laser-driven ICF) have lower cost as a key objective, where experiments along the development path are envisioned to have much lower costs than NIF and ITER\@.  Secondly, the mature fusion and plasma scientific
understanding and computational tools, as well as many fusion-engineering technologies, developed over 65+ years of
controlled-fusion research do not need to be reinvented and need only
be leveraged in the development of the alternate and privately funded approaches.

High values of Lawson parameter and
triple product, which are required for energy gain, are a necessary but not sufficient condition for commercial fusion energy.
Additional necessary conditions include attractive economics and social acceptance,
including but not limited to considerations of
RAMI (reliability, accessibility, maintainability, and inspectability) and the ability to be licensed under an appropriate regulatory
framework. These
necessary conditions require additional technological attributes beyond
high energy gain, e.g.,
(1)~a fusion plasma core that is compatible with both surrounding
materials and subsystems that survive the extreme fusion particle, heat, and radiation flux, and
(2)~a sustainable fuel cycle (e.g., tritium breeding, separation, and processing technologies for D-T fusion).
Therefore, while this paper's primary objective
is to explain and highlight the achieved Lawson
parameters (and triple products) of many fusion concepts
and experiments as a measure of fusion's progress toward
energy breakeven and gain, these are not the only criteria
for justifying continued pursuit of and investment into a given fusion concept, including concepts using advanced fusion fuels.

\appendix

\section{Plot of triple products vs.\ $T_i$}
\label{sec:other_formulations}
\begin{figure*}[p]
\centerline{
\includegraphics[width=17.5cm]{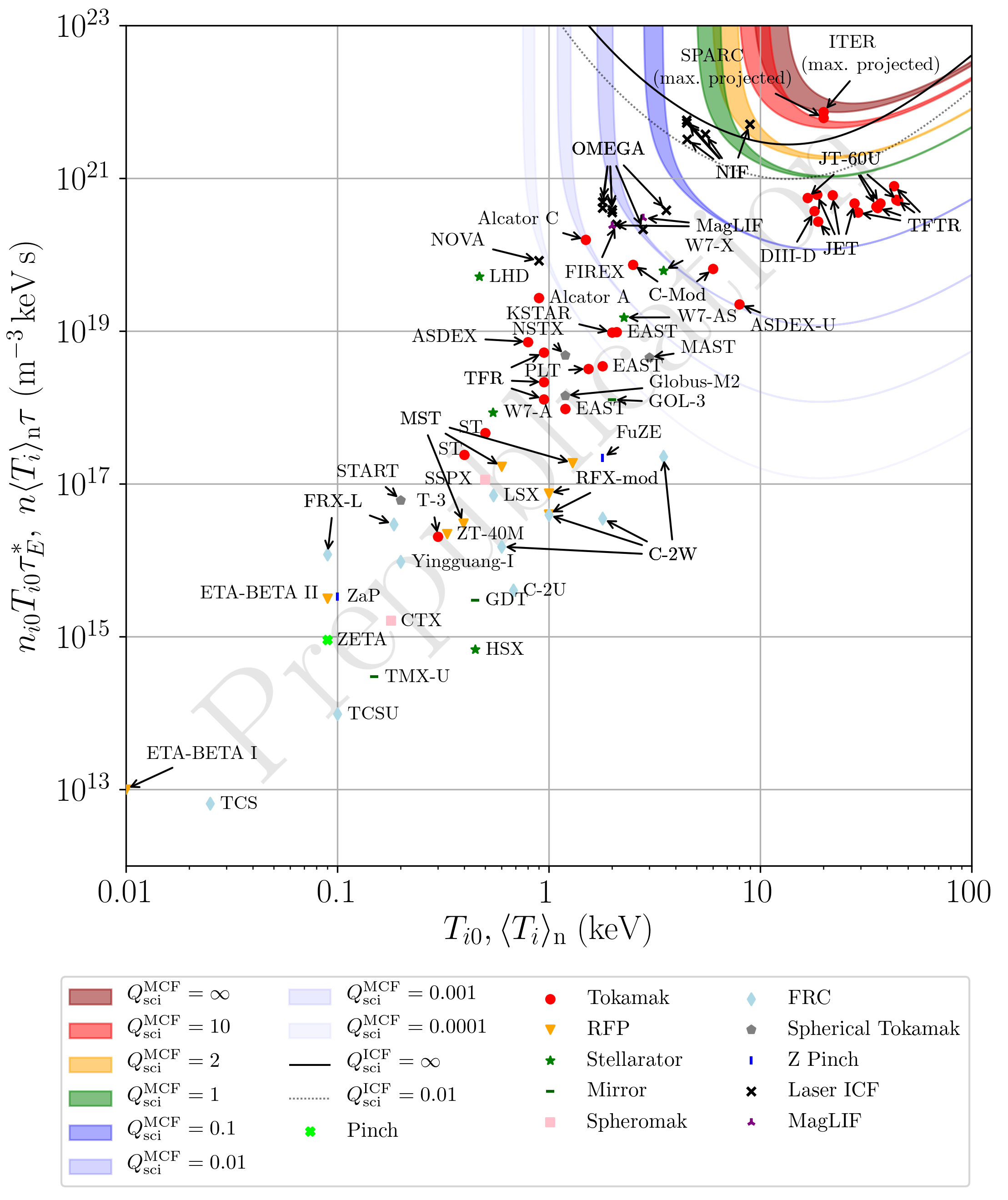}}
\caption{Experimentally inferred, peak triple products of fusion 
experiments vs.\ ion temperature, extracted from published literature. See the caption of Fig.~\ref{fig:scatterplot_ntauE_vs_T} for more details.}
\label{fig:scatterplot_nTtauE_vs_T}
\end{figure*}

Figure~\ref{fig:scatterplot_nTtauE_vs_T} shows achieved triple products
versus $T_i$, based on the same data points used in Fig.~\ref{fig:scatterplot_ntauE_vs_T}.

\section{Data tables}
Table~\ref{tab:mainstream_mcf_data_table} provides numerical values of the data for tokamaks, spherical tokamaks, and stellarators.
Table~\ref{tab:alternates_mcf_data_table} provides numerical values of the data for ``alternate'' MCF concepts, i.e., not tokamaks or stellarators. Table~\ref{tab:icf_mif_data_table} provides numerical values of the data for ICF and MIF experiments.
We group lower-density and higher-density MIF approaches
with MCF alternate concepts 
(Table~\ref{tab:alternates_mcf_data_table}) and ICF (Table~\ref{tab:icf_mif_data_table}), respectively.

\input{table_6}
\input{table_7}
\input{table_8}

\section{Effect of mitigating bremsstrahlung losses}
\label{sec:mitigating_brems}

\begin{figure}[!tb]
\begin{flushleft}(a)\end{flushleft}
    \includegraphics[width=8cm]{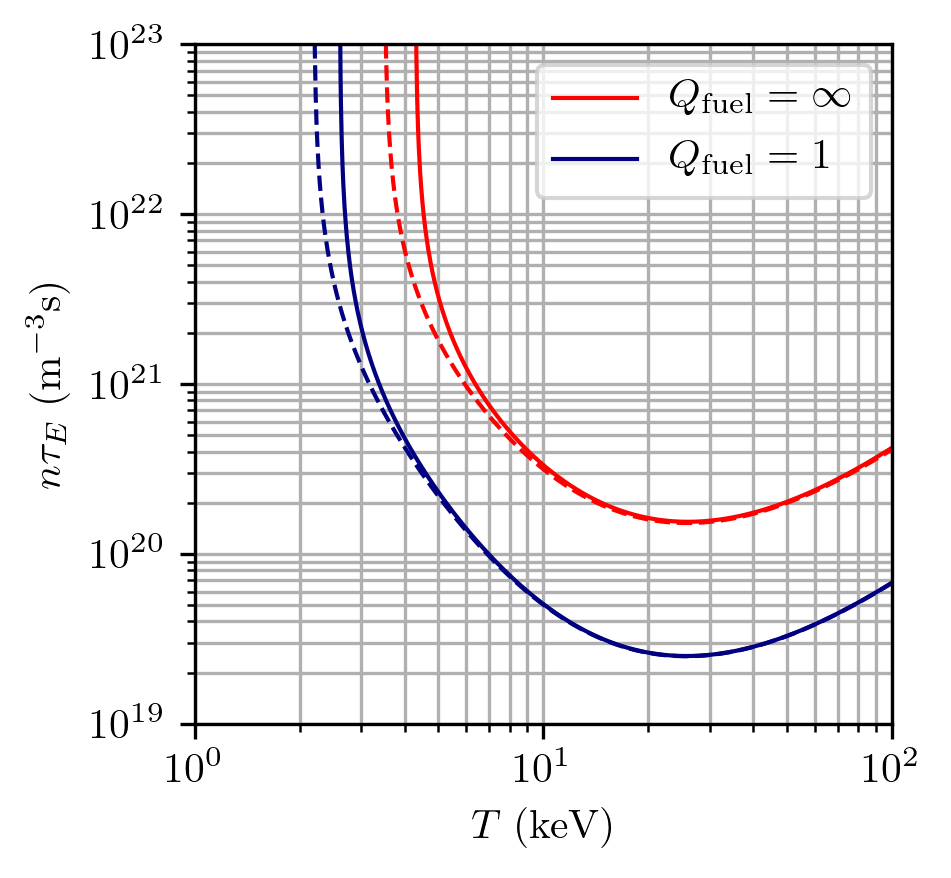}
\begin{flushleft}(b)\end{flushleft}
    \includegraphics[width=8cm]{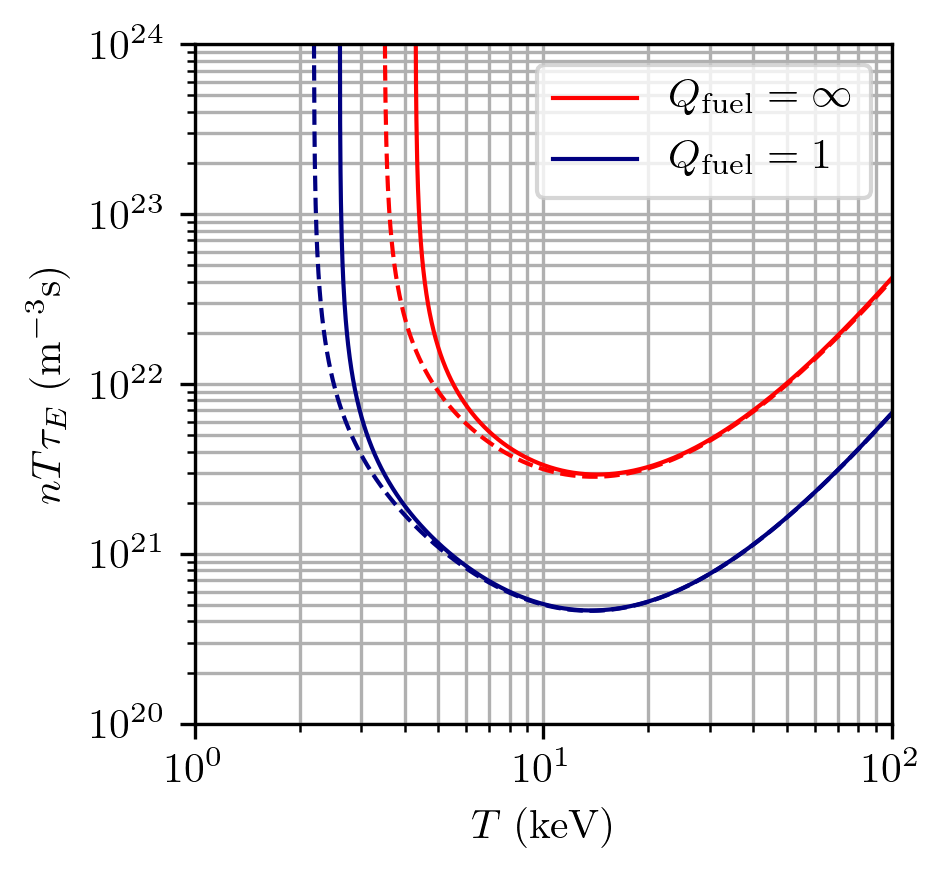}
    \caption{Contours of $Q_{\rm fuel}$ plotted vs.\ $T$ and (a) $n\tau_E$ and (b) $nT\tau_E$ for D-T fusion (assuming $T=T_i=T_e$)\@. Dashed lines represent arbitrarily
    reducing bremsstrahlung losses by a factor of 2, i.e., replacing
    $C_B$ by $C_B/2$ in Eqs.~(\ref{eq:MCF_Lawson_parameter_Q_fuel}) and (\ref{eq:triple_product_steady_state}).}
\label{fig:effect_of_bremsstrahlung}
\end{figure}

If bremsstrahlung radiation losses are mitigated, e.g., in 
pulsed ICF\cite{Atzeni04}
or MIF\cite{Kirkpatrick95,Wurden16}
approaches with an optically thick pusher,\cite{Kirkpatrick81,Kirkpatrick12} then the $Q_{\rm fuel}$ and $Q_{\rm sci}$ contours of Figs.~\ref{fig:MCF_ntau_contours_q_fuel_q_sci} and \ref{fig:ICF_ntau_contours_q_fuel_q_sci} can be modified.
Figure~\ref{fig:effect_of_bremsstrahlung} illustrates
the effect of arbitrarily reducing
$P_B$ by a factor of 2, i.e., by replacing $C_B$ with $C_B/2$ in
Eqs.~(\ref{eq:MCF_Lawson_parameter_Q_fuel}) and (\ref{eq:triple_product_steady_state}).

\section{Lawson parameters for advanced fusion fuels}
\label{sec:advanced_fuels}

The main body of this paper focuses on D-T fusion
because it has the highest maximum reactivity occurring at the lowest temperature compared to all known fusion fuels. As a result, the required D-T Lawson
parameters and triple products to reach high $Q_{\rm fuel}$
are the lowest and most accessible.
However, D-T fusion has two major drawbacks:  (i)~it
produces 14-MeV neutrons that carry 80\% of the fusion energy,
and (ii)~the tritium must be bred (because it does not occur abundantly in nature due
to a 12.3-year half life) and be continuously processed and handled safely.

Advanced fuels, such as D-$^3$He, D-D, and p-$^{11}$\!B, mitigate these drawbacks to different extents.\cite{Nevins_1998}
However, because their peak reactivities are all lower and occur at higher temperatures compared to D-T, the required Lawson parameters and triple products for these advanced fuels to achieve equivalent values of $Q_{\rm fuel}$ are much higher.

Furthermore, at the high temperatures required for advanced fuels, relativistic bremsstrahlung effects become significant. We utilize the relativistic-correction approximation to Eq.~(\ref{eq:generalized_bremsstrahlung_power_density}) from Ref.~\onlinecite{Putvinski_2019},
\begin{equation}
\label{eq:relativistic_bremsstrahlung_power_density}
    P_B = C_B n_e T_e^{1/2} \gamma(Z_{\rm eff}),
\end{equation}
where
\begin{equation}
\label{eq:putvinski_relativistic_correction}
    \gamma(Z_{\rm eff}) = Z_{\rm eff}(1+1.78 t ^ {1.34}) + 2.12t(1+1.1t+t^2-1.25t^{2.5})
\end{equation}
and $t=T_e / m_e c^2$.

To quantify the Lawson-parameter and triple-product requirements for advanced fuels with non-identical reactants and reaction products that are immediately removed from the plasma (e.g., D-$^3$He and p-$^{11}$\!B without ash buildup or subsequent reactions), we first generalize the expression for $n\tau_E$ [Eq.~(\ref{eq:MCF_Lawson_parameter_Q_fuel})] to account for the effect of relativistic bremsstrahlung and the reaction of two ion species with charge per ion $Z_1$ and $Z_2$, ion number densities $n_1$ and $n_2$, and relative densities $k_1 = n_1 / n_e$ and $k_2 = n_2 / n_e$, respectively.

A more detailed treatment of advanced fuels would need to consider scenarios in which $T_e < T_i$ and account for an additional term in the power-balance equation for ion energy transfer to electrons. Maintaining $T_e \ll T_i$ has the advantage of reduced bremsstrahlung (especially at high $T_i$) and lower plasma pressure for a given $T_i$\@. The challenge of such a scenario is maintaining $T_i>T_e$ for a sufficient duration of time
and with acceptable additional input power. In this section, we only
consider $T=T_i=T_e$, except in the discussion of Fig.~\ref{fig:pB11_vs_bremsstrahlung}.
Accounting for the above,
\begin{equation}
\label{eq:confinement_parameter_generalized}
n_e\tau_E = \frac{(3T/2)(k_1 + k_2 + 1)}  {(f_c + Q_{\rm fuel}^{-1})k_1 k_2 \langle \sigma v \rangle_{1,2} E_{1,2}  - C_B T^{1/2} \gamma(Z_{\rm eff})},
\end{equation}
where $Z_{\rm eff} = \Sigma_j n_j Z_j^2/n_e$, and $j$ is summed over
the different reactant species.

The relative density for each ion species $j$ that maximizes
\footnote{A proof of this statement follows. The fusion power density $S_F = n_1 n_2 \langle \sigma v \rangle_{1,2} E_{1,2}  = k_1 k_2 n_e^2 \langle \sigma v \rangle_{1,2} E_{1,2}$ is maximized when 
$k_1 k_2$ is maximized. Quasi-neutrality requires that $Z_1 n_1 + Z_2 n_2 = n_e$, which, after some further algebra,
results in the expression $k_1 k_2 = k_2/Z_1 - k_2^2 Z_2/Z_1$.  This is maximized when $k_2 = 1/(2Z_2)$ and $k_1 = 1/(2Z_1)$. Rather than maximizing fusion power, other optimizations are possible, e.g., minimizing the required Lawson parameter to achieve a certain value of $Q_{\rm fuel}$ by minimizing the entire right-hand side of Eq.~(\ref{eq:confinement_parameter_generalized}) with respect to $k_1$ and $k_2$. However, maximizing fusion power is the simplest choice.}
fusion power for a fixed value of $n_e^2$ is $k_j = 1/(2Z_j)$ 
and $Z_{\rm eff}=(Z_1+Z_2)/2$.
Assuming this condition, Eq.~(\ref{eq:triple_product_steady_state}) becomes
\begin{equation}
\begin{split}
\label{eq:triple_product_optimal_mix_electron}
& n_e T \tau_E = \\ &\frac{(3T^2/2)[(2Z_1)^{-1} + (2Z_2)^{-1} + 1]} {(f_c + Q_{\rm fuel}^{-1})\langle \sigma v \rangle_{1,2} E_{1,2}/(4 Z_1 Z_2)  - C_B T^{1/2} \gamma(Z_{\rm eff}) },
\end{split}
\end{equation}
or equivalently,
\begin{equation}
\begin{split}
\label{eq:triple_product_optimal_mix_ion}
& n_i T \tau_E = \\ &\frac{(3T^2/2)[(2Z_1)^{-1} + (2Z_2)^{-1}][(2Z_1)^{-1} + (2Z_2)^{-1} + 1]}
{(f_c + Q_{\rm fuel}^{-1})\langle \sigma v \rangle_{1,2} E_{1,2}/(4 Z_1 Z_2)  - C_B T^{1/2} \gamma(Z_{\rm eff}) },
\end{split}
\end{equation}
where we have multiplied both sides of Eq.~(\ref{eq:triple_product_optimal_mix_electron})
by $(k_1+k_2)=(2Z_1)^{-1} + (2Z_2)^{-1}$.
This expression ignores synchrotron radiation losses, which may become important
at the very high temperatures required to reach 
Lawson conditions for advanced fuels in magnetically confined systems.

\subsection{D-$^3$He}
\begin{figure}[!tb]
\begin{flushleft}(a)\end{flushleft}
\includegraphics[width=8cm]{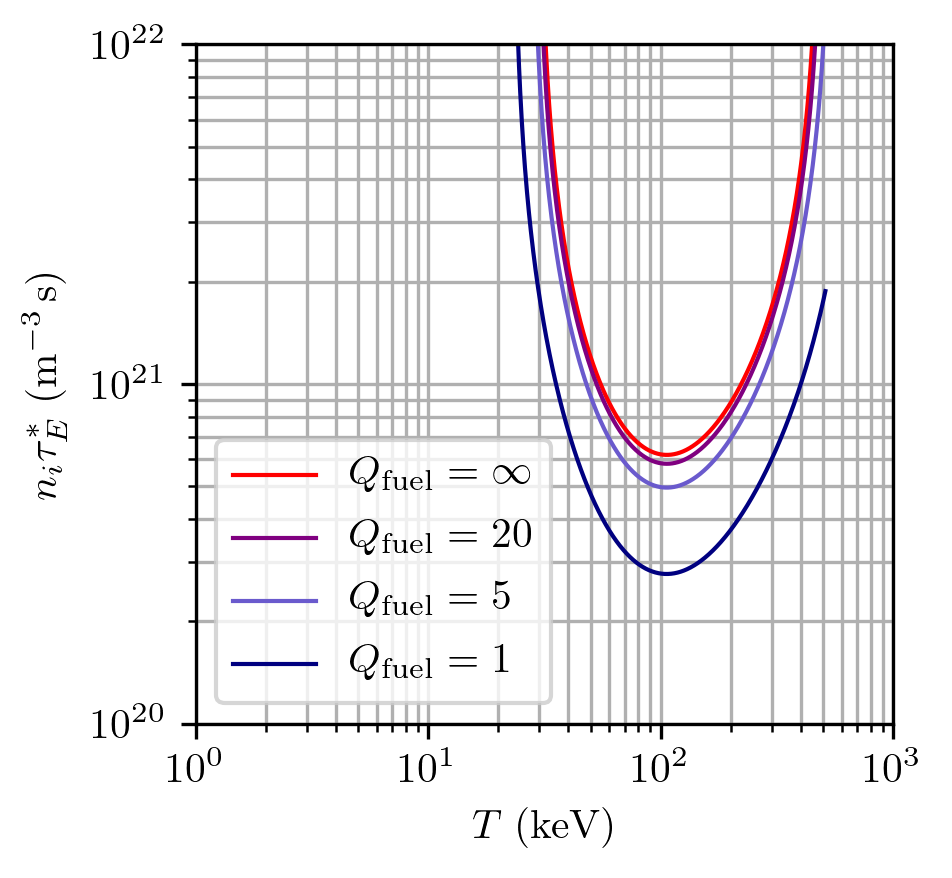}\\
\begin{flushleft}(b)\end{flushleft}
\includegraphics[width=8cm]{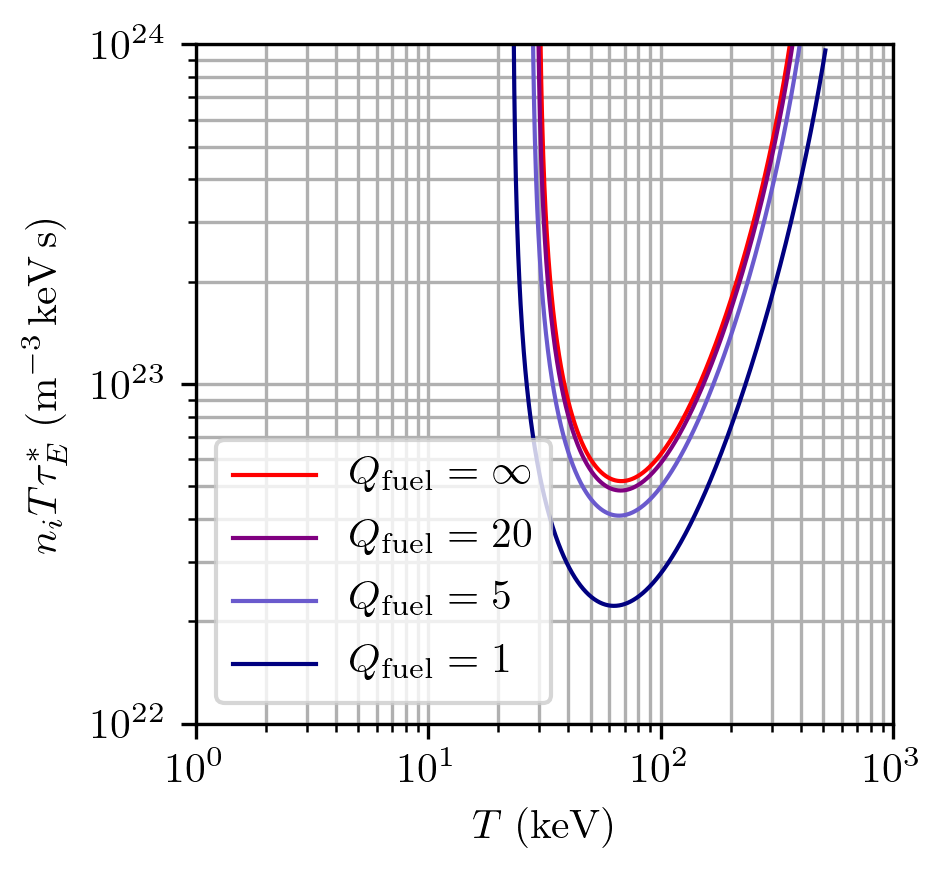}
\caption{Required (a) Lawson parameters and (b) triple products vs.\ $T_i$ to achieve the indicated values of $Q_{\rm fuel}$ for D-$^3$He (assuming $T=T_e=T_i$).}
\label{fig:D-3He}
\end{figure}

The D-$^3$He fusion reaction has the advantage that its primary reaction,
\begin{align}
   \mathrm{D}+^3\!\!\mathrm{He} & \rightarrow \alpha + \mathrm{p}~(\mathrm{18.3~MeV}),
\end{align}
is aneutronic, where the $\alpha$ is a $^4$He ion. However, $^3$He is not abundant on earth and must be bred via other reactions or mined from the moon, both of which involve additional complexity and cost. Also, D-$^3$He will not be completely aneutronic because of D-D reactions. 
The requirement for ignition of D-$^3$He ignoring side D-D reactions is $n_i T \tau_E^* \ge 5.2 \times 10^{22}$~m$^{-3}$\,keV\,s at 68~keV (see Fig.~\ref{fig:D-3He}), 18 times higher than for D-T.

\subsection{p-$^{11}$B}
\begin{figure}[tb!]
\includegraphics[width=8cm]{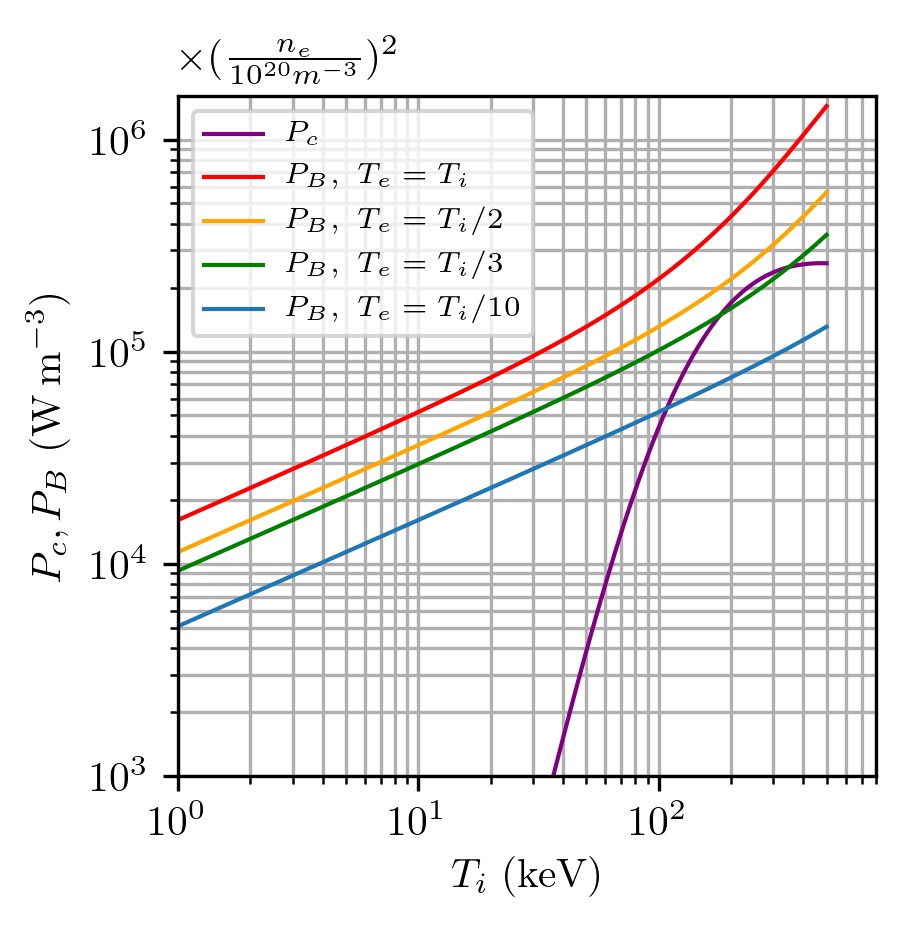}
\caption{Charged-particle fusion power density
$P_c$ (purple line) and bremsstrahlung power density $P_B$ for various
ratios of $T_e/T_i$ vs.\ $T_i$ for p-$^{11}$B, showing
that $P_B$ always exceeds $P_c$ when $T_e \gtrsim T_i/3$.  This plot
uses the parameterized p-$^{11}$B reactivity in Ref.~\onlinecite{Nevins_2000}.  Updated, higher p-$^{11}$\!B fusion cross sections\cite{Sikora_2016} suggest that ignition may be possible for p-$^{11}$\!B.\cite{Putvinski_2019}}
\label{fig:pB11_vs_bremsstrahlung}
\end{figure}

\begin{figure}[!tb]
\begin{flushleft}(a)\end{flushleft}
\includegraphics[width=8cm]{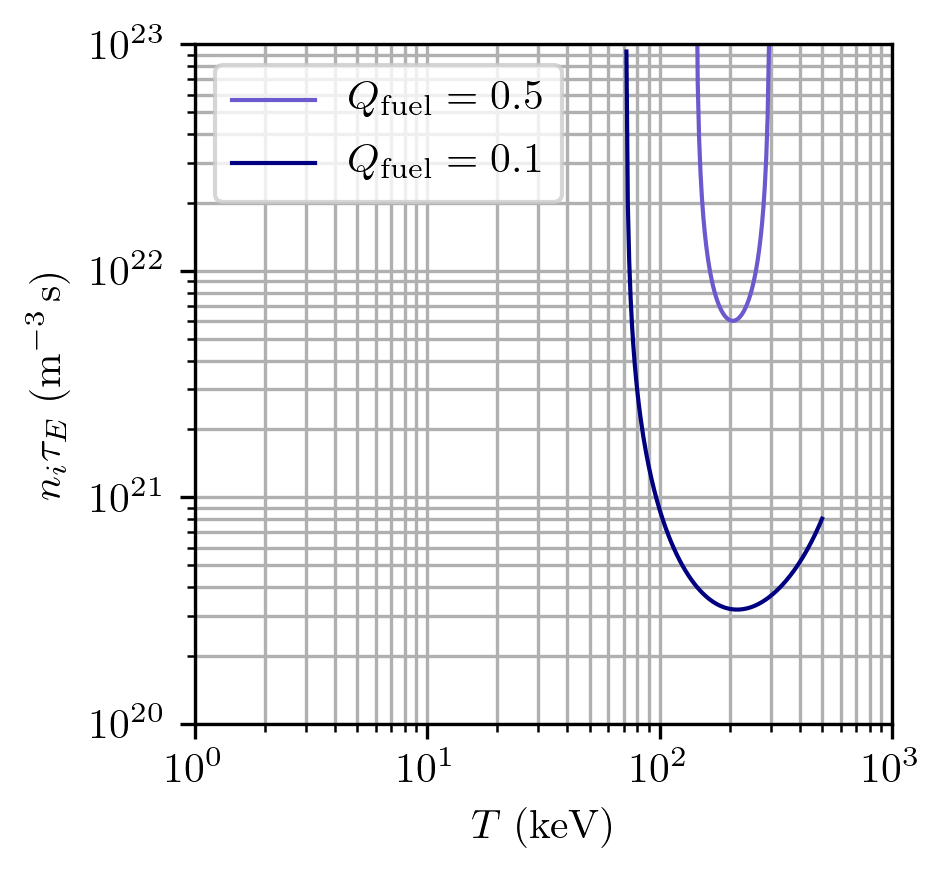}
\begin{flushleft}(b)\end{flushleft}
\includegraphics[width=8cm]{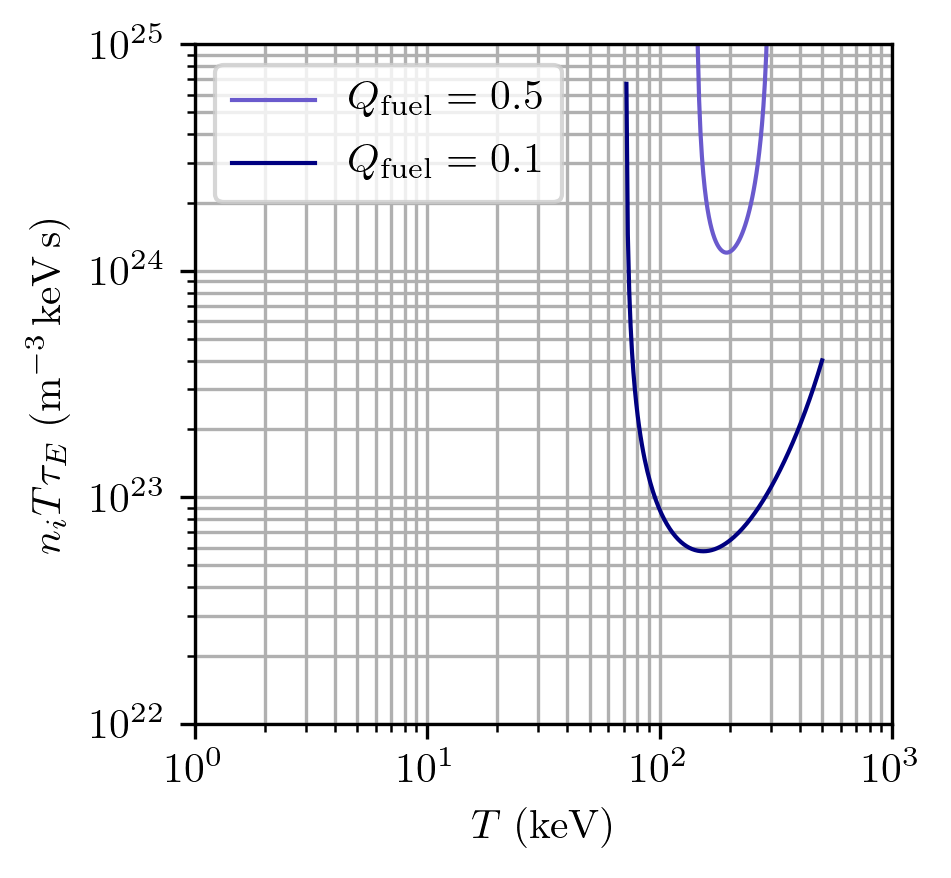}
\caption{Required (a) Lawson parameters and (b) triple products vs.\ $T_i$ to achieve values of $Q_{\rm fuel}$ assuming $T=T_i=T_e$,
for p-$^{11}$\!B based on the p-$^{11}$\!B fusion reactivity from Ref.~\onlinecite{Nevins_2000}.}
\label{fig:pB11}
\end{figure}

The p-$^{11}$\!B fusion reaction has the advantage that its reactants are abundant on earth, and the reaction products are three electrically
charged $\alpha$ particles, potentially allowing for direct energy conversion to electricity. However, this reaction requires temperatures around 100~keV, at which bremsstrahlung radiation losses per unit volume exceed fusion power density, and ignition is not possible for a p-$^{11}$\!B plasma where $T_e=T_i$, as shown in Fig.~\ref{fig:pB11_vs_bremsstrahlung}, which uses the parametrized p-$^{11}$\!B fusion reactivity from Ref.~\onlinecite{Nevins_2000}. The boron and
proton concentrations are set to maximize fusion power for a fixed electron density as described earlier in this section. Also shown is the effect of reduced bremsstrahlung if $T_e$ is maintained at levels below $T_i$. We are neglecting the issue of the ion-electron thermal equilibration time here.
Figure~\ref{fig:pB11} shows that only modest values of $Q_{\rm fuel}$ are physically possible
for $T_e=T_i$, at triple products three orders of magnitude higher than that of D-T.

However recent work\cite{Sikora_2016} points to a higher reactivity, and given certain assumptions, high-$Q_{\rm fuel}$ operation up to and including ignition may be theoretically possible.

\subsection{Fully catalyzed D-D}
The D-D fusion reaction has the advantage that its sole reactant is abundant on earth. In the fully catalyzed D-D reaction,\cite{Mills_1971,Stott_2005} the T and $^3$He produced as reaction products undergo subsequent reactions with D, releasing more energy. The reaction paths are
\begin{align}
    \mathrm{D} + \mathrm{D} & \rightarrow {\mathrm{T}}~(\mathrm{1.01~MeV}) +\mathrm{p}~(\mathrm{3.02~MeV})\\[4pt] 
     & \mathrm{D} + {\mathrm{T}}  \rightarrow \alpha~(\mathrm{3.5~MeV}) + \mathrm{n}~(\mathrm{14.1~MeV})\\[4pt]
    \mathrm{D}+\mathrm{D} & \rightarrow {^3\!\mathrm{He}}~(0.82~\mathrm{MeV}) + \mathrm{n}~(2.45~\mathrm{MeV}) \\[4pt]
   & \mathrm{D} + {^3\!\mathrm{He}} \rightarrow \mathrm{\alpha~(\mathrm{3.6~MeV})+ p}~(\mathrm{14.7~MeV}), 
\end{align}
with 62\% of the 43.2~MeV released in charged particles (compared with only 20\% for D-T)\@.

Note that there are other forms of ``catalyzed D-D'' which go by different names in different contexts. For example extraction of tritium before the subsequent D-T reaction occurs is sometimes called ``$^3\!\mathrm{He}$ double-catalyzed D-D''.\cite{Stott_2005} Here we only consider the steady-state reaction path where $^3\!$He and T react with D at the same rate as they are created in each branch of the D-D reaction. Furthermore, we assume an idealized scenario without synchrotron radiation and that the ``ash'' $\alpha$ particles and protons immediately exit after depositing their energy and comprise a negligible fraction of ions in the plasma. Lastly, we assume that D is added at the same rate as it is consumed and that $T=T_i=T_e$.

The ion number density is the sum of the constituent ion number densities, 
\begin{equation}
    n_i = n_{\rm D} + n_{\rm ^3He} + n_{\rm T},
\end{equation}
and the electron density is,
\begin{equation}
    n_e = n_{\rm D} + 2 n_{\rm ^3He} + n_{\rm T}.
\end{equation}

Requiring that the rate of production of $^3$\!He and T are consumed at the same rate as they are produced, 
\begin{align}
    \frac{1}{2}n_{\rm D}^2 \langle \sigma v \rangle_{\rm DD,^3\!He} & = n_{\rm D} n_{\rm ^3\!He} \langle \sigma v \rangle_{\rm D ^3\!He},\\
    \frac{1}{2} n_{\rm D}^2\langle \sigma v \rangle_{\rm DD,T} & = n_{\rm D} n_{\rm T} \langle \sigma v \rangle_{\rm DT}.
\end{align}
Rearranging gives the $T$-dependent, steady-state number density of $^3$He and T ions,
respectively,
\begin{equation}
    n_{\rm ^3\!He} = \frac{1}{2} \frac{\langle \sigma v \rangle_{\rm DD,^3\!He}}{\langle \sigma v \rangle_{\rm D ^3\!He}} n_{\rm D},\\
\label{eq:cat-dd_density_h}
\end{equation}
\begin{equation}
    n_{\rm T} = \frac{1}{2}  \frac{\langle \sigma v \rangle_{\rm DD,T}}{\langle \sigma v \rangle_{\rm DT}} n_{\rm D}.
\label{eq:cat-dd_density_t}
\end{equation}
The total fusion power density is the sum of the power released in its four constituent reactions,
\begin{equation}
\begin{split}
S_F = & \frac{n_{\rm D}^2}{2}\langle \sigma v \rangle_{\rm DD,^3He}E_{\rm DD,^3He} + n_D n_{\rm ^3He} \langle \sigma v \rangle_{\rm D^3He} E_{\rm D^3He} + \\
& \frac{n_{\rm D}^2}{2} \langle \sigma v \rangle_{\rm DD,T} E_{\rm DD,T} + n_{\rm D} n_{\rm T} \langle \sigma v \rangle_{\rm DT} E_{\rm DT}.
\end{split}
\label{eq:total_power_density_cat_dd}
\end{equation}
The bremsstrahlung power density is
\begin{equation}
S_B = C_B n_e^2 T_e^{1/2} \gamma(Z_{\rm eff}),   
\end{equation}
and from Eq.~(\ref{eq:Z_eff}),
\begin{equation}
Z_{\rm eff} = \frac{1}{n_e}(n_{\rm D} + 4n_{\rm ^3He} + n_{\rm T}).
\end{equation}
The power lost to thermal conduction per unit volume is
\begin{equation}
    S_\kappa = \frac{(3/2)T(n_e + n_i)}{\tau_E}.
\end{equation}
Defining $\chi_h$ and $\chi_t$ as the number density ratios of $n_{\rm ^3\!He}$ to $n_{\rm D}$ and $n_{\rm T}$ to $n_{\rm D}$ respectively, 
\begin{align}
    \chi_{h} & \equiv \frac{n_{\rm ^3\!He}}{n_{\rm D}} = \frac{1}{2} \frac{\langle \sigma v \rangle_{\rm DD,^3\!He}}{\langle \sigma v \rangle_{\rm D ^3\!He}},\\
    \chi_{t} & \equiv \frac{n_{\rm T}}{n_{\rm D}} = \frac{1}{2} \frac{\langle \sigma v \rangle_{\rm DD,T}}{\langle \sigma v \rangle_{\rm DT}}.
\end{align}
From the steady-state power balance of Eq.~(\ref{eq:power_balance_steady_state}) and the above, the Lawson parameter required to achieve fuel gain $Q_{\rm fuel}$ at $T_i$ is,

\begin{equation}
\begin{split}
    & n_i \tau_E = \\ & \frac{T(3 + 9\chi_{h}/2 + 3\chi_{t})(1+\chi_{h}+\chi_{t})} {(f_c + Q_{\rm fuel}^{-1}) \langle \sigma v \rangle_{\rm DD} E_{\rm tot}/ 4 - C_B (1 + 2\chi_{h} + \chi_{t})^2 T^{1/2}\gamma(Z_{\rm eff})}
\end{split}
\end{equation}
with
\begin{equation}
    Z_{\rm eff} = (1 + 4\chi_h + \chi_t) / (1 + 2 \chi_h + \chi_t),
\end{equation}
\begin{equation}
    \langle \sigma v \rangle_{\rm DD} = \langle \sigma v \rangle_{\rm DD,^3\!He} + \langle \sigma v \rangle_{\rm DD,T},
\end{equation}
and
\begin{equation}
  E_{\rm tot} = E_{\rm DD,T} + E_{\rm DT} + E_{\rm DD,^3He} + E_{\rm D^3He}.
\end{equation}

\begin{figure}[!tb]
\begin{flushleft}(a)\end{flushleft}
\includegraphics[width=8cm]{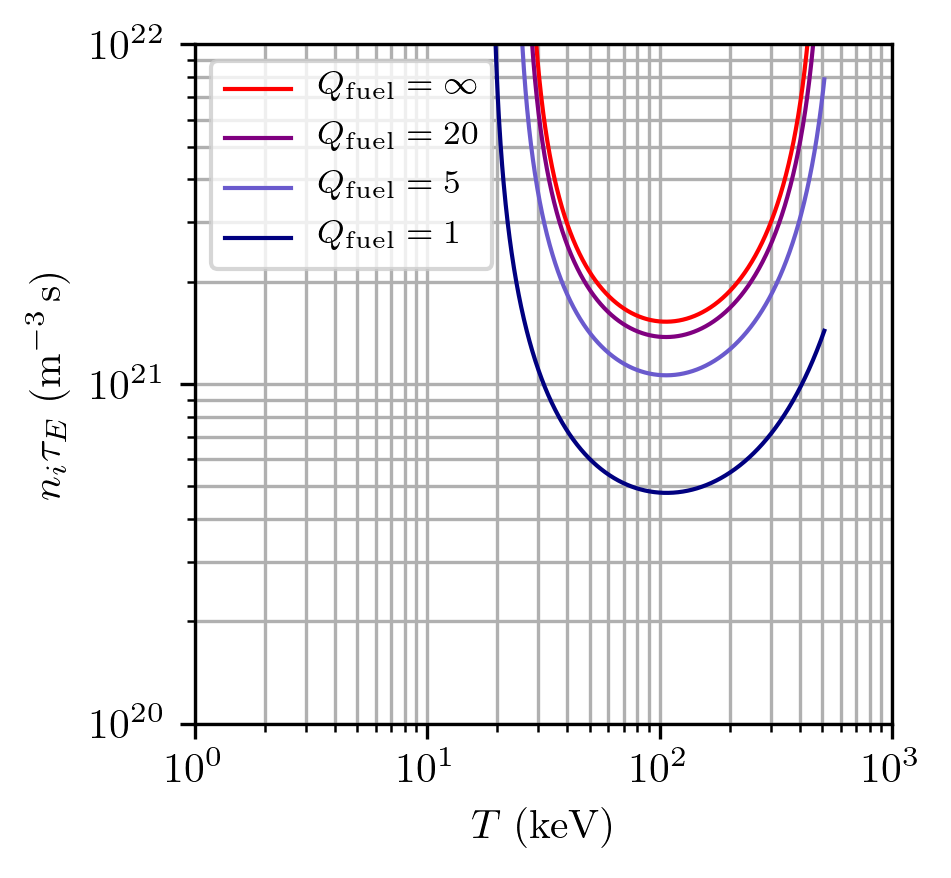}
\begin{flushleft}(b)\end{flushleft}
\includegraphics[width=8cm]{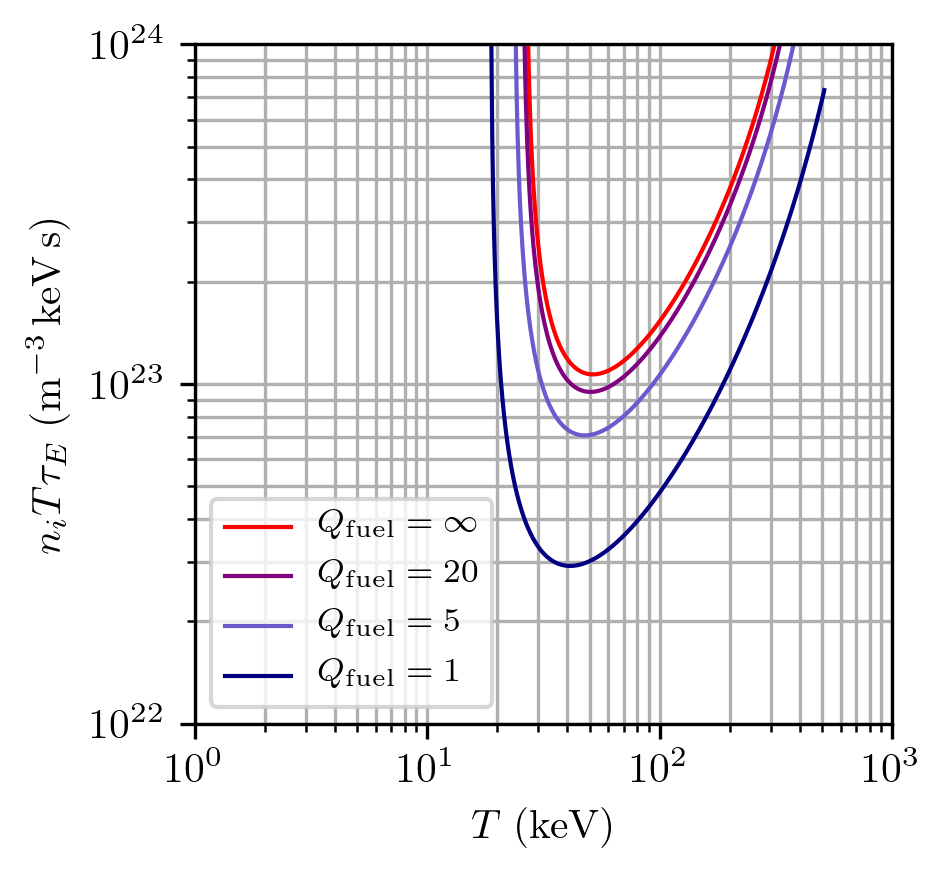}
\caption{Required (a) Lawson parameters and (b) triple products vs.\ $T$ to achieve the indicated values of $Q_{\rm fuel}$ for catalyzed D-D (assuming $T=T_e=T_i$).}
\label{fig:CAT_D-D}
\end{figure}

The requirement for ignition of catalyzed D-D is $n_i T \tau_E^* \ge 1.1\times10^{23}$~m$^{-3}$\,keV\,s at $T=52$~keV (see Fig.~\ref{fig:CAT_D-D}), 38 times higher than required for D-T\@. 

\subsection{Advanced-fuels summary}
\begin{figure}[!tb]
\begin{flushleft}(a)\end{flushleft}
\includegraphics[width=8cm]{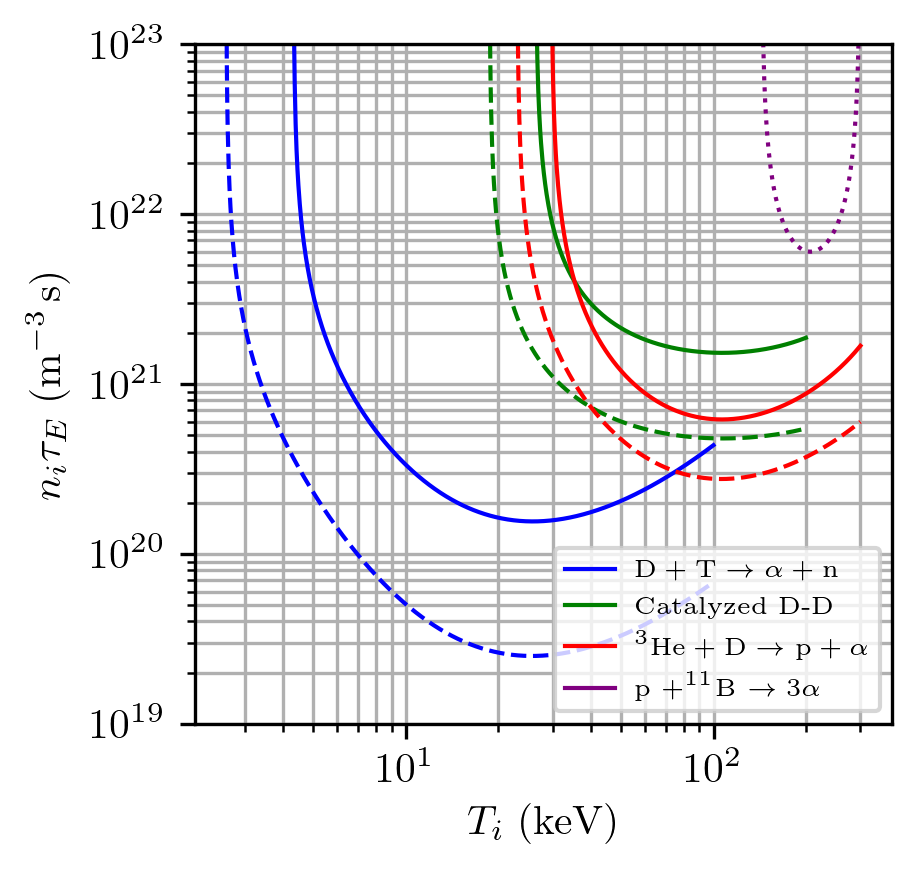}
\begin{flushleft}(b)\end{flushleft}
\includegraphics[width=8cm]{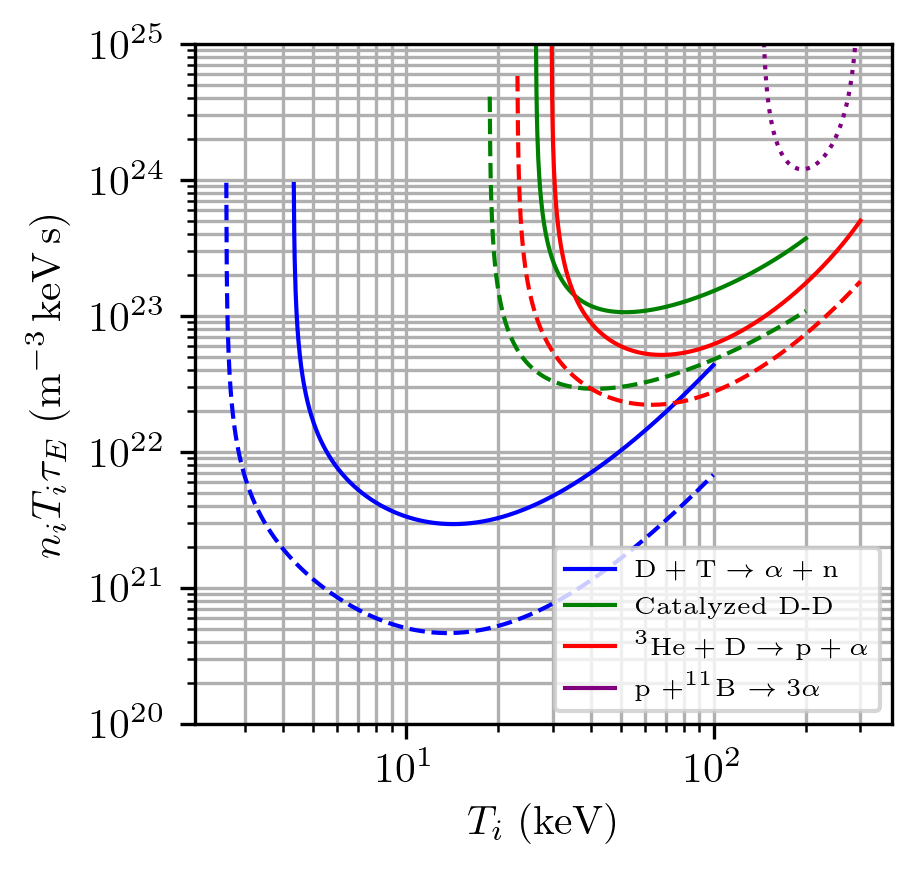}
\caption{Required (a) Lawson parameters and (b) triple products vs.\ $T$ to achieve $Q_{\rm fuel}=\infty$ (solid lines), $Q_{\rm fuel}=1$ (dashed lines), and $Q_{\rm fuel}=0.5$ (dotted line, p-$^{11}$B only) for the indicated fuels, assuming $T=T_e=T_i$. Neither fuel breakeven ($Q_{\rm fuel}=1$) nor ignition ($Q=\infty$) appears to be possible for p-$^{11}$B if $T_e=T_i$.}
\label{fig:all_reactions}
\end{figure}

The extreme requirements for advanced fuels compared to D-T are illustrated in Fig.~\ref{fig:all_reactions}, which shows the required Lawson parameters and triple products vs.\ $T_i$ required to achieve $Q_{\rm fuel}=1$ (dashed lines) and $Q_{\rm fuel}=\infty$ (solid lines) for all of the reactions discussed in this appendix. For all reactions except p-$^{11}$\!B, $T_i=T_e$ is assumed. For p-$^{11}$\!B, neither fuel breakeven nor ignition appears possible when $T_i=T_e$.

\section{Conceptual power plants with non-electrical recirculating power}
\label{sec:mechanical_recirculating power}
\begin{figure}[!b]
    \includegraphics[width=8cm]{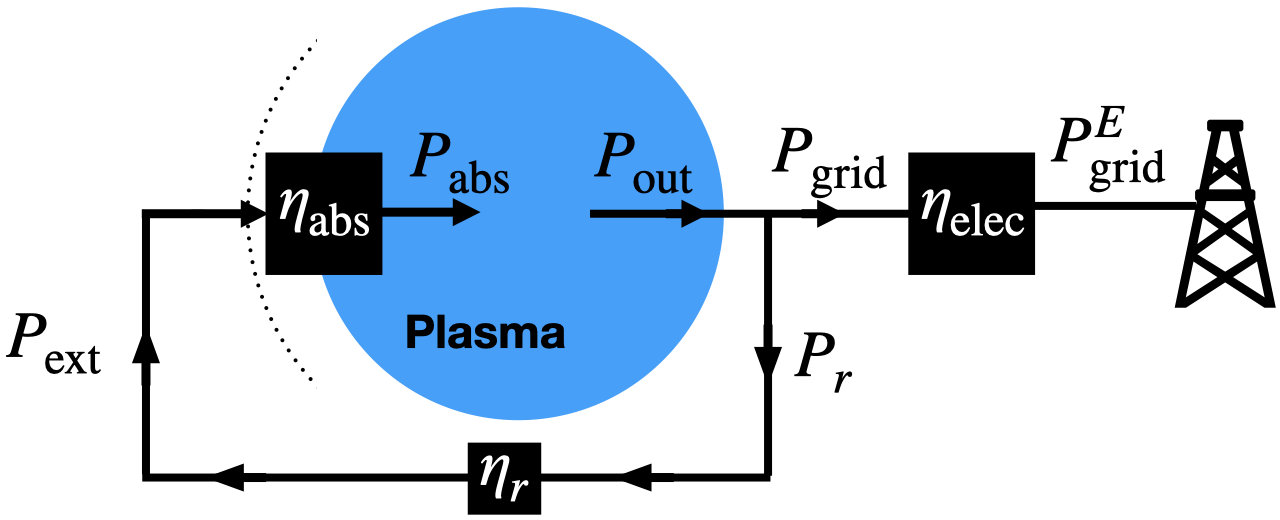}
    \caption{Conceptual schematic of a fusion power plant that recirculates mechanical power with efficiency $\eta_r$. In this system, engineering gain is defined as $Q_{\rm eng} =P_{\rm grid}^{E}/P_{r}$.}
    \label{fig:conceptual_plant_non_electrical_recirculating}
\end{figure}

Some fusion designs do not recirculate electrical power but rather capture a portion of the thermal $P_{\rm out}$ via mechanical means and use it with efficiency $\eta_r$ as $P_{\rm ext}$. This is illustrated in Fig.~\ref{fig:conceptual_plant_non_electrical_recirculating}.
An example of this approach is the compression of plasma by an imploding liquid-metal vortex driven by compressed-gas pistons,\cite{laberge19jfe} which recapture a fraction of $P_{\rm out}$ to re-energize the pistons with efficiency $\eta_r$ for the next pulse. If we define engineering gain in this system as the ratio of electrical power to the grid to recirculating mechanical power, then $Q_{\rm eng} =P_{\rm grid}^{E}/P_{r}$, and it is straightforward to show that
\begin{equation}
    \label{eq:engineering_gain_appendix}
    Q_{\rm eng} = \eta_{\rm elec}\eta_r(Q_{\rm sci} +1) - \eta_{\rm elec}.
\end{equation}

\begin{figure}[!tb]
    \includegraphics[width=8cm]{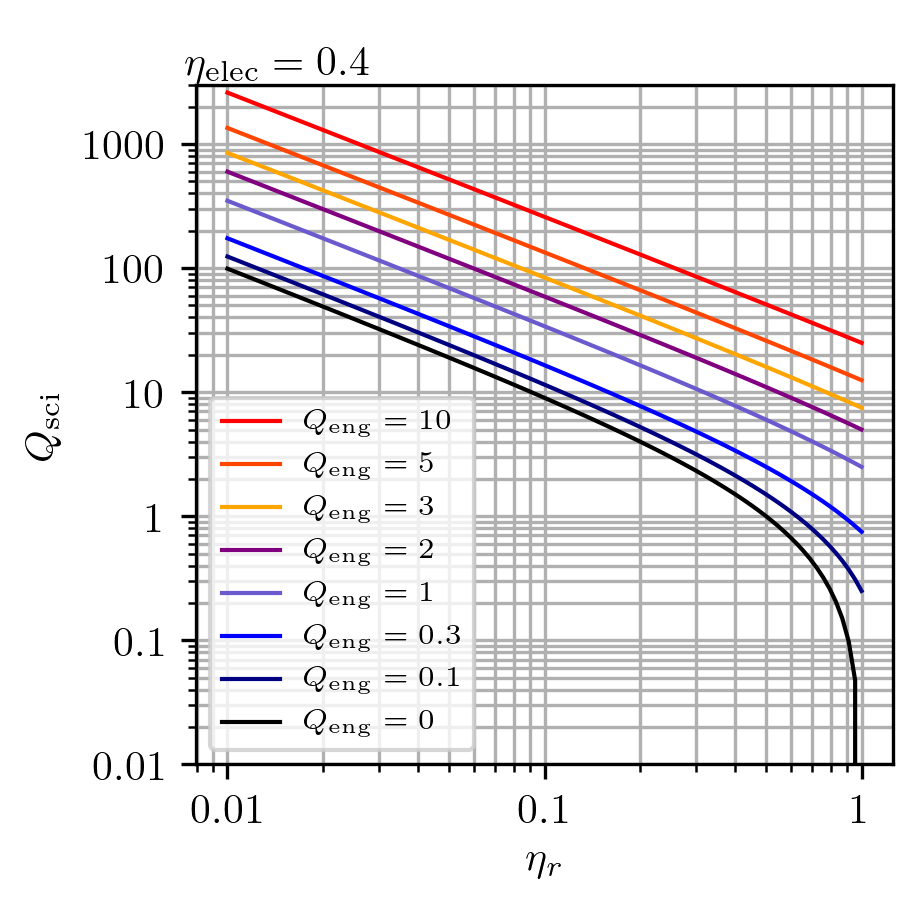}
    \caption{Required combinations of $Q_{\rm sci}$ and $\eta_r$ in the system shown in Fig.~\ref{fig:conceptual_plant_non_electrical_recirculating} to permit values of $Q_{\rm eng}$ ranging from zero (i.e., $P_{\rm grid}^{E}=0$) to ten (i.e., $P_{\rm grid}^{E}=10P_{r}$), where $\eta_{\rm elec}=0.4$ is assumed. Note that at high $\eta_r$, net electricity ($Q_{\rm eng} > 0$) is possible with $Q_{\rm sci}<1$ even though $\eta_{\rm elec}$ is only 0.4, corresponding to D-T fuel and a standard steam cycle.}
    \label{fig:Qeng_appendix}
\end{figure}
This approach has the advantage that net electricity can be generated ($Q_{\rm eng} > 0$) with $Q_{\rm sci} < 1$ if the recirculating efficiency $\eta_r$ is sufficiently high, without advanced fuels or direct conversion (i.e., assuming D-T fuel and a standard steam cycle $\eta_{\rm elec}=0.4$). This is due to the fact that the recirculating power bypasses the conversion to electricity. 

\section{Relationships between peak and volume-averaged quantities for MCF}
\label{sec:volume_averaging}

In this appendix, we describe the equations
used for volume averaging of plasma parameters
for MCF, for
the purpose of relating peak values (variables denoted
with a subscript of `0') to their volume-averaged quantities
(denoted with $\langle ...\rangle$) to, ultimately,
relating the peak $n_0 T_0 \tau_E$ to an overall $Q_{\rm fuel}$
that accounts for profile effects in $n$ and $T$.  We
denote this as $\langle Q \rangle$, even though $Q_{\rm fuel}$
is inherently a volume-averaged quantity.

For any quantity $f(x,y)$, such as $n$ or $T$,
the volume average of $f$ over the plasma cross-sectional
surface $S$ (in the $x$-$y$ plane) is
\begin{equation}
    \langle f \rangle = \frac{\int\!\!\int_S f(x,y)\,\mathrm{d}S}{A},
    \label{eq:<f>}
\end{equation}
where $A=\int\!\!\int_S \mathrm{d}S$ is the area
(inside the separatrix or last closed flux surface),
and axisymmetry is assumed.  

\subsection{Cylinder or large-aspect-ratio torus}

For a circular cylinder
with radius $a$ or a 
torus with inverse aspect ratio $\epsilon=a/R\ll 1$ (where $a$ and $R$ are the minor and major
radii, respectively), and $f(x,y)=f(r)$ (i.e., circular, concentric
flux surfaces with no Shafranov shift), Eq.~(\ref{eq:<f>}) becomes
\begin{equation}
    \langle f \rangle = \frac{2\int_0^a r f(r)\,\mathrm{d}r}{a^2}.
    \label{eq:<f>cylinder}
\end{equation}
For the particular profile
\begin{equation} 
    f(x,y)=f(r)=f_0\left[1-\left(\frac{r}{a}\right)^2
    \right]^{S_f},
    \label{eq:f(r)}
\end{equation}
where $r=(x^2+y^2)^{1/2}$,
Eq.~(\ref{eq:<f>cylinder}) becomes
\begin{equation}
     \langle f \rangle = \frac{2f_0\int_0^a r \left[1-(r/a)^2\right]^{S_f}\mathrm{d}r}{a^2} = \frac{f_0}{1+S_f}.
\end{equation}
If $n=n_0[1-(r/a)^2]^{\nu_{n}}$ and $T=T_0[1-(r/a)^2]^{\nu_{T}}$,
then it follows that
\begin{equation}
    \langle nT \rangle = \frac{n_0 T_0}{1+\nu_{n} + \nu_{T}}.
\end{equation}

\subsection{Arbitrary aspect-ratio torus}

\begin{figure}[tb!]
\includegraphics[width=8cm]{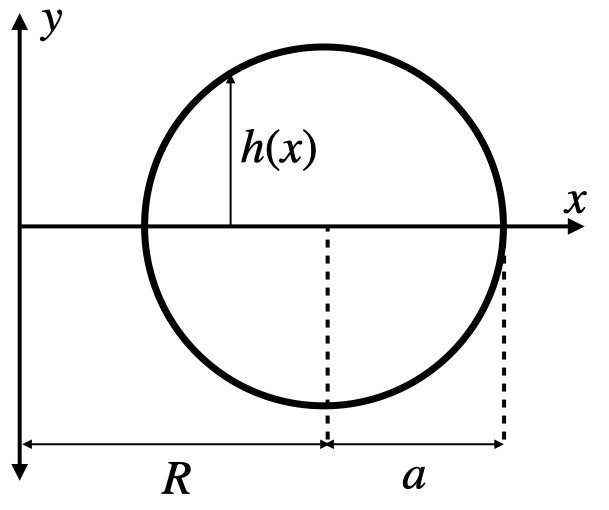}
\caption{Cross section of up-down symmetric torus with upper boundary defined by $h(x)$ (shown here as a semi-circle).}
\label{fig:torus_cross_section}
\end{figure}

For an up/down-symmetric torus with arbitrary $\epsilon$ and $f(x,y)$, Eq.~(\ref{eq:<f>}) becomes
\begin{equation}
    \langle f \rangle = \frac{\int_{R-a}^{R+a} \int_0^{h(x)} x f(x,y)\,\mathrm{d}y\,\mathrm{d}x}{\int_{R-a}^{R+a} x h(x)\,\mathrm{d}x},
    \label{eq:<f>torus}
\end{equation}
where $h(x)$ is the half height of the plasma cross section at
horizontal position $x$ as shown in Figure \ref{fig:torus_cross_section}.
If $h(x)$ and $f(x,y)=f_0 \bar{f}(x,y)$ are specified, where
$f_0$ is the peak value of $f$ and
$\mathrm{max}(\bar{f})=1$, then Eq.~(\ref{eq:<f>torus}) can
be numerically integrated to provide a quantitative
relationship between $\langle f \rangle$ and $f_0$.  The
function $h(x)$ allows for any plasma cross-sectional shape, e.g., the highly elongated, D-shaped flux surfaces of high-performance tokamaks.

For the particular case of an up/down-symmetric torus with circular cross section and $f(x,y)$ as given in Eq.~(\ref{eq:f(r)}),
where $r=[(x-R)^2 + y^2)^{1/2}$, Eq.~(\ref{eq:<f>torus}) becomes
\begin{equation}
    \langle f \rangle = \frac{f_0\int_{R-a}^{R+a}\!\int_{0}^{h(x)} x \{1-[(x-R)^2 + y^2]/a^2\}^{S_f}\,\mathrm{d}y\,\mathrm{d}x}{\int_{R-a}^{R+a}xh(x)\,\mathrm{d}x},
\end{equation}
where $h(x)=[a^2-(x-R)^2]^{1/2}$. 
Again, this can be integrated numerically to provide a relationship
between $\langle f \rangle$ and $f_0$.

\begin{acknowledgements}
Most of the first author's contributions were performed while affiliated with Fusion Energy Base prior to joining ARPA-E\@.
We are grateful for feedback on drafts of this paper provided by Riccardo Betti, Rob Goldston, Rich Hawryluk, Omar Hurricane, Harry McLean, Dale Meade, Bob Mumgaard, Brian Nelson, Kyle Peterson, Uri Shumlak, and Glen Wurden.  Responsibility for all content in the paper lies with the authors.
Reference herein to any specific non-federal person or commercial entity, product, process, or service by trade name, trademark, manufacturer, or otherwise, does not necessarily constitute or imply its endorsement, recommendation, or favoring by the U.S. Government or any agency thereof or its contractors or subcontractors.
\end{acknowledgements}


\bibliography{lawson-bibliography}   

\end{document}

%% file: table_1.tex
\begin{longtable}{l p{6cm}}
\caption{Definitions of variables used in this paper.}\label{tab:glossary}\\
\hline\noalign{\smallskip}
                        Variable &                                                                                                                                                                         Definition \\
\noalign{\smallskip}\hline\noalign{\smallskip}
\endhead
\noalign{\smallskip}\hline\noalign{\smallskip}
\multicolumn{2}{r}{{Continued on next page}} \\
\noalign{\smallskip}\hline\noalign{\smallskip}
\endfoot

\noalign{\smallskip}\hline
\endlastfoot
                           $T_i$ &                                                                                                                                                                    Ion temperature \\
                           $T_e$ &                                                                                                                                                               Electron temperature \\
                        $T_{i0}$ &                                                                                                                                                            Central ion temperature \\
                        $T_{e0}$ &                                                                                                                                                       Central electron temperature \\
   $\langle T_i \rangle_{\rm n}$ &                                                                                                                                                   Neutron-averaged ion temperature \\
                             $T$ &                                                                                            Generic temperature, used to refer to either ion or electron temperature when $T_i=T_e$ \\
                           $n_i$ &                                                                                                                                                                        Ion density \\
                           $n_e$ &                                                                                                                                                                   Electron density \\
                        $n_{i0}$ &                                                                                                                                                                Central ion density \\
                        $n_{e0}$ &                                                                                                                                                           Central electron density \\
                             $n$ &                                                                        Generic density, used to refer to either ion or electron density when $n_i=n_e$ in a pure hydrogenic plasma \\
                          $\tau$ &                                                                                                                                                                     Pulse duration \\
                        $\tau_E$ &                                                                                                                                                            Energy confinement time \\
                $\tau_{\rm eff}$ &                                           Effective characteristic time combining pulse duration and energy confinement time, see Sec.~\ref{sec:extending_lawsons_second_scenario} \\
                      $\tau_E^*$ &                                                    Modified energy confinement time, which accounts for for transient heating, see Sec.~\ref{sec:accounting_for_transient_effects} \\
                             $p$ &                                                                                                                                                            Plasma thermal pressure \\
                             $V$ &                                                                                                                                                                      Plasma volume \\
 $\langle \sigma v \rangle_{ij}$ &  Temperature-dependent fusion reactivity between species $i$ and $j$ (cross section $\sigma$ times relative velocity $v$ of ions averaged over a Maxwellian velocity distribution) \\
                    $\epsilon_F$ &                                                                                                                                          Total energy released per fusion reaction \\
             $\epsilon_{\alpha}$ &                                                                                                                       Energy released in $\alpha$-particle per D-T fusion reaction \\
                           $f_c$ &                                                                                                                            Energy fraction of fusion products in charged particles \\
                         $P_{F}$ &                                                                                                                                                                       Fusion power \\
                         $S_{F}$ &                                                                                                                                                               Fusion power density \\
                         $P_{c}$ &                                                                                                                                          Fusion power emitted as charged particles \\
                         $S_{c}$ &                                                                                                                                          Fusion power density in charged particles \\
                         $P_{n}$ &                                                                                                                                                   Fusion power emitted as neutrons \\
                         $P_{B}$ &                                                                                                                                                               Bremsstrahlung power \\
                         $S_{B}$ &                                                                                                                                                       Bremsstrahlung power density \\
                   $P_{\rm ext}$ &                                                                                                                                                   Externally applied heating power \\
                   $P_{\rm abs}$ &                                                                                                                                          Externally applied power absorbed by fuel \\
                   $E_{\rm abs}$ &                                                                                                                                         Externally applied energy absorbed by fuel \\
                   $P_{\rm out}$ &                                                                                                                                                Sum of all power exiting the plasma \\
                             $Z$ &                                                                                                                                                             Charge state of an ion \\
                       $\bar{Z}$ &                                                                                                Mean charge state, i.e., ratio of electron to ion density in a quasi-neutral plasma \\
                   $Z_{\rm eff}$ &                                         Effective value of charge state. Factor by which bremsstrahlung is increased as compared to a hydrogenic plasma, see Eq.~(\ref{eq:Z_eff}). \\
                          $\eta$ &                                                                 Efficiency of recapturing thermal energy at the conclusion of the confinement duration in Lawson's second scenario \\
                      $\eta_{E}$ &                                                                                        Efficiency of converting electrical recirculating power to externally applied heating power \\
                $\eta_{\rm abs}$ &                                                                                                                        Efficiency of coupling externally applied power to the fuel \\
                 $\eta_{\rm hs}$ &                                                                                      Efficiency of coupling shell kinetic energy to hotspot thermal energy in laser ICF implosions \\
                 $\eta_{\rm th}$ &                                                                                                                         Efficiency of converting total output power to electricity \\
                  $Q_{\rm fuel}$ &                                                                                                                     Fuel gain. Ratio of fusion power to power absorbed by the fuel \\
  $\langle Q_{\rm fuel} \rangle$ &                                                                                                                      Volume-averaged fuel gain in the case of non-uniform profiles \\
                   $Q_{\rm sci}$ &                                                                                                         Scientific gain. Ratio of fusion power to externally applied heating power \\
   $\langle Q_{\rm sci} \rangle$ &                                                                                                                Volume-averaged scientific gain in the case of non-uniform profiles \\
                   $Q_{\rm eng}$ &                                                                                                     Engineering gain. Ratio of electrical power to the grid to recirculating power \\
                    $Q_{\rm wp}$ &                                                                                                      Wall-plug gain. Ratio of fusion power to input electrical power from the grid \\
                             $Q$ &                                                            Generic energy gain. For MCF, this can refer to $Q_{\rm fuel}$ or $Q_{\rm sci}$. For ICF, this refers to $Q_{\rm sci}$. \\
\end{longtable}

%% file: table_2.tex
\begin{table}
\centering
\caption{Values of minimum $n_i \tau_E$ and corresponding $T$ for $Q_{\rm fuel}=1$ and $Q_{\rm fuel}=\infty$, for different fusion fuels assuming $T=T_i=T_e$, based on Eq.~(\ref{eq:MCF_Lawson_parameter_Q_fuel}) for D-T (see Appendix~\ref{sec:advanced_fuels} for advanced fuels).}
\label{tab:minimum_lawson_parameter_table}
\begin{tabular}{lrll}
\hline\noalign{\smallskip}
              Reaction & $Q_{\rm fuel}$ & \thead{$T$ \\ (\si{keV})} & \thead{$n_i \tau_E$ \\ (\si{m^{-3} s})} \\
\noalign{\smallskip}\hline\noalign{\smallskip}
      $\mathrm{D + T}$ &              1 &                        26 &                           \num{2.5e+19} \\
      $\mathrm{D + T}$ &       $\infty$ &                        26 &                           \num{1.6e+20} \\
         Catalyzed D-D &              1 &                       107 &                           \num{4.8e+20} \\
         Catalyzed D-D &       $\infty$ &                       106 &                           \num{1.5e+21} \\
 $\mathrm{D + ^{3}He}$ &              1 &                       106 &                           \num{2.8e+20} \\
 $\mathrm{D + ^{3}He}$ &       $\infty$ &                       106 &                           \num{6.2e+20} \\
 $\mathrm{p + ^{11}B}$ &              1 &                         – &                                       – \\
 $\mathrm{p + ^{11}B}$ &       $\infty$ &                         – &                                       – \\
\noalign{\smallskip}\hline
\end{tabular}
\end{table}

%% file: table_3.tex
\begin{table}
\centering
\caption{Values of minimum $n_i T \tau_E$ and corresponding $T$ for $Q_{\rm fuel}=1$ and $Q_{\rm fuel}=\infty$ for different fusion fuels assuming $T=T_i=T_e$, based on Eq.~(\ref{eq:triple_product_steady_state}) for D-T (see Appendix~\ref{sec:advanced_fuels} for advanced fuels).}
\label{tab:minimum_triple_product_table}
\begin{tabular}{lrll}
\hline\noalign{\smallskip}
              Reaction & $Q_{\rm fuel}$ & \thead{$T$ \\ (\si{keV})} & \thead{$n_i T \tau_E$ \\ (\si{m^{-3} keV s})} \\
\noalign{\smallskip}\hline\noalign{\smallskip}
        $\mathrm{D+T}$ &              1 &                        14 &                                 \num{4.6e+20} \\
        $\mathrm{D+T}$ &       $\infty$ &                        14 &                                 \num{2.9e+21} \\
         Catalyzed D-D &              1 &                        41 &                                 \num{2.9e+22} \\
         Catalyzed D-D &       $\infty$ &                        52 &                                 \num{1.1e+23} \\
 $\mathrm{D + ^{3}He}$ &              1 &                        63 &                                 \num{2.2e+22} \\
 $\mathrm{D + ^{3}He}$ &       $\infty$ &                        68 &                                 \num{5.2e+22} \\
 $\mathrm{p + ^{11}B}$ &              1 &                         – &                                             – \\
 $\mathrm{p + ^{11}B}$ &       $\infty$ &                         – &                                             – \\
\noalign{\smallskip}\hline
\end{tabular}
\end{table}

%% file: table_4.tex
\begin{table}
\centering
\caption{Typical efficiency values $\eta_{E}$, $\eta_{\rm abs}$, $\eta_{\rm hs}$, and $\eta_{\rm elec}$
                              for different classes of fusion concepts. Note $\eta_{\rm hs}$ is only defined for ICF concepts
                              pursuing hot spot ignition. Approximate values of $\eta_{\rm abs}$ and $\eta_{\rm hs}$ for
                              direct and indirect drive ICF are from Ref. \onlinecite{Craxton2015} and
                              Ref. \onlinecite{Zylstra_2021}, respectively.
                              }
\label{tab:efficiency_table}
\begin{tabular}{lllll}
\hline\noalign{\smallskip}
                      Class &           $\eta_{E}$ &     $\eta_{\rm abs}$ &      $\eta_{\rm hs}$ &    $\eta_{\rm elec}$ \\
\noalign{\smallskip}\hline\noalign{\smallskip}
                        MCF &                  0.7 &                  0.9 &                    - &                  0.4 \\
                        MIF &                  0.9 &                  0.1 &                    - &                  0.4 \\
   Laser ICF (direct drive) &                  0.1 &                 0.06 &                  0.4 &                  0.4 \\
 Laser ICF (indirect drive) &                  0.1 &                0.009 &                  0.7 &                  0.4 \\
\noalign{\smallskip}\hline
\end{tabular}
\end{table}

%% file: table_5.tex
\begin{table}
\centering
\caption{Peaking values required to convert reported volume-averaged quantities to peak value quantities.}
\label{tab:mcf_peaking_values_table}
\begin{tabular}{lrrl}
\hline\noalign{\smallskip}
           Concept & $T_0 / \langle T \rangle$ & $n_0 / \langle n \rangle$ &                                 Reference \\
\noalign{\smallskip}\hline\noalign{\smallskip}
           Tokamak &                       2.0 &                       1.5 &                 \onlinecite{Angioni_2009} \\
       Stellarator &                       3.0 &                       1.0 &               \onlinecite{Sheffield_2016} \\
 Spherical Tokamak &                       2.1 &                       1.7 &                  \onlinecite{Buxton_2019} \\
               FRC &                       1.0 &                       1.3 &  \onlinecite{Slough_1995,Steinhauer_2018} \\
               RFP &                       1.2 &                       1.2 &                 \onlinecite{Chapman_2002} \\
         Spheromak &                       2.0 &                       1.5 &                    \onlinecite{Hill_2000} \\
\noalign{\smallskip}\hline
\end{tabular}
\end{table}

%% file: table_6.tex
\begin{sidewaystable*}[p]
\centering
\caption{Data for tokamaks, spherical tokamaks, and stellarators.}
\label{tab:mainstream_mcf_data_table}
\begin{tabular*}{\textwidth}{@{\extracolsep{\fill}}llrlllllllll}
\hline\noalign{\smallskip}
   Project &            Concept &  Year &                     Shot Identifier &                                                                          Reference & \thead{$T_{i0}$ \\ (\si{keV})} & \thead{$T_{e0}$ \\ (\si{keV})} & \thead{$n_{i0}$ \\ (\si{m^{-3}})} & \thead{$n_{e0}$ \\ (\si{m^{-3}})} & \thead{$\tau_{E}^{*}$ \\ (\si{s})} & \thead{$n_{i0} \tau_{E}^{*}$ \\ (\si{m^{-3}~s})} & \thead{$n_{i0} T_{i0} \tau_{E}^{*}$ \\ (\si{keV~m^{-3}~s})} \\
\noalign{\smallskip}\hline\noalign{\smallskip}
       T-3 &            Tokamak &  1969 &  $H_z=25$kOe, $I_z=85$kA discharges &                                  \onlinecite{Peacock_1969,Mirnov_1969,Mirnov_1970} &                      \num{0.3} &                     \num{1.05} &       \num{2.25e+19}$^{\ddagger}$ &                    \num{2.25e+19} &                        \num{0.003} &                                    \num{6.8e+16} &                                               \num{2.0e+16} \\
        ST &            Tokamak &  1971 &                 12cm limiter, 56 kA &                                                           \onlinecite{Dimock_1971} &                      \num{0.5} &                     \num{1.14} &        \num{9.2e+19}$^{\ddagger}$ &                     \num{9.2e+19} &                       \num{0.0101} &                                    \num{9.3e+17} &                                               \num{4.6e+17} \\
        ST &            Tokamak &  1972 &                             Unknown &                                                          \onlinecite{Stodiek_1985} &                      \num{0.4} &                      \num{0.8} &          \num{6e+19}$^{\ddagger}$ &                       \num{6e+19} &                         \num{0.01} &                                    \num{6.0e+17} &                                               \num{2.4e+17} \\
       TFR &            Tokamak &  1974 &                  Molybdenum limiter &                                                         \onlinecite{TFRGroup_1985} &                     \num{0.95} &                      \num{1.8} &        \num{7.1e+19}$^{\ddagger}$ &                     \num{7.1e+19} &                        \num{0.019} &                                    \num{1.3e+18} &                                               \num{1.3e+18} \\
       PLT &            Tokamak &  1976 &                           22149-231 &                                                            \onlinecite{Grove_1976} &                     \num{1.54} &                     \num{1.86} &        \num{5.2e+19}$^{\ddagger}$ &                     \num{5.2e+19} &                         \num{0.04} &                                    \num{2.1e+18} &                                               \num{3.2e+18} \\
 Alcator A &            Tokamak &  1978 &                             Unknown &                                                      \onlinecite{Gondhalekar_1978} &                      \num{0.9} &                      \num{0.9} &          \num{1e+21}$^{\ddagger}$ &                       \num{1e+21} &                         \num{0.03} &                                    \num{3.0e+19} &                                               \num{2.7e+19} \\
      W7-A &        Stellarator &  1980 &                        Zero current &                                                         \onlinecite{Bartlett_1980} &                    \num{0.545} &                    \num{0.316} &        \num{9.6e+19}$^{\ddagger}$ &                     \num{9.6e+19} &                       \num{0.0165} &                                    \num{1.6e+18} &                                               \num{8.6e+17} \\
       TFR &            Tokamak &  1981 &                      Iconel limiter &                                                         \onlinecite{TFRGroup_1985} &                     \num{0.95} &                      \num{1.2} &       \num{1.61e+20}$^{\ddagger}$ &                    \num{1.61e+20} &                        \num{0.034} &                                    \num{5.5e+18} &                                               \num{5.2e+18} \\
       TFR &            Tokamak &  1982 &                      Carbon limiter &                                                         \onlinecite{TFRGroup_1985} &                     \num{0.95} &                      \num{1.5} &          \num{9e+19}$^{\ddagger}$ &                       \num{9e+19} &                        \num{0.025} &                                    \num{2.2e+18} &                                               \num{2.1e+18} \\
 Alcator C &            Tokamak &  1984 &                             Unknown &                                                        \onlinecite{Greenwald_1984} &                      \num{1.5} &                      \num{1.5} &                       \num{2e+21} &                       \num{2e+21} &                        \num{0.052} &                                    \num{1.0e+20} &                                               \num{1.6e+20} \\
     ASDEX &            Tokamak &  1988 &                            23349-57 &                                                          \onlinecite{Soldner_1998} &                      \num{0.8} &                      \num{1.0} &     \num{7.50e+19}$^{\ddagger *}$ &                                 – &                         \num{0.12} &                                    \num{9.0e+18} &                                               \num{7.2e+18} \\
       JET &            Tokamak &  1991 &                               26087 &                                                         \onlinecite{1992_Jet_Team} &                     \num{18.6} &                     \num{10.5} &                     \num{4.1e+19} &                     \num{5.1e+19} &                   \num{0.8}$^{\#}$ &                                    \num{3.3e+19} &                                               \num{6.1e+20} \\
       JET &            Tokamak &  1991 &                               26095 &                                                         \onlinecite{1992_Jet_Team} &                     \num{22.0} &                     \num{11.9} &                     \num{3.4e+19} &                     \num{4.5e+19} &                   \num{0.8}$^{\#}$ &                                    \num{2.7e+19} &                                               \num{6.0e+20} \\
       JET &            Tokamak &  1991 &                               26148 &                                                         \onlinecite{1992_Jet_Team} &                     \num{18.8} &                      \num{9.9} &                     \num{2.4e+19} &                     \num{3.6e+19} &                   \num{0.6}$^{\#}$ &                                    \num{1.4e+19} &                                               \num{2.7e+20} \\
      TFTR &            Tokamak &  1994 &                               76778 &                                                         \onlinecite{Hawryluk_1999} &                     \num{44.0} &                     \num{11.5} &                     \num{6.3e+19} &                     \num{8.5e+19} &                  \num{0.19}$^{\#}$ &                                    \num{1.2e+19} &                                               \num{5.3e+20} \\
    JT-60U &            Tokamak &  1994 &                               17110 &                                                             \onlinecite{Mori_1994} &                     \num{37.0} &                     \num{12.0} &                     \num{4.2e+19} &                     \num{5.5e+19} &                   \num{0.3}$^{\#}$ &                                    \num{1.3e+19} &                                               \num{4.7e+20} \\
      TFTR &            Tokamak &  1994 &                               80539 &                                                         \onlinecite{Hawryluk_1999} &                     \num{36.0} &                     \num{13.0} &                     \num{6.7e+19} &                    \num{1.02e+20} &                  \num{0.17}$^{\#}$ &                                    \num{1.1e+19} &                                               \num{4.1e+20} \\
      TFTR &            Tokamak &  1994 &                               68522 &                                                         \onlinecite{Hawryluk_1999} &                     \num{29.0} &                     \num{11.7} &                     \num{6.8e+19} &                     \num{9.6e+19} &                  \num{0.18}$^{\#}$ &                                    \num{1.2e+19} &                                               \num{3.5e+20} \\
      TFTR &            Tokamak &  1995 &                               83546 &                                                         \onlinecite{Hawryluk_1999} &                     \num{43.0} &                     \num{12.0} &                     \num{6.6e+19} &                     \num{8.5e+19} &                  \num{0.28}$^{\#}$ &                                    \num{1.8e+19} &                                               \num{7.9e+20} \\
    JT-60U &            Tokamak &  1996 &                              E26949 &                                                         \onlinecite{Ushigusa_1996} &                     \num{35.5} &                     \num{11.0} &                     \num{4.3e+19} &                    \num{5.85e+19} &                  \num{0.28}$^{\#}$ &                                    \num{1.2e+19} &                                               \num{4.3e+20} \\
    JT-60U &            Tokamak &  1996 &                              E26939 &                                                         \onlinecite{Ushigusa_1996} &                     \num{45.0} &                     \num{10.6} &                    \num{4.35e+19} &                       \num{6e+19} &                  \num{0.26}$^{\#}$ &                                    \num{1.1e+19} &                                               \num{5.1e+20} \\
       JET &            Tokamak &  1997 &                               42976 &                                                       \onlinecite{Keilhacker_1999} &                     \num{28.0} &                     \num{14.0} &                     \num{3.3e+19} &                     \num{4.1e+19} &                  \num{0.51}$^{\#}$ &                                    \num{1.7e+19} &                                               \num{4.7e+20} \\
    DIII-D &            Tokamak &  1997 &                               87977 &                                                          \onlinecite{Lazarus_1997} &                     \num{18.1} &                      \num{7.5} &                     \num{8.5e+19} &                       \num{1e+20} &                  \num{0.24}$^{\#}$ &                                    \num{2.0e+19} &                                               \num{3.7e+20} \\
     START &  Spherical Tokamak &  1998 &                               35533 &                                                            \onlinecite{Sykes_1999} &          \num{0.2}$^{\dagger}$ &                      \num{0.2} &     \num{1.02e+20}$^{\ddagger *}$ &                                 – &                        \num{0.003} &                                    \num{3.1e+17} &                                               \num{6.1e+16} \\
    JT-60U &            Tokamak &  1998 &                              E31872 &                                                           \onlinecite{Fujita_1999} &                     \num{16.8} &                      \num{7.2} &                     \num{4.8e+19} &                     \num{8.5e+19} &                  \num{0.69}$^{\#}$ &                                    \num{3.3e+19} &                                               \num{5.6e+20} \\
     W7-AS &        Stellarator &  2002 &                          H-NBI mode &                                                           \onlinecite{Wagner_2005} &                 \num{2.27}$^*$ &                              – &     \num{1.10e+20}$^{\ddagger *}$ &                                 – &                         \num{0.06} &                                    \num{6.6e+18} &                                               \num{1.5e+19} \\
       HSX &        Stellarator &  2005 &                   QHS configuration &                                                         \onlinecite{Anderson_2006} &         \num{0.45}$^{\dagger}$ &                     \num{0.45} &        \num{2.5e+18}$^{\ddagger}$ &                     \num{2.5e+18} &                       \num{0.0006} &                                    \num{1.5e+15} &                                               \num{6.8e+14} \\
      MAST &  Spherical Tokamak &  2006 &                               14626 &                                                            \onlinecite{Lloyd_2007} &                      \num{3.0} &                      \num{2.0} &          \num{3e+19}$^{\ddagger}$ &                       \num{3e+19} &                         \num{0.05} &                                    \num{1.5e+18} &                                               \num{4.5e+18} \\
       LHD &        Stellarator &  2008 &                 High triple product &                                                           \onlinecite{Komori_2010} &                     \num{0.47} &                     \num{0.47} &          \num{5e+20}$^{\ddagger}$ &                       \num{5e+20} &                         \num{0.22} &                                    \num{1.1e+20} &                                               \num{5.2e+19} \\
      NSTX &  Spherical Tokamak &  2009 &                              129041 &                                                        \onlinecite{Mansfield_2009} &                      \num{1.2} &                      \num{1.2} &          \num{5e+19}$^{\ddagger}$ &                       \num{5e+19} &                         \num{0.08} &                                    \num{4.0e+18} &                                               \num{4.8e+18} \\
     KSTAR &            Tokamak &  2014 &                                7081 &                                                              \onlinecite{Kim_2014} &                      \num{2.0} &                              – &     \num{4.80e+19}$^{\ddagger *}$ &                                 – &                          \num{0.1} &                                    \num{4.8e+18} &                                               \num{9.6e+18} \\
      EAST &            Tokamak &  2015 &                               41079 &                                                               \onlinecite{Hu_2015} &          \num{1.2}$^{\dagger}$ &                      \num{1.2} &          \num{2e+19}$^{\ddagger}$ &                       \num{2e+19} &                         \num{0.04} &                                    \num{8.0e+17} &                                               \num{9.6e+17} \\
     C-Mod &            Tokamak &  2016 &                          1160930042 &                                                           \onlinecite{Hughes_2018} &          \num{6.0}$^{\dagger}$ &                      \num{6.0} &          \num{2e+20}$^{\ddagger}$ &                       \num{2e+20} &                        \num{0.054} &                                    \num{1.1e+19} &                                               \num{6.5e+19} \\
     C-Mod &            Tokamak &  2016 &                          1160930033 &                                                           \onlinecite{Hughes_2018} &          \num{2.5}$^{\dagger}$ &                      \num{2.5} &        \num{5.5e+20}$^{\ddagger}$ &                     \num{5.5e+20} &                        \num{0.054} &                                    \num{3.0e+19} &                                               \num{7.4e+19} \\
   ASDEX-U &            Tokamak &  2016 &                               32305 &                                                             \onlinecite{Bock_2017} &                      \num{8.0} &                      \num{5.0} &          \num{5e+19}$^{\ddagger}$ &                       \num{5e+19} &                        \num{0.056} &                                    \num{2.8e+18} &                                               \num{2.2e+19} \\
      EAST &            Tokamak &  2017 &                               56933 &                                                             \onlinecite{Yang_2017} &                      \num{2.1} &                      \num{1.8} &        \num{8.5e+19}$^{\ddagger}$ &                     \num{8.5e+19} &                        \num{0.054} &                                    \num{4.6e+18} &                                               \num{9.6e+18} \\
      W7-X &        Stellarator &  2017 &                    W7X 20171207.006 &                                \onlinecite{Wolf_2019,Bozhenkov_2020,Baldzuhn_2020} &                      \num{3.5} &                      \num{3.5} &          \num{8e+19}$^{\ddagger}$ &                       \num{8e+19} &                         \num{0.22} &                                    \num{1.8e+19} &                                               \num{6.2e+19} \\
      EAST &            Tokamak &  2018 &                               71320 &                                                              \onlinecite{Gao_2018} &                      \num{1.8} &                      \num{1.8} &        \num{5.5e+19}$^{\ddagger}$ &                     \num{5.5e+19} &                        \num{0.035} &                                    \num{1.9e+18} &                                               \num{3.5e+18} \\
 Globus-M2 &  Spherical Tokamak &  2019 &                               37873 &                                                         \onlinecite{Bakharev_2019} &                      \num{1.2} &                              – &     \num{1.19e+20}$^{\ddagger *}$ &                                 – &                         \num{0.01} &                                    \num{1.2e+18} &                                               \num{1.4e+18} \\
     SPARC &            Tokamak &  2025 &                           Projected &                                  \onlinecite{Rodriguez-Fernandez_2020,Creely_2020} &                     \num{20.0} &                     \num{22.0} &          \num{4e+20}$^{\ddagger}$ &                       \num{4e+20} &                         \num{0.77} &                                    \num{3.1e+20} &                                               \num{6.2e+21} \\
      ITER &            Tokamak &  2035 &                           Projected &  \onlinecite{Mukhovatov_2003,Wagner_2009,Singh_2017,Meneghini_2016,Meneghini_2020} &                     \num{20.0} &                              – &          \num{1e+20}$^{\ddagger}$ &                       \num{1e+20} &                          \num{3.7} &                                    \num{3.7e+20} &                                               \num{7.4e+21} \\
\noalign{\smallskip}\hline
\end{tabular*}
                              \raggedright
                              \footnotesize{\\$*$ Peak value of density or temperature has been inferred from volume-averaged value as described in Sec.~\ref{sec:inferring_peak_from_average}.\\
$\dagger$ Ion temperature has been inferred from electron temperature as described in Sec.~\ref{sec:inferring_ion_quantities_from_electron_quantities}.\\
$\ddagger$ Ion density has been inferred from electron density as described in Sec.~\ref{sec:inferring_ion_quantities_from_electron_quantities}.\\
$\#$ Energy confinement time $\tau_E^*$ (TFTR/Lawson method) has been inferred from a measurement of the energy confinement time $\tau_E$ (JET/JT-60) method as described in Sec.~\ref{sec:accounting_for_transient_effects}.}
\end{sidewaystable*}

%% file: table_7.tex
\begin{sidewaystable*}[p]
\centering
\caption{Data for magnetic alternate concepts.}
\label{tab:alternates_mcf_data_table}
\begin{tabular*}{\textwidth}{@{\extracolsep{\fill}}llrlllllllll}
\hline\noalign{\smallskip}
     Project &    Concept &  Year &           Shot Identifier &                          Reference & \thead{$T_{i0}$ \\ (\si{keV})} & \thead{$T_{e0}$ \\ (\si{keV})} & \thead{$n_{i0}$ \\ (\si{m^{-3}})} & \thead{$n_{e0}$ \\ (\si{m^{-3}})} & \thead{$\tau_{E}^{*}$ \\ (\si{s})} & \thead{$n_{i0} \tau_{E}^{*}$ \\ (\si{m^{-3}~s})} & \thead{$n_{i0} T_{i0} \tau_{E}^{*}$ \\ (\si{keV~m^{-3}~s})} \\
\noalign{\smallskip}\hline\noalign{\smallskip}
        ZETA &      Pinch &  1957 &    140ka-180ka discharges &             \onlinecite{Butt_1959} &                     \num{0.09} &                              – &          \num{1e+20}$^{\ddagger}$ &                       \num{1e+20} &                       \num{0.0001} &                                    \num{1.0e+16} &                                               \num{9.0e+14} \\
  ETA-BETA I &        RFP &  1977 &                   Summary &         \onlinecite{Ortolani_1985} &                     \num{0.01} &                              – &                       \num{1e+21} &                                 – &                        \num{1e-06} &                                    \num{1.0e+15} &                                               \num{1.0e+13} \\
 ETA-BETA II &        RFP &  1984 &                     59611 &           \onlinecite{Bassan_1985} &         \num{0.09}$^{\dagger}$ &                     \num{0.09} &        \num{3.5e+20}$^{\ddagger}$ &                     \num{3.5e+20} &                       \num{0.0001} &                                    \num{3.5e+16} &                                               \num{3.2e+15} \\
       TMX-U &     Mirror &  1984 &                2/2/84 S21 &          \onlinecite{Dimonte_1987} &                     \num{0.15} &                    \num{0.045} &                       \num{2e+18} &                       \num{2e+18} &                        \num{0.001} &                                    \num{2.0e+15} &                                               \num{3.0e+14} \\
      ZT-40M &        RFP &  1987 &                   Unknown &           \onlinecite{Cayton_1987} &         \num{0.33}$^{\dagger}$ &                     \num{0.33} &     \num{9.60e+19}$^{\ddagger *}$ &                                 – &                       \num{0.0007} &                                    \num{6.7e+16} &                                               \num{2.2e+16} \\
         CTX &  Spheromak &  1990 &      Solid flux conserver &           \onlinecite{Jarboe_1990} &                     \num{0.18} &                     \num{0.18} &     \num{4.50e+19}$^{\ddagger *}$ &                                 – &                       \num{0.0002} &                                    \num{9.0e+15} &                                               \num{1.6e+15} \\
         LSX &        FRC &  1993 &                       s~2 &           \onlinecite{Slough_1995} &                    \num{0.547} &                    \num{0.253} &                \num{1.30e+21}$^*$ &                                 – &                       \num{0.0001} &                                    \num{1.3e+17} &                                               \num{7.1e+16} \\
         MST &        RFP &  2001 &                  390 $kA$ &          \onlinecite{Chapman_2002} &                    \num{0.396} &                    \num{0.792} &     \num{1.20e+19}$^{\ddagger *}$ &                                 – &                \num{0.0064}$^{\#}$ &                                    \num{7.7e+16} &                                               \num{3.0e+16} \\
         ZaP &    Z Pinch &  2003 &                   Unknown &          \onlinecite{Shumlak_2003} &                      \num{0.1} &                              – &          \num{9e+22}$^{\ddagger}$ &                       \num{9e+22} &                      \num{3.7e-07} &                                    \num{3.3e+16} &                                               \num{3.3e+15} \\
       FRX-L &        FRC &  2003 &                      2027 &         \onlinecite{Intrator_2004} &                     \num{0.09} &                     \num{0.09} &                       \num{4e+22} &                       \num{4e+22} &                      \num{3.3e-06} &                                    \num{1.3e+17} &                                               \num{1.2e+16} \\
         TCS &        FRC &  2005 &                      9018 &              \onlinecite{Guo_2008} &                    \num{0.025} &                    \num{0.025} &     \num{6.50e+18}$^{\ddagger *}$ &                                 – &                        \num{4e-05} &                                    \num{2.6e+14} &                                               \num{6.5e+12} \\
       FRX-L &        FRC &  2005 &                      3684 &            \onlinecite{Zhang_2006} &                 \num{0.18}$^*$ &                              – &     \num{4.81e+22}$^{\ddagger *}$ &                                 – &                      \num{3.3e-06} &                                    \num{1.6e+17} &                                               \num{2.9e+16} \\
        SSPX &  Spheromak &  2007 &                     17524 &           \onlinecite{Hudson_2008} &          \num{0.5}$^{\dagger}$ &                      \num{0.5} &     \num{2.25e+20}$^{\ddagger *}$ &                                 – &                        \num{0.001} &                                    \num{2.2e+17} &                                               \num{1.1e+17} \\
       GOL-3 &     Mirror &  2007 &                   Unknown &         \onlinecite{Burdakov_2007} &                      \num{2.0} &                      \num{2.0} &                       \num{7e+20} &                       \num{7e+20} &                       \num{0.0009} &                                    \num{6.3e+17} &                                               \num{1.3e+18} \\
     RFX-mod &        RFP &  2008 &                     24063 &           \onlinecite{Valisa_2008} &          \num{1.0}$^{\dagger}$ &                      \num{1.0} &          \num{3e+19}$^{\ddagger}$ &                       \num{3e+19} &                       \num{0.0025} &                                    \num{7.5e+16} &                                               \num{7.5e+16} \\
     RFX-mod &        RFP &  2008 &                     23962 &         \onlinecite{Piovesan_2009} &          \num{1.0}$^{\dagger}$ &                      \num{1.0} &          \num{1e+19}$^{\ddagger}$ &                       \num{1e+19} &                        \num{0.004} &                                    \num{4.0e+16} &                                               \num{4.0e+16} \\
        TCSU &        FRC &  2008 &                     21214 &              \onlinecite{Guo_2008} &                      \num{0.1} &                      \num{0.1} &     \num{1.30e+19}$^{\ddagger *}$ &                                 – &                      \num{7.5e-05} &                                    \num{9.7e+14} &                                               \num{9.7e+13} \\
         MST &        RFP &  2009 &               w/o pellets &          \onlinecite{Chapman_2009} &                      \num{1.3} &                      \num{1.9} &        \num{1.2e+19}$^{\ddagger}$ &                     \num{1.2e+19} &                        \num{0.012} &                                    \num{1.4e+17} &                                               \num{1.9e+17} \\
         MST &        RFP &  2009 &                w/ pellets &          \onlinecite{Chapman_2009} &                      \num{0.6} &                      \num{0.7} &          \num{4e+19}$^{\ddagger}$ &                       \num{4e+19} &                        \num{0.007} &                                    \num{2.8e+17} &                                               \num{1.7e+17} \\
 Yingguang-I &        FRC &  2015 &                 150910-01 &              \onlinecite{Sun_2017} &                  \num{0.2}$^*$ &                              – &     \num{4.81e+22}$^{\ddagger *}$ &                                 – &                        \num{1e-06} &                                    \num{4.8e+16} &                                               \num{9.6e+15} \\
        C-2U &        FRC &  2017 &                     46366 &  \onlinecite{Baltz_2017,Gota_2017} &                 \num{0.68}$^*$ &                              – &     \num{2.47e+19}$^{\ddagger *}$ &                                 – &                      \num{0.00024} &                                    \num{5.9e+15} &                                               \num{4.0e+15} \\
        FuZE &    Z Pinch &  2018 &  Multiple identical shots &            \onlinecite{Zhang_2019} &                      \num{1.8} &                              – &        \num{1.1e+23}$^{\ddagger}$ &                     \num{1.1e+23} &                      \num{1.1e-06} &                                    \num{1.2e+17} &                                               \num{2.2e+17} \\
         GDT &     Mirror &  2018 &  Multiple identical shots &         \onlinecite{Yakovlev_2018} &         \num{0.45}$^{\dagger}$ &                     \num{0.45} &        \num{1.1e+19}$^{\ddagger}$ &                     \num{1.1e+19} &                       \num{0.0006} &                                    \num{6.6e+15} &                                               \num{3.0e+15} \\
        C-2W &        FRC &  2019 &                    107322 &             \onlinecite{Gota_2019} &                  \num{0.6}$^*$ &                              – &     \num{2.08e+19}$^{\ddagger *}$ &                                 – &                       \num{0.0012} &                                    \num{2.5e+16} &                                               \num{1.5e+16} \\
        C-2W &        FRC &  2019 &                    104989 &             \onlinecite{Gota_2019} &                  \num{1.0}$^*$ &                              – &     \num{1.30e+19}$^{\ddagger *}$ &                                 – &                        \num{0.003} &                                    \num{3.9e+16} &                                               \num{3.9e+16} \\
        C-2W &        FRC &  2020 &                    114534 &             \onlinecite{Gota_2021} &                  \num{1.8}$^*$ &                              – &     \num{1.30e+19}$^{\ddagger *}$ &                                 – &                       \num{0.0015} &                                    \num{2.0e+16} &                                               \num{3.5e+16} \\
        C-2W &        FRC &  2021 &                    118340 &            \onlinecite{Roche_2021} &                  \num{3.5}$^*$ &                              – &     \num{1.30e+19}$^{\ddagger *}$ &                                 – &                        \num{0.005} &                                    \num{6.5e+16} &                                               \num{2.3e+17} \\
\noalign{\smallskip}\hline
\end{tabular*}
                              \raggedright
                              \footnotesize{\\$*$ Peak value of density or temperature has been inferred from volume-averaged value as described in Sec.~\ref{sec:inferring_peak_from_average}.\\
$\dagger$ Ion temperature has been inferred from electron temperature as described in Sec.~\ref{sec:inferring_ion_quantities_from_electron_quantities}.\\
$\ddagger$ Ion density has been inferred from electron density as described in Sec.~\ref{sec:inferring_ion_quantities_from_electron_quantities}.\\
$\#$ Energy confinement time $\tau_E^*$ (TFTR/Lawson method) has been inferred from a measurement of the energy confinement time $\tau_E$ (JET/JT-60) method as described in Sec.~\ref{sec:accounting_for_transient_effects}.}
\end{sidewaystable*}

%% file: table_8.tex
\begin{sidewaystable*}[p]
\centering
\caption{Data for ICF and MIF concepts.}
\label{tab:icf_mif_data_table}
\begin{tabular*}{\textwidth}{@{\extracolsep{\fill}}llrllrrrrrllll}
\hline\noalign{\smallskip}
Project &    Concept &  Year & Shot Identifier &                                 Reference & \thead{$\langle T_i \rangle_{\rm n}$ \\ (\si{keV})} & \thead{$T_e$ \\ (\si{keV})} & \thead{$\rho R_{tot(n)}^{no (\alpha)}$ \\ (\si{g/cm^{-2}})} &  YOC & \thead{$p_{stag}$ \\ (\si{Gbar})} & \thead{$\tau_{stag}$ \\ (\si{s})} & \thead{$P\tau$ \\ (\si{atm~s})} & \thead{$n\tau$ \\ (\si{m^{-3}~s})} & \thead{$n \langle T \rangle_{\rm n} \tau$ \\ (\si{keV~m^{-3}~s})} \\
\noalign{\smallskip}\hline\noalign{\smallskip}
   NOVA &  Laser ICF &  1994 &    100 atm fill &                   \onlinecite{Cable_1994} &                                                0.90 &                           – &                                                           – &    – &                             16.00 &                       \num{5e-11} &                      \num{0.26} &                      \num{9.2e+19} &                                                     \num{8.3e+19} \\
  OMEGA &  Laser ICF &  2007 &           47206 &  \onlinecite{Sangster_2008,Sangster_2010} &                                                2.00 &                           – &                                                       0.202 &  0.1 &                                 – &                                 – &                      \num{1.23} &                      \num{1.9e+20} &                                                     \num{3.9e+20} \\
  OMEGA &  Laser ICF &  2007 &           47210 &  \onlinecite{Sangster_2008,Sangster_2010} &                                                2.00 &                           – &                                                       0.182 &  0.1 &                                 – &                                 – &                      \num{1.13} &                      \num{1.8e+20} &                                                     \num{3.6e+20} \\
  OMEGA &  Laser ICF &  2009 &         Unknown &                \onlinecite{Sangster_2010} &                                                1.80 &                           – &                                                       0.240 &  0.1 &                                 – &                                 – &                      \num{1.29} &                      \num{2.3e+20} &                                                     \num{4.1e+20} \\
  OMEGA &  Laser ICF &  2009 &           55468 &                \onlinecite{Sangster_2010} &                                                1.80 &                           – &                                                       0.300 &  0.1 &                                 – &                                 – &                      \num{1.55} &                      \num{2.7e+20} &                                                     \num{4.9e+20} \\
  OMEGA &  Laser ICF &  2013 &           69236 &               \onlinecite{Goncharov_2014} &                                                2.80 &                           – &                                                           – &    – &                             18.00 &                    \num{1.15e-10} &                      \num{0.68} &                      \num{7.7e+19} &                                                     \num{2.1e+20} \\
 MagLIF &     MagLIF &  2014 &           z2613 &                   \onlinecite{Gomez_2019} &                                                2.00 &                           – &                                                           – &    – &                              0.56 &                    \num{1.38e-09} &                      \num{0.76} &                      \num{1.2e+20} &                                                     \num{2.4e+20} \\
    NIF &  Laser ICF &  2014 &         N140304 &                  \onlinecite{LePape_2018} &                                                5.50 &                           – &                                                           – &    – &                            222.00 &                    \num{1.63e-10} &                     \num{11.86} &                      \num{6.8e+20} &                                                     \num{3.8e+21} \\
 MagLIF &     MagLIF &  2015 &           z2850 &                   \onlinecite{Gomez_2019} &                                                2.80 &                           – &                                                           – &    – &                              0.60 &                    \num{1.62e-09} &                      \num{0.96} &                      \num{1.1e+20} &                                                     \num{3.0e+20} \\
  OMEGA &  Laser ICF &  2015 &           77068 &                   \onlinecite{Regan_2016} &                                                3.60 &                           – &                                                           – &    – &                             56.00 &                     \num{6.6e-11} &                      \num{1.21} &                      \num{1.1e+20} &                                                     \num{3.8e+20} \\
    NIF &  Laser ICF &  2017 &         N170601 &                  \onlinecite{LePape_2018} &                                                4.50 &                           – &                                                           – &    – &                            320.00 &                     \num{1.6e-10} &                     \num{16.78} &                      \num{1.2e+21} &                                                     \num{5.3e+21} \\
    NIF &  Laser ICF &  2017 &         N170827 &                  \onlinecite{LePape_2018} &                                                4.50 &                           – &                                                           – &    – &                            360.00 &                    \num{1.54e-10} &                     \num{18.17} &                      \num{1.3e+21} &                                                     \num{5.7e+21} \\
  FIREX &  Laser ICF &  2019 &           40558 &                  \onlinecite{Matsuo_2020} &                                                   – &                         2.1 &                                                           – &    – &                              2.00 &                       \num{4e-10} &                      \num{0.79} &                      \num{1.2e+20} &                                                     \num{2.5e+20} \\
    NIF &  Laser ICF &  2019 &         N191007 &                 \onlinecite{Zylstra_2021} &                                                4.52 &                           – &                                                           – &    – &                            206.00 &                    \num{1.51e-10} &                     \num{10.20} &                      \num{7.1e+20} &                                                     \num{3.2e+21} \\
    NIF &  Laser ICF &  2021 &         N210808 &                 \onlinecite{2021_APS-DPP} &                                                8.94 &                           – &                                                           – &    – &                            550.00 &                     \num{8.9e-11} &                     \num{16.05} &                      \num{5.7e+20} &                                                     \num{5.1e+21} \\
\noalign{\smallskip}\hline
\end{tabular*}
\end{sidewaystable*}